\documentclass[twocolumn,citeautoscript,superscriptaddress,8pt]{revtex4-2}

\usepackage{graphicx} 
\usepackage{kotex}
\usepackage{amsmath}
\usepackage{amssymb}
\usepackage{upgreek}
\usepackage[table]{xcolor}
\definecolor{mygrey}{rgb}{0.5,0.5,0.5} 
\usepackage{hyperref}
\usepackage{sidecap}
\hypersetup{urlcolor=blue, colorlinks=true  , linkcolor=blue, citecolor=blue}

\graphicspath{ {./figures} }

\begin{document}

\title{Sub-microsecond conformational dynamics in an optical nanocavity}

\author{Heehun~Sung}
\affiliation{School of Mathematics and Physics, The University of Queensland, St. Lucia, 4072, Australia}
\affiliation{ARC Centre of Excellence in Quantum Biotechnology, St. Lucia, Queensland 4067, Australia}

\author{Sam~C~Scholten}
\affiliation{School of Mathematics and Physics, The University of Queensland, St. Lucia, 4072, Australia}
\affiliation{ARC Centre of Excellence in Quantum Biotechnology, St. Lucia, Queensland 4067, Australia}

\author{Pavlina~Sasheva}
\affiliation{School of Mathematics and Physics, The University of Queensland, St. Lucia, 4072, Australia}
\affiliation{ARC Centre of Excellence in Quantum Biotechnology, St. Lucia, Queensland 4067, Australia}

\author{Igor~Marinković}
\affiliation{School of Mathematics and Physics, The University of Queensland, St. Lucia, 4072, Australia}
\affiliation{ARC Centre of Excellence in Quantum Biotechnology, St. Lucia, Queensland 4067, Australia}

\author{Warwick P Bowen}
\email{w.bowen@uq.edu.au}
\affiliation{School of Mathematics and Physics, The University of Queensland, St. Lucia, 4072, Australia}
\affiliation{ARC Centre of Excellence in Quantum Biotechnology, St. Lucia, Queensland 4067, Australia}

\begin{abstract}

Microsecond conformational changes underlie many protein functions, but ensemble averaging has been needed to observe them without labels. Single-shot measurements on individual proteins have had insufficient speed, while protein Brownian-motion has obscured signals. Here, we report a fibre-integrated silicon-photonic sensor that overcomes these barriers,  resolving protein dynamics at sub-microsecond speeds
in continuous single-shot measurements that can extend over minutes. 
This is achieved by combining far-sub-wavelength optical field confinement with high optical field uniformity that suppresses protein Brownian-motion by a factor of sixty. 
In 
single-shot 
measurements on ferritin molecules, we observe tens of thousands of transitions consistent with conformational fluctuations of the ferritin shell, resolving them over timescales as short as $400$\,ns.
The ability to continuously monitor transitions over long times reveals switching kinetics, memory effects and molecular heterogeneity hidden in ensemble averages. This opens a new path to improved mechanistic understanding of protein function.

\end{abstract}
\maketitle 




Protein structural changes that occur over microsecond timescales underpin critical biological processes including molecular recognition~\cite{molrecogni}, enzyme catalysis~\cite{enzycatalysis}, signaling~\cite{signaling}, and regulation~\cite{allosregula}. 
Improved understanding of them is important for applications ranging from vaccine and drug development~\cite{drugdis, vaccine} to cancer therapy~\cite{cancer}, enzyme engineering~\cite{enzymeengi}, and protein design~\cite{proteindesign}. 
%
However, their short timescales and small amplitude make direct measurements exceptionally difficult. 
Instead, fluorescent labelling is generally used~\cite{Lerner2018}, but can perturb dynamics~\cite{SanchezRico2017} and cannot readily capture both rapid motions and their slower modulation by global structural changes~\cite{Miao2025, sch2021single}. Averaging is also often required~\cite{fluoesem}, obscuring heterogeneity and the stochastic fluctuations that dominate protein dynamics~\cite{hetero}.

Label-free single molecule sensing promises to overcome these challenges~\cite{wbook}, but has been unable to continuously monitor protein dynamics at the necessary speeds. Structural changes of individual unlabelled proteins have been measured using double nanohole optical tweezers~\cite{doublenanoholePSD, ferritinPH2, BSAenergy}, but with temporal resolution of only around a tenth of a millisecond~\cite{doublenanoholePSD}. 
By contrast, optoplasmonic approaches have achieved sub-microsecond resolution~\cite{nano}, but only for short lived events ($\le0.1$~ms),
relying
on ensemble averaging to infer conformational changes~\cite{WGMen}.

Here, we introduce a new label-free single molecule sensor that enables both fast measurements and continuous monitoring. A key conceptual advance is the recognition that Brownian motion of the protein is a dominant source of noise, and that this is amplified by the large optical field gradients of previous label-free single molecule sensors~\cite{ferritinPH2, nano}. 
We overcome this limitation by developing silicon-chip-based optical nanocavities that provide highly uniform optical fields, suppressing Brownian motion noise by a factor of sixty, while retaining
far-sub-wavelength field confinement.
We employ weak interactions with a protein monolayer to hold single proteins in place. 
This allows continuous observation for minutes at a time while further suppressing Brownian motion. 

We apply the sensor to probe the dynamics of ferritin, an iron-storage protein~\cite{ferrhistory}.
Our measurements reveal previously unobserved stepping dynamics that occur at sub-microsecond timescales -- well beyond the temporal resolution of previous methods~\cite{doublenanoholePSD, ferritinPH2}. 
By tracking single ferritins continuously over extended periods, we are able to measure tens of thousands of steps in a single shot. This reveals  heterogeneous dynamics that has previously been hidden by ensemble averaging, including switching kinetics and memory effects introduced by the fluctuations of the protein structure. We find that the dynamics depend strongly on both ionic concentration and the presence or absence of a ferrihydrite core. This is consistent with expectations for conformational fluctuations that are expected to arise from structural instability of the ferritin shell~\cite{ferritinPH1, ferritinPH2}.
We expect this new capability to study the dynamics of single proteins at their native speed, in single shot, unperturbed by fluorescent labels, and over long durations to provide new insights into the fast conformational processes that govern function and misfunction, with potential applications in drug discovery, molecular engineering, and disease diagnostics.
The sensors are lithographically fabricated in arrays of hundreds of devices and are fibre-integrated, making them naturally suited to scalable high-throughput measurements  such as parallelized protein detection and drug screening.

\section*{Results} 

\subsection*{Slotted photonic crystal cavity for single protein sensing}

\begin{figure*}[ht!]
    \centering
    \includegraphics[width=0.92\textwidth]{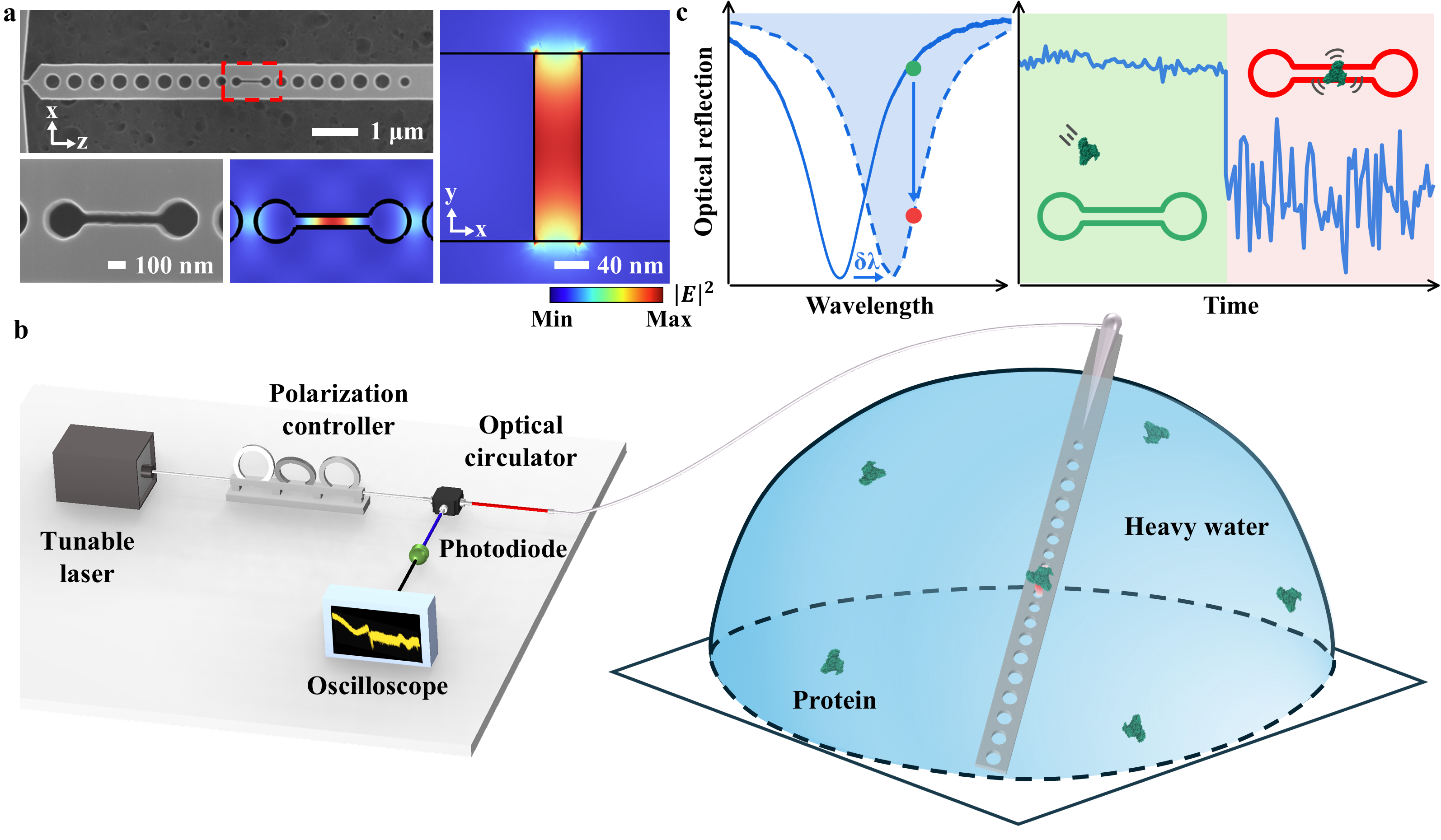} \\
    \caption{
        \textbf{Photonic crystal cavity for single molecule sensing.}
        \textbf{a},~Scanning electron microscope image (greyscale) and finite element method simulation (color) of the silicon nanobeam cavities.
        The design combines strong and uniform field confinement within the slot region.
        \textbf{b},~Experimental configuration for single protein measurements.
        A tapered fibre is bonded to individual cavities, before cleaving from the chip for measurement in solution.
        A tunable laser (1550\,nm) supplies light to the cavity, and the reflection is monitored via an optical circulator, photodiode and digital oscilloscope. 
        \textbf{c},~The cavity resonance shifts when a protein enters the slot (left).
        By tuning the laser to one side of the cavity resonance, single proteins binding events can be detected as a step in reflection (right).
        Consequent motion of the protein in the optical field whilst bound can then be measured.
        }
    \label{fig:design}
\end{figure*}

To sense single proteins, we employ the dispersive sensing mechanism~\cite{shift1}, in which the optically induced dipole moment of a protein results in a measurable angular frequency shift $\delta \omega$ of the cavity. Relative to the cavity linewidth $\Gamma$, this shift is given by~\cite{shift1, QVequ} 
\begin{equation}
{\delta \omega \over \Gamma} \sim - {Q \over 2 V}  \left ( \frac{E({\bf r})}{E({\bf r}_{\rm max})} \right )^2 \alpha \label{dispersive_eqn}
\end{equation}
where $\alpha$ is the excess polarizability of the protein, $V$, $Q$ and $E$
are the cavity volume, quality factor, and optical electric field strength respectively, 
$\bf r=(x,y,z)$ is the position of the protein, and ${\bf r}_{\rm max}$ is the position of maximum electric field. Since the structure of the protein influences its excess polarizability~\cite{Larnii}, this sensing mechanism is capable of probing structural changes. 

Motivated by Eq.~(\ref{dispersive_eqn}), much previous research on unlabelled single molecule sensing has focused on increasing $Q/V$~\cite{phcSi, mminter}.
However, this approach breaks down for the small, high-quality cavities most suited to single protein measurements.
It has recently been shown that once cavity thermal noise is larger than optical shot noise, the signal-to-noise ratio becomes independent of $Q$~\cite{TRN, detectV}.
Here, we identify a further departure due to the translational Brownian motion of the protein. This intrinsic thermal noise is typically much larger than cavity thermal noise~\cite{doublenanoholePSD}, and directly obscures structural dynamics.
When it dominates, neither increasing $Q$ nor decreasing $V$ improves signal-to-noise. 
This leads to a fundamentally different set of design principles for optical protein sensors: they should be designed to maximise optical field uniformity because, as can be seen from Eq. (\ref{dispersive_eqn}), field gradients transduce translational Brownian motion into the measured signal. 

While widely employed to detect single proteins~\cite{nano, ferritinPH2, WGMen}, the extreme field confinement provided by plasmonic and optoplasmonic sensors produces large field gradients.
Here, we instead employ photonic crystal cavities. When previously applied to protein detection, these have also suffered from high field gradients, while having weaker light--protein coupling due to their lower field confinement and because the optical field maximum was within the cavity dielectric rather than the surrounding protein-containing solution~\cite{phcSi}.
We overcome these limitations by introducing a nanoscale slot at the centre of a one-dimensional photonic crystal cavity, following Refs.~\cite{slot1, slotBis}.
Fig.~\ref{fig:design}a shows an example of one such {\it slot photonic crystal}, together with finite-element simulations of its optical field distribution. 
As can be seen, the optical field is strongly confined within the slot, enabling mode volumes much smaller than other photonic crystals~\cite{phcSi} and allowing proteins to access the optical field maximum. 
Crucially, it is highly uniform, suppressing noise due to protein Brownian motion.

\subsection*{Design and fabrication of slot photonic crystals}

\begin{figure*}[ht!]
    \centering
    \includegraphics[width=\textwidth]{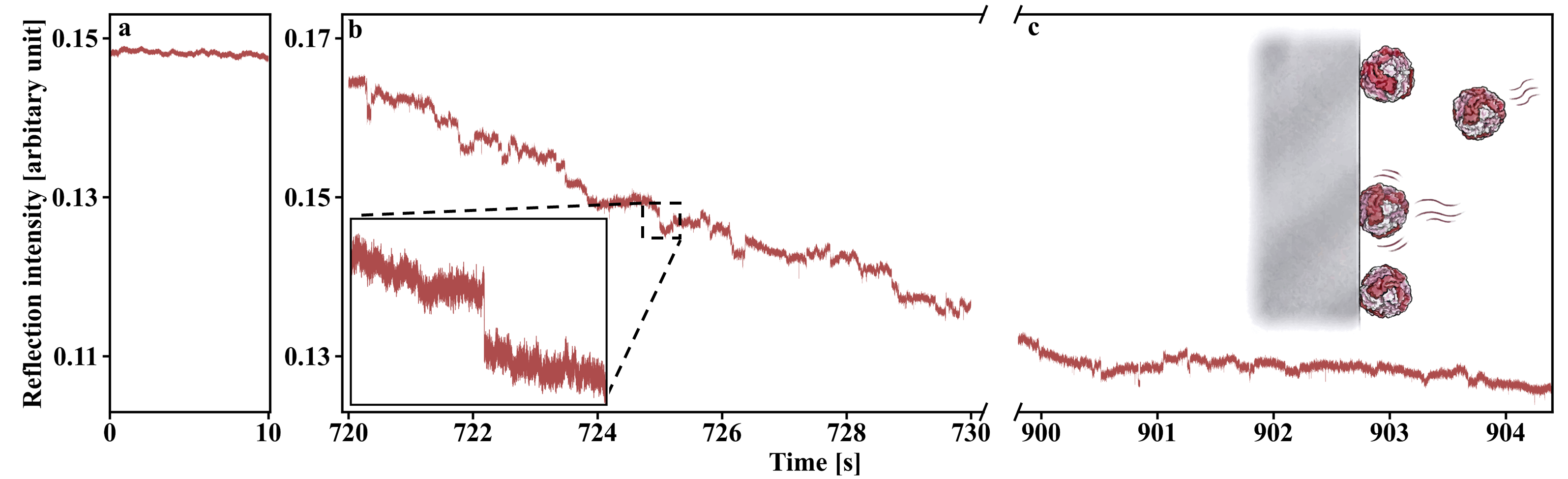}
    \caption{
    \textbf{Ferritin monolayer formation via adsorption onto the cavity surface.} 
    \textbf{a},~Signal measured prior to the addition of proteins to the solution (150\,mM NaCl). 
    \textbf{b},~After the addition of proteins (1\,nM apo-ferritin), a gradual shift in cavity resonance is observed as a monolayer of the protein is formed across the nanobeam surface.
    The inset highlights a single ferritin adsorption event within the slot region.
    \textbf{c},~Stabilized signal observed after several minutes.
    Once the monolayer has been formed and stabilised, the laser is retuned to the side of the cavity resonance for subsequent measurements.
    }
    \label{fig:mono}
\end{figure*}

We fabricate slot photonic crystals from a silicon device layer on a silicon-on-insulator wafer (Fig.~\ref{fig:design}a), using silicon for its high refractive index and scalability. 
They consist of a 220\,nm thick silicon beam with Bragg mirrors formed by a lattice of holes, and a nanoscale slot at the centre.

To optimise the  cavities for detection of protein structural changes, we perform extensive simulations (Methods, Supplementary Section I.B.a). 
From electrostatic simulations, we find an optimum slot width of around 55\,nm. 
This balances enhanced protein-light interactions for smaller slots against more effective protein entry for larger slots, with electrostatic repulsion from the charged silicon surfaces~\cite{dark} preventing entry if the slot is too small. 
We note that electrostatic screening can be used to mitigate this repulsion, but at the necessary ionic concentrations this suppresses polarizability enhancements found to increase protein signals~\cite{dark}.
For a 55\,nm slot width, finite-element modelling provides a volume of $V \approx 0.04\,(\lambda/n)^3$, where $\lambda$ is the resonance wavelength and $n$ is the refractive index of the surrounding solution.
This is more than an order of magnitude improved compared to the only previous photonic crystal cavities used for single protein sensing~\cite{phcSi}.

We perform further finite element simulations to assess the uniformity of the field within the slot (Supplementary Section I.B.b). To evaluate Brownian motion induced signal fluctuations, we define the Brownian coupling strength $g_{\text{B}}=|d(\ln |E|^2) /dx|$, which quantifies the local fractional signal variation per unit displacement of the protein. For vertical sidewalls, the calculated $g_{\text{B}}$ over the slot sidewall surface has a mean value of 0.3$\%$ per nanometre, with local values ranging from below 0.01$\%$ to 0.6$\%$ per nanometre. This is more than  a factor of sixty lower than in plasmonic nanostructures ($\approx$ 20$\%$ per nanometre)~\cite{nano}. At the point of maximum $g_{\text{B}}$ on the cavity wall surface, $g_{\text{B}}$ has a weak linear dependence on the sidewall tilt angle $\theta$ of approximately $0.1\%/\mathrm{deg}$.

We use standard silicon-on-insulator fabrication processes (Methods, Supplementary Section I.C), including an anisotropic dry etching process that minimizes lateral etching and enables the formation of vertical sidewalls. The etching conditions are optimized to eliminate residual silicon and improve sidewall quality (see Methods). After removing the underlying SiO$_2$ layer with vapor HF, we obtain a suspended slot photonic crystal with vertical sidewalls (Fig.~\ref{fig:design}a).

Silicon chip-based approaches offer attractive scaling advantages, with the ability to fabricate thousands of cavities on a single chip. 
However, a significant barrier to realising this is that, to date, measurements with any single cavity have involved immersing the entire sensing chip in solution~\cite{phcSi, onchip1}, with contamination from the solution degrading all cavities on the chip. To address this, we adopt a new approach that allows detachment of individual fibre-coupled cavities from a chip containing thousands of devices suspended by thin connection tethers~\cite{pickup}.
Individual slot photonic crystals are bonded to a tapered fibre using optical adhesive, after which the tether is mechanically cleaved  to release them (Methods, Fig.~\ref{fig:design}b). 
This approach enables measurements in a separately prepared droplet of clean solution, preventing contamination of the chip. 
Moreover, it allows the selection of high quality devices and ensures stable and efficient fibre-coupling throughout the measurement. 

\subsection*{Dynamics of weakly bound ferritin}

We employ the slot photonic crystals to observe single protein dynamics using the apparatus shown in Fig.~\ref{fig:design}b (see Methods).
Proteins are prepared dissolved in a heavy water droplet (D$_2$O) surrounding the sensor as described in the Methods. Heavy water is chosen to 
to reduce thermal effects, as its absorption coefficient at the laser wavelength of 1500\,nm is nearly two orders of magnitude lower than that of regular water (H$_2$O)~\cite{water}. Proteins induce shifts of the cavity resonance frequency both when they enter the slot and when their structure changes.
We resolve these shifts by tuning the laser onto the red side of the resonance, mapping them onto a change in reflected intensity (Fig.~\ref{fig:design}c).

Figure~\ref{fig:mono}a-c show an example trace of the measured intensity a short time after introducing ferritin into the droplet. A downwards drift is observed
as a monolayer forms on the cavity surface, with occasional discrete step-like shifts when individual ferritin molecules adsorb within the slot. 
The drift eventually stops once a full monolayer is formed (Fig.~\ref{fig:mono}c), consistent with previous experiments~\cite{corona1}. 
We term this monolayer the `hard layer' in analogy to the `hard protein corona' that forms on nanoparticles~\cite{hardcorona}. 
Proteins within the hard layer are typically irreversibly bound, denatured, and biologically inactivated, as strong interactions induce protein flattening on the surface and consequent conformational modifications~\cite{coronaReview}.

\begin{figure*}[ht!]
    \centering
    \includegraphics[width=\textwidth]{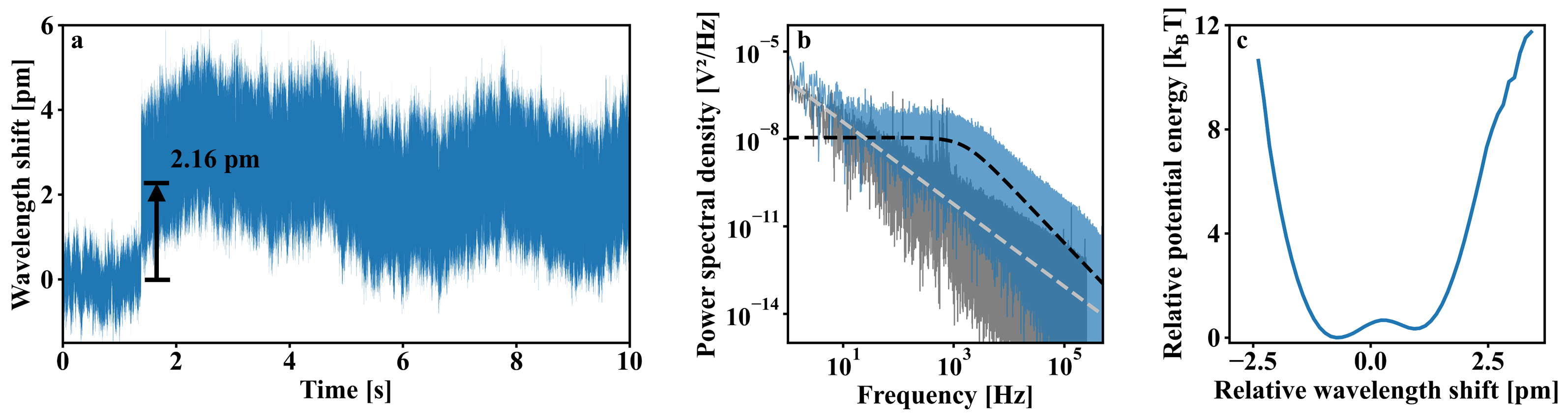} \\
    \caption{
        \textbf{Optical tracking of fast ferritin motions.}
         \textbf{a}, Recording of a single holoferritin attaching to a holoferritin monolayer.
         Compared with direct binding to the cavity, these events are rare, transient, and feature an increase in signal fluctuations.
         \textbf{b},~The corresponding power spectral density of post-binding fluctuations (blue), compared with the bare device signal (grey).
         The protein signal is fit to a Lorentzian model (black dashed), and the baseline noise to a linear curve (grey dashed).
         \textbf{c},~The wavelength shift histogram is plot as a relative potential landscape, displaying a bimodal distribution.
         A 10\,Hz high-pass filter is applied to remove measurement drift.
        }
    \label{fig:bindings}
\end{figure*}

After the signal stabilizes following hard layer formation, we observe occasional distinct step-like resonance shifts followed by enhanced fluctuations over extended durations as single proteins enter the slot, are weakly confined by their interactions with the hard layer, and are probed continuously by the optical cavity. Fig.~\ref{fig:bindings}a shows a typical event, observed for holoferritin which includes a ferrihydrite core.
Converting the measured reflected optical intensities to cavity resonance wavelength shifts via the known optical cavity transfer function, we find that the wavelength shift is $2.16 \pm 0.01$ nm, where the uncertainty represents the standard error.
Since the cavity field is highly uniform, the protein can be well-approximated as located at its electric field maximum. Eq.~(\ref{dispersive_eqn}) can then be applied to find the excess polarizability of the protein from this wavelength shift.
We find that $\alpha= 29700 \pm 100~\mathrm{\AA}^3$, consistent with expectations (Supplementary Section I.A.a).



The weak confinement
we observe is consistent with previous observations of proteins within the `soft corona' around nanoparticles, characterized using techniques such as click-capture proteomics~\cite{surfacespe2, coronaReview}. 
By analogy we term the 
layer in which  this occurs the `soft layer'~\cite{hardcorona}.
Protein confinement within this layer conveys three significant advantages. First, we find that it allows continuous observation of single protein dynamics over periods of minutes.
Second, proteins in the soft corona are known to retain a native-like state due to the weakness of interactions~\cite{coronaReview}. 
As such, the observations are minimally invasive, without the strong optical forces present in plasmonic sensing~\cite{doublenanohole} or strong chemical binding to the surface~\cite{chemicalb}. Third, it suppresses the Brownian motion of the protein relative to internal structural dynamics. We estimate a factor of five in suppression by performing comparative measurements on protein-scale silica nanoparticles optically trapped at the centre of the slot (Supplementary Section III.C).

The power spectral density (PSD) of the cavity resonance wavelength fluctuations 
exhibits an approximately Lorentzian shape (Fig.~\ref{fig:bindings}b, blue trace), characterized by a flat plateau at low frequencies and roughly $1/f^2$ decay above the corner frequency, as expected for internal dynamics or confined Brownian motion (Supplementary Section III.B and IV.C). It remains above the
background noise measured under identical conditions (gray trace) 
up to frequencies beyond 1\,MHz, enabling the resolution of microsecond protein dynamics (Supplementary Section III.E).
This bandwidth exceeds that of previous measurements 
by two orders of magnitude~\cite{doublenanohole, doublenanoholePSD}.

To understand the underlying energy landscape of the protein dynamics, we derive the effective potential $U$ from the measured probability distribution $p$ of the wavelength shift using the Boltzmann relation~\cite{thermalbook}, $U/k_{\text{B}}T = -\ln p$, where $k_{\text{B}}$ is the Boltzmann constant and $T$ is the temperature. As shown in Fig.~\ref{fig:bindings}c, this exhibits an asymmetric double well potential with inter-state barrier height of approximately $k_\mathrm{B}T$, suggesting that dynamical transitions are occurring between two local minima.

As a control, we make similar measurements on two additional types of protein, catalase and bovine serum albumin (Supplementary Section III.D). These also form monolayers and exhibit weak confinement of individual proteins in the soft layer. However, their potentials are single-welled, agreeing well with the theoretical model for loosely bound proteins on the hard layer (Supplementary Section III.B).





\begin{figure*}[ht!]
    \centering
    \includegraphics[width=0.95\textwidth]{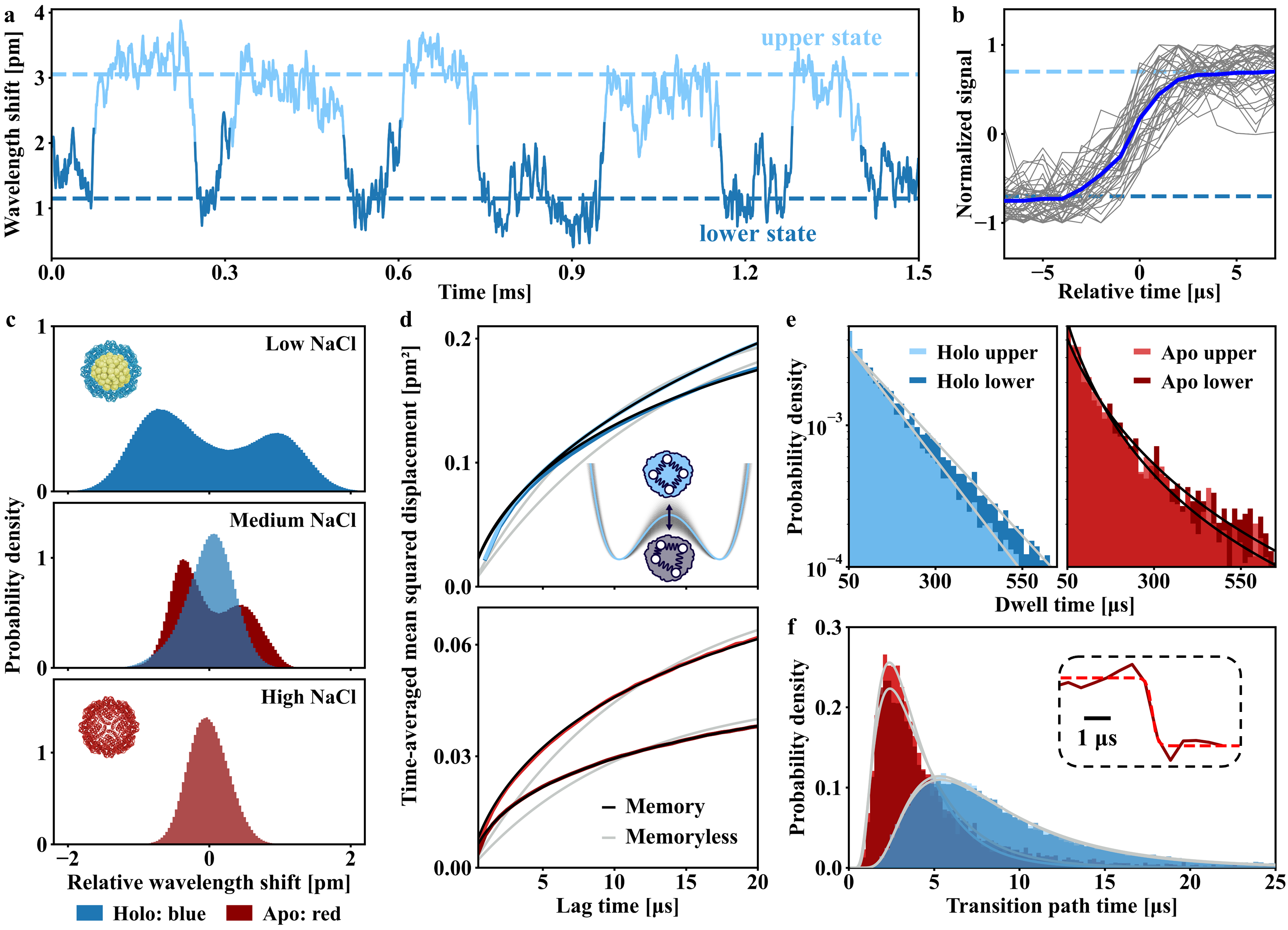} \\
    \caption{
        \textbf{Analysis of stepping dynamics in ferritin.}
        \textbf{a},~Short time scale view of single-holoferritin signal from Fig.~\ref{fig:bindings}a, displaying stepping between an `upper' and `lower' state.
        \textbf{b},~Forty different upward transition of holoferritin in \textbf{a}, with the blue solid curve showing their average.
        \textbf{c},~Histograms of the wavelength shift for holoferritin at low NaCl (4\,mM, top), holoferritin and apoferritin at medium NaCl (10\,mM, middle), and apoferritin at high NaCl (150\,mM, bottom). 
        A 10\,Hz high-pass filter is applied before data analysis to remove measurement drift.
        In this figure, blue-toned data represent holoferritin, while red-toned data represent apoferritin. 
        \textbf{d},~Time-averaged mean squared displacement calculated from individual trajectory segments and averaged separately for the upper (light) and lower (dark) states in holoferritin at low NaCl and apoferritin at medium NaCl.
        The inset graphic illustrates fluctuations of the double well landscape along the observable coordinate due to slow hidden structural changes.
        \textbf{e},~The corresponding dwell time histograms for the upper and lower states.
        Dwell times below 50\,$\mu$s are excluded, as spike-like noise could be misclassified as transition events and thereby skew the distribution.
        Holo-ferritin displays no memory dynamics in its dwell time distributions, whilst apo-ferritin does (see text).
        \textbf{f},~The corresponding transition path time histograms for the upward (light) and downward (dark) transitions.
        The inset displays the fastest transition (normalised) for apo-ferritin.
        In \textbf{d}-\textbf{f}, black and grey curves represent fits to memory dependent and memoryless behaviour models, respectively.
        }
    \label{fig:double}
\end{figure*}

\subsection*{Ferritin stepping dynamics}

%
To investigate the behavior of holoferritin within its double-well potential,
we examine its temporal dynamics over shorter time scales,  observing 
unmistakable stepping between
two distinct states (Fig.~\ref{fig:double}a). The upper state exhibits nearly twice the wavelength shift of the lower state. This twofold change 
is consistent in magnitude with polarizability fluctuations that have been reported in previous studies of ferritin~\cite{ferritinPH2}. Those studies interpreted the fluctuations as due to conformational dynamics, but had insufficient time resolution to observe distinct steps. The mechanism underlying the magnitude of the signals, and similar anomalously large signals from other optical-polarization based single molecule measurements, is not yet fully understood~\cite{ferritinPH2, doublenanoholePSD}.
Given that the electric double layer surrounding the protein enhances its polarizability~\cite{dark}, one possibility is that conformational changes modify the ferritin surface charge distribution, perturbing the surrounding electric double layer and thereby amplifying the change in polarizability~\cite{amplify}.

The ability to resolve real-time trajectories of the steps allows us to analyse their heterogeneity.
Figure~\ref{fig:double}b shows, for example, forty rising steps (grey lines) from a single measurement trace for one holoferritin molecule. Sigmoid fits to them indicate a high degree of heterogeneity in the transition time, with its standard deviation of 2.3\,$\mu$s being more than  half its mean (4.3\,$\pm$ 0.4\,$\mu$s).

To determine whether the stepping between transient states is mechanically tunable, we vary the stability of the protein shell using two established handles:
ionic strength, which modulates the electrostatic screening of interactions between subunits of the ferritin shell~\cite{FerritinSalt}; and the ferrihydrite core, which stiffens the shell structure~\cite{doublenanoholePSD}.
Stepping due to instability of the ferritin shell should vanish under stabilising conditions and reappear under destabilising ones~\cite{ferritinPH1, ferritinPH2}.
As a first set of experiments, we repeat the measurements on holoferritin with ionic concentration raised from 4 to 10\,mM of NaCl, enhancing the screening and therefore the stability of the ferritin shell.
Consistent with expectations, we find that the  potential landscape collapses to a single well, quenching the stepping dynamics (Fig.~\ref{fig:double}c).
Thus, the instability can be switched on and off by electrostatic means.

To isolate the role of the ferrihydrite core, we then measure apoferritin (ferritin with no ferrihydrite core) at selected ionic concentrations.
At similar ionic concentration (10\,mM NaCl) to that which suppresses dynamics in holoferritin, apoferritin retains clear stepping dynamics and an asymmetric double well potential (Fig.~\ref{fig:double}c middle). It is therefore less stable than holoferritin, as expected given the removal of the stabilising ferrihydrite core~\cite{ferritinPH2, doublenanoholePSD}. 
Similar to holoferritin, at higher ionic strength (150\,mM NaCl) increased screening causes the apoferritin to become stable, converging to a single well (Fig.~\ref{fig:double}c bottom).
Together, the stepping dynamics appear exclusively under conditions of known shell destabilisation and vanish when either stabilising factor is restored. 

\subsection*{Dynamics within and between states}

The ability to clearly distinguish steps between states allows us to explore differences in dynamics when the protein is in the upper and lower states, as well as the statistics of transitions between states. 
To do this systematically, 
we first classify regions of each trace into the upper or lower states with a hidden Markov model (Fig.~\ref{fig:double}a; Supplementary Section V.A).
We focus on apoferritin with 10\,mM NaCl and holoferritin with 4\,mM of NaCl, because they exhibit the clearest double well potentials, allowing the states to be most accurately distinguished.
The model identifies a total of 71,718 transitions between states from a single 8.5\,s holoferritin trace and 12,767 transitions from a single 1\,s apoferritin trace, allowing accurate statistical analysis. 


The potential landscape of protein conformational changes is modulated by the global protein structure as it changes over time. This is known to introduce memory effects in the diffusion of the protein within each conformational state~\cite{multiScience, memory1}. To explore this, we calculate the time-averaged mean squared displacement (tMSD) for both upper and lower states. Shown in Fig.~\ref{fig:double}d, these exhibit dynamics that is consistent with memory-dependent models (black lines) and inconsistent with the memory-less dynamics expected for the states of a simple double well potential (grey lines). Specifically, rather than undergoing diffusive motion at short times followed by an exponential approach to a plateau; they exhibit subdiffusive motion and approach the plateau as a power law (Supplementary Section IV.A). This is qualitatively consistent with previous observations attributed to 
protein structural dynamics~\cite{multiScience, memory1}.

To further understand the dynamics, 
we analyse the heterogeneity of the dwell time within each state and the transition path time between states.
These quantities probe complementary aspects of the underlying energy landscape: dwell times are governed primarily by the stability of a state and the surrounding free-energy barriers~\cite{BSAenergy}, whereas transition path times reflect the effective friction and diffusivity encountered during barrier crossing~\cite{internal2}. 


The observed dwell time distributions are shown in Fig.~\ref{fig:double}e for holo- (blue) and apo- (red) ferritin. For holoferritin, the mean dwell times  are $102.3 \pm 0.7$\,$\mu$s and $134.8 \pm 0.9$\,$\mu$s in the upper and lower states, respectively; while for apoferritin they are $139 \pm 4$\,$\mu$s and $174 \pm 5$\,$\mu$s. 
In both cases, they are longer in the lower state, indicating that it is more stable.
Both sets of distributions exhibit
significant heterogeneity,  which would be obscured by ensemble or time averaging, with observed values ranging over more than an order-of-magnitude.

While holoferritin exhibits exponential distributions (Fig.~\ref{fig:double}e {\it left}), 
the distributions for 
apoferritin are
non-exponential and heavy-tailed (Fig.~\ref{fig:double}e {\it right}). 
%
This likely arises from memory effects.
For holoferritin, the mean dwell times are around a factor of two longer than the typical time over which memory of the initial protein structure in one state is lost (Supplementary Section IV.B.b). This is consistent with memory-less Markovian-like transitions~\cite{markovianP} and the observed exponential distribution. By contrast, the mean dwell times for apoferritin are both well within the memory dominated regime (Supplementary Section IV.B.b) so that memory-less dynamics are not expected. In this case, we find that the dwell-time distributions agree well with Mittag-Leffler-type distributions (black lines, Fig.~\ref{fig:double}e {\it right}) characteristic of dynamics with long-lived, power-law memory~\cite{memoryless}.

Figure~\ref{fig:double}f shows the transition path times for both upwards and downwards transitions. Similar to previous measurements of protein dynamics~\cite{tptmeasure}, memory effects appear not to be significant during transitions, with good agreement with memory-less models (grey lines).
Faster transitions are observed for apoferritin, even under higher ionic concentrations where electrostatic repulsion is weaker. This suggests it has lower internal friction~\cite{internal2}, consistent with earlier reports that the presence of a ferrihydritecore results in a more compact conformation~\cite{doublenanoholePSD}.
The distributions of upwards and downwards transitions within each form of ferritin are near identical,
as expected from microscopic reversibility (each transition path has a time-reversed counterpart with the same duration). For holoferritin, the modal values for the upward and downward transitions are $5.3\pm 0.5$\,$\mu$s and $5.3\pm 0.2$\,$\mu$s, respectively; while for apoferritin, they are $2.1\pm 0.3$\,$\mu$s and $2.4\pm 0.2$\,$\mu$s. Similar to the dwell times, the distributions exhibit high heterogeneity. Some transitions occur over timescales as short as $400$\,ns (Fig.~\ref{fig:double}f {\it inset}), limited by our data acquisition rate, while the tails of the distributions extend beyond 25~$\mu$s.

\section*{Discussion} 

Fluorescence-based single-protein sensors have yielded major insights into protein conformational dynamics and their relationship to function~\cite{Lerner2018}. However, they face two important limitations. First, fluorophore attachment can alter protein conformation and dynamics~\cite{SanchezRico2017}, creating uncertainty as to how faithfully the measured trajectories represent those of the unlabelled protein. Second, the finite photon budget before photobleaching imposes a fundamental trade-off between temporal resolution and observation duration~\cite{Miao2025, sch2021single}. At microsecond temporal resolution, they have been constrained to millisecond durations~\cite{fluor2}. Our label-free approach overcomes these barriers, combining sub-microsecond temporal resolution with continuous trajectories that can last for minutes. This allows rapid conformational transitions to be examined over long timescales, providing access to slow modulation, dynamic heterogeneity and memory effects, as demonstrated here for ferritin. These capabilities may help resolve longstanding questions, including whether protein conformational dynamics are ergodic~\cite{li2022non}, how hidden states govern transitions~\cite{grossman2015single}, and the role of rare pathways and long-lived intermediates~\cite{ye2018conformational}.

The fast stepping dynamics that we observe in ferritin are likely associated with conformational gating of its 3-fold channels that are predicted by simulations but have previously been unresolvable experimentally~\cite{ferritinPH1, ferritinPH2}.
Further exploration of these dynamics
may contribute towards establishing a mechanistic understanding of how specific variants disrupt iron homeostasis in disorders such as hypoferritinemia~\cite{ferritina1} and neuroferritinopathy~\cite{ferritina2}.
They may also inform the optimization of ferritin nanocages for drug delivery applications, where channel gating dynamics, pH-dependent dissociation, and thermal breathing modes directly impact cargo loading capacity and triggered release kinetics~\cite{releakinetics, releakinetics2}.

A general challenge for label-free single-protein sensing is that the measured signal lacks intrinsic molecular specificity, making it difficult to determine whether fluctuations arise from conformational changes, or from centre-of-mass motion or interaction with the sensor surface. Several observations nevertheless support our interpretation of the ferritin signal as conformational dynamics. First, the observed memory-dependent kinetics are consistent with expectations for protein dynamics and measurements on other proteins~\cite{multiScience, memory1}. Second, the dependence of the dynamics on ionic concentration and on the presence of a ferrihydrite core agrees with expectations for conformational fluctuations of the ferritin shell~\cite{FerritinSalt, doublenanoholePSD}. Third, the observations are consistent with earlier measurements of conformational fluctuations in ferritin~\cite{ferritinPH2}, though they lacked the temporal resolution to resolve the discrete transitions observed here.

A possible alternative is that ferritin hops between two locations on the hard layer that experience different optical electric-field strengths. Although our modelling predicts an exceptionally uniform field, localised hotspots could arise from silicon surface roughness~\cite{roughnesss}. This mechanism, however, does not readily explain the memory-dependent dynamics we observe. Moreover, our simulations show that surface heterogeneity could account for only a fraction of the measured wavelength shift (Supplementary Section II.C). The absence of analogous stepping in catalase and BSA (Supplementary Section VI.A.a) further argues against a generic artifact.

Our use of interactions with a protein hard layer to weakly confine individual proteins conveys significant advantages, allowing unlabeled measurements over long durations. 
While proteins in the soft layer have been found to retain near-native conformations because of weak surface interactions~\cite{coronaReview}, proximity to the hard layer may nevertheless influence their dynamics~\cite{coronaReview}. Future work should assess the extent of such effects.
Beyond conformational dynamics, the capacity to vary the molecular composition of the hard layer, and resolve molecular binding events and structure at sub-microsecond resolution in a scalable, multiplexed sensor may allow a new platform to study molecular interactions~\cite{surfacespe2}.



Many protein structural changes occur on timescales faster than we resolve~\cite{signaling, Lerner2018} or involve spatially localised, low-amplitude motions~\cite{Lerner2018, Larnii}. Resolving such dynamics will require suppression of two distinct noise sources.


At frequencies below a megahertz, our measurements are dominated by ferritin motion (Fig.~\ref{fig:bindings}b), likely including residual translational Brownian motion that obscures subtle dynamics. This residual noise could be reduced by increasing the uniformity of the optical electric field. The electric field at the vertical centre of the slot is approximately a factor of thirty more uniform than its average across the sidewalls, suggesting that preferentially localising ferritin near this position would substantially increase sensitivity to subtle structural changes.
The protein position could be determined through postselection using multiple cavity modes~\cite{hungerNa} or measurements of the cavity decay rate~\cite{zhu}, or controlled by patterning a polymer layer that restricts binding to regions of near-zero field gradient~\cite{peg}.
Further improvements could come from modifying the cavity geometry and reducing surface roughness. Together, this could plausibly suppress translational Brownian motion by a further factor of one hundred. Additional suppression may be possible by changing the composition of the hard layer to increase confinement, although this could also perturb the protein dynamics.

Above 1 MHz, protein motion falls below our measurement noise floor, so resolving faster dynamics would require stronger light–protein interactions or lower noise. The noise floor is likely dominated by cavity thermorefractive noise due to silicon’s large thermo-optic coefficient~\cite{TRN}. 
Replacing silicon with silicon nitride or tantalum pentoxide would increase the mode volume because of their lower refractive indices, but their smaller thermo-optic coefficients~\cite{TOC} could nevertheless yield an estimated tenfold gain in signal-to-noise ratio.
Smaller mode-volume cavities~\cite{slot1, atomicphc}, designed to retain low field gradients, could extend this further.
Combined with the sharp decline in thermorefractive noise above its corner frequency in the tens-of-megahertz range~\cite{TRN}, these advances could make temporal resolution in the nanosecond-to-tens-of-nanosecond regime accessible.

\section*{Methods} 

\subsection*{Simulation} 

To evaluate the optical performance of the slot photonic crystal, finite element method simulations were performed using the Electromagnetic Waves, Frequency Domain Interface (COMSOL Multiphysics version 6.1, COMSOL Inc.). 
The device has a cross-sectional dimension of $220 \times 595.2\,\text{nm}^2$, with a central slot of $342.3 \times 55.8\,\text{nm}^2$. 
The cavity consists of ten mirror holes on the right and eight mirror holes on the left of the central slot. 
The eight holes around the slot are tapered with a linear gradient to optimize the mode profile. 
The holes immediately adjacent to the slot have a centre to centre spacing of 536.3\,nm. 
Moving away from the slot, the hole diameters increase from 189.9\,nm to 279.3\,nm in steps of 22.3\,nm, while the lattice spacing increases from 268.1\,nm to 394.3\,nm in steps of 31.5\,nm. 
Finally, to achieve critical coupling to the waveguide, the diameters of two coupling holes at the left end are scaled down by 41.9\,nm and 83.8\,nm, respectively. 

\subsection*{Fabrication}

The devices were fabricated on a silicon-on-insulator (SOI) chip ($15\times 15\,\text{mm}^2$) comprising a 220\,nm thick silicon layer and a 2\,$\mu$m buried oxide (SiO$_2$) layer (Supplementary Section I.C). 
For electron-beam lithography, an electron-beam resist (ARP 6200.09, Allresist) was spin-coated at 1000\,rpm and baked at $\text{180}^{\circ}$C for two minutes. 
The device patterns were written using an electron-beam pattern generator (EBPG 5150, Raith GmbH) operated at an accelerating voltage of 100\,kV with a dose of $360\,\mu\text{C/cm}^2$. 
After exposure, the patterns were developed in an amyl acetate based developer (AR 600-546, Allresist) for 60\,s and rinsed in o-xylene and isopropyl alcohol (IPA). 
The silicon layer was subsequently etched using an inductively coupled plasma reactive ion etching (ICP-RIE) system (PlasmaPro 100 Cobra, Oxford Instruments) with a CHF$_3$/SF$_6$ gas mixture. 
The residual resist was removed using O$_2$ plasma cleaning. 
Finally, the devices were released by removing the underlying SiO$_2$ layer using vapor HF etching. 

\subsection*{Device pickup}

To pick up the device, a tapered optical fibre tip was coated with optical adhesive (NOA86H, Norland Products) over a length of 20\,$\mu$m.
The optical device was detached from the silicon layer by breaking the connection tether ($\approx 40 \times 10\,\text{nm}^2$) through a lateral swing motion of the fibre. 
The captured device was then immersed in a protein-containing heavy water droplet (Supplementary Section I.C).  

\subsection*{Experimental set-up}
An optical field was generated with a red sideband detuning relative to the cavity resonance.
A tunable laser source (TSL-770, Santec) provides 5\,mW of input power to the tapered fibre through an optical circulator. 
The reflected optical signal was converted to photocurrent via a photodetector (PDA 10CS-EC, Thorlabs) with a bandwidth of 775\,kHz and recorded by an oscilloscope (RTE1000, Rohde $\&$ Schwarz). 
Data acquisition was performed at sampling rates of 1-2\,MS/s with a bandwidth of 500\,kHz. 
Wavelength shift calibration was achieved using the laser's internal reference.

\subsection*{Protein preparation and measurement}

Holoferritin (F4503) and apoferritin (A3641) were purchased from Sigma-Aldrich and stored according to the manufacturer's instructions. Holoferritin and apoferritin from equine spleen were supplied in a saline solution. Fresh working salt (NaCl, SA046, Chem-supply) and protein (100\,nM) solutions were prepared at the start of each experiment using filter-sterilized heavy water (D$_2$O, heavy water, 99.9$\%$, Novachem). 
In a typical experiment, 250-300\,$\mu$L of the prepared salt solution was pipetted onto a clean coverslip. 
The fibre-coupled photonic crystal cavity was immersed in this prepared solution to record the background noise. 
The proteins were then added to the solution to achieve final concentrations of 1\,nM. 
All sample preparation and measurement were conducted in an enclosed environment with continuous HEPA filtration to minimize contamination from external particulates.



\section*{Data availability} 

All data in the main text or Supplementary Information are available from the corresponding author upon reasonable request.

\section*{References} 
\bibliography{reference}

@article{molrecogni,
  title={Recognition dynamics up to microseconds revealed from an RDC-derived ubiquitin ensemble in solution},
  author={Lange, Oliver F and Lakomek, Nils-Alexander and Fares, Christophe and Schroder, Gunnar F and Walter, Korvin FA and Becker, Stefan and Meiler, Jens and Grubmuller, Helmut and Griesinger, Christian and De Groot, Bert L},
  journal={Science},
  volume={320},
  number={5882},
  pages={1471--1475},
  year={2008},
  publisher={American Association for the Advancement of Science}
}

@article{Lerner2018,
  author  = {Lerner, Eitan and Cordes, Thorben and Ingargiola, Antonino
             and Alhadid, Yazan and Chung, SangYoon and Michalet, Xavier
             and Weiss, Shimon},
  title   = {Toward dynamic structural biology: Two decades of
             single-molecule Förster resonance energy transfer},
  journal = {Science},
  volume  = {359},
  number  = {6373},
  pages   = {eaan1133},
  year    = {2018},
  doi     = {10.1126/science.aan1133}
}

@article{SanchezRico2017,
  author  = {Sánchez-Rico, Carolina and Voith von Voithenberg, Lena
             and Warner, Lisa R. and Lamb, Don C. and Sattler, Michael},
  title   = {Effects of Fluorophore Attachment on Protein Conformation
             and Dynamics Studied by spFRET and NMR Spectroscopy},
  journal = {Chemistry -- A European Journal},
  volume  = {23},
  pages   = {14267--14277},
  year    = {2017},
  doi     = {10.1002/chem.201702423}
}

@article{Miao2025,
  author  = {Miao, Yu and Cheng, Yuxiao and Xia, Yushi and others},
  title   = {Supervised multi-frame dual-channel denoising enables
             long-term single-molecule FRET under extremely low photon budget},
  journal = {Nature Communications},
  volume  = {16},
  pages   = {74},
  year    = {2025},
  doi     = {10.1038/s41467-024-54652-w}
}

@article{enzycatalysis,
  title={Enzyme dynamics during catalysis},
  author={Eisenmesser, Elan Zohar and Bosco, Daryl A and Akke, Mikael and Kern, Dorothee},
  journal={Science},
  volume={295},
  number={5559},
  pages={1520--1523},
  year={2002},
  publisher={American Association for the Advancement of Science}
}

@article{signaling,
  title={The dynamic process of $\beta$2-adrenergic receptor activation},
  author={Nygaard, Rie and Zou, Yaozhong and Dror, Ron O and Mildorf, Thomas J and Arlow, Daniel H and Manglik, Aashish and Pan, Albert C and Liu, Corey W and Fung, Juan Jos{\'e} and Bokoch, Michael P and others},
  journal={Cell},
  volume={152},
  number={3},
  pages={532--542},
  year={2013},
  publisher={Elsevier}
}

@article{allosregula,
  title={Tunable microsecond dynamics of an allosteric switch regulate the activity of a AAA+ disaggregation machine},
  author={Mazal, Hisham and Iljina, Marija and Barak, Yoav and Elad, Nadav and Rosenzweig, Rina and Goloubinoff, Pierre and Riven, Inbal and Haran, Gilad},
  journal={Nature Communications},
  volume={10},
  number={1},
  pages={1438},
  year={2019},
  publisher={Nature Publishing Group UK London}
}

@article{drugdis,
  title={Leveraging conformational ensembles in allosteric drug discovery},
  author={Nussinov, Ruth and Regev, Clil and Jang, Hyunbum},
  journal={Trends in Pharmacological Sciences},
  year={2026},
  volume={47},
  pages={276-289},
  publisher={Elsevier}
}

@article{vaccine,
  title={Microsecond dynamics control the HIV-1 Envelope conformation},
  author={Bennett, Ashley L and Edwards, Robert and Kosheleva, Irina and Saunders, Carrie and Bililign, Yishak and Williams, Ashliegh and Bubphamala, Pimthada and Manosouri, Katayoun and Anasti, Kara and Saunders, Kevin O and others},
  journal={Science Advances},
  volume={10},
  number={5},
  pages={eadj0396},
  year={2024},
  publisher={American Association for the Advancement of Science}
}

@article{cancer,
  title={Unveiling the “invisible” druggable conformations of GDP-bound inactive Ras},
  author={Liu, Dan and Mao, Yunyun and Gu, Xue and Zhou, Yang and Long, Dong},
  journal={Proceedings of the National Academy of Sciences},
  volume={118},
  number={11},
  pages={e2024725118},
  year={2021},
  publisher={National Academy of Sciences}
}

@article{peg,
  title={Lithographic techniques and surface chemistries for the fabrication of PEG-passivated protein microarrays},
  author={Kannan, Balaji and Castelino, Kenneth and Chen, Fanqing Frank and Majumdar, Arun},
  journal={Biosensors and Bioelectronics},
  volume={21},
  number={10},
  pages={1960--1967},
  year={2006},
  publisher={Elsevier}
}

@article{zhu,
  title={On-chip single nanoparticle detection and sizing by mode splitting in an ultrahigh-Q microresonator},
  author={Zhu, Jiangang and Ozdemir, Sahin Kaya and Xiao, Yun-Feng and Li, Lin and He, Lina and Chen, Da-Ren and Yang, Lan},
  journal={Nature Photonics},
  volume={4},
  number={1},
  pages={46--49},
  year={2010},
  publisher={Nature Publishing Group UK London}
}

@article{mminter,
  title={Single-molecule nucleic acid interactions monitored on a label-free microcavity biosensor platform},
  author={Baaske, Martin D and Foreman, Matthew R and Vollmer, Frank},
  journal={Nature Nanotechnology},
  volume={9},
  number={11},
  pages={933--939},
  year={2014},
  publisher={Nature Publishing Group UK London}
}

@article{fluor2,
  title={Rate-engineered plasmon-enhanced fluorescence for real-time microsecond dynamics of single biomolecules},
  author={Nooteboom, Sjoerd W and Okholm, Kasper R and Lamberti, Vincenzo and Oomen, Bas and Sutherland, Duncan S and Zijlstra, Peter},
  journal={Nano Letters},
  volume={24},
  number={37},
  pages={11641--11647},
  year={2024},
  publisher={ACS Publications}
}

@article{Larnii,
  title={Modelling of the dynamic polarizability of macromolecules for single-molecule optical biosensing},
  author={Booth, Larnii S and Browne, Eloise V and Mauranyapin, Nicolas P and Madsen, Lars S and Barfoot, Shelley and Mark, Alan and Bowen, Warwick P},
  journal={Scientific Reports},
  volume={12},
  number={1},
  pages={1995},
  year={2022},
  publisher={Nature Publishing Group UK London}
}

@article{doublenanoholePSD,
  title={Optical monitoring of in situ iron loading into single, native ferritin proteins},
  author={Yousefi, Arman and Ying, Cuifeng and Parmenter, Christopher DJ and Assadipapari, Mahya and Sanderson, Gabriel and Zheng, Ze and Xu, Lei and Zargarbashi, Saaman and Hickman, Graham J and Cousins, Richard B and others},
  journal={Nano Letters},
  volume={23},
  number={8},
  pages={3251--3258},
  year={2023},
  publisher={ACS Publications}
}

@article{ferritinPH2,
  title={Structural flexibility and disassembly kinetics of single ferritin molecules using optical nanotweezers},
  author={Yousefi, Arman and Zheng, Ze and Zargarbashi, Saaman and Assadipapari, Mahya and Hickman, Graham J and Parmenter, Christopher DJ and Bueno-Alejo, Carlos J and Sanderson, Gabriel and Craske, Dominic and Xu, Lei and others},
  journal={ACS Nano},
  volume={18},
  number={24},
  pages={15617--15626},
  year={2024},
  publisher={ACS Publications}
}

@article{WGMen,
  title={Label-free optical detection of single enzyme-reactant reactions and associated conformational changes},
  author={Kim, Eugene and Baaske, Martin D and Schuldes, Isabel and Wilsch, Peter S and Vollmer, Frank},
  journal={Science Advances},
  volume={3},
  number={3},
  pages={e1603044},
  year={2017},
  publisher={American Association for the Advancement of Science}
}

@article{doublenanohole,
  title={Optical trapping of a single protein},
  author={Pang, Yuanjie and Gordon, Reuven},
  journal={Nano Letters},
  volume={12},
  number={1},
  pages={402--406},
  year={2012},
  publisher={ACS Publications}
}

@article{BSAenergy,
  title={Energy landscape of conformational changes for a single unmodified protein},
  author={Peters, Matthew and Zhao, Tianyu and George, Sherin and Truong, Viet Giang and Nic Chormaic, S{\'\i}le and Ying, Cuifeng and Nome, Ren{\'e} A and Gordon, Reuven},
  journal={npj Biosensing},
  volume={1},
  number={1},
  pages={14},
  year={2024},
  publisher={Nature Publishing Group UK London}
}

@article{nano,
  title={Nanosecond time scale transient optoplasmonic detection of single proteins},
  author={Baaske, Martin D and Asgari, Nasrin and Punj, Deep and Orrit, Michel},
  journal={Science Advances},
  volume={8},
  number={2},
  pages={eabl5576},
  year={2022},
  publisher={American Association for the Advancement of Science}
}

@article{pickup,
  title={Coupling a single trapped atom to a nanoscale optical cavity},
  author={Thompson, Jeffrey Douglas and Tiecke, TG and de Leon, Nathalie P and Feist, J and Akimov, AV and Gullans, M and Zibrov, Alexander S and Vuleti{\'c}, V and Lukin, Mikhail D},
  journal={Science},
  volume={340},
  number={6137},
  pages={1202--1205},
  year={2013},
  publisher={American Association for the Advancement of Science}
}

@article{hardcorona,
  title={Biomolecular coronas provide the biological identity of nanosized materials},
  author={Monopoli, Marco P and {\AA}berg, Christoffer and Salvati, Anna and Dawson, Kenneth A},
  journal={Nature Nanotechnology},
  volume={7},
  number={12},
  pages={779--786},
  year={2012},
  publisher={Nature Publishing Group UK London}
}

@article{corona1,
  title={Real-time optical tracking of protein corona formation on single nanoparticles in serum},
  author={Dolci, Mathias and Wang, Yuyang and Nooteboom, Sjoerd W and Soto Rodriguez, Paul Eduardo David and Sanchez, Samuel and Albertazzi, Lorenzo and Zijlstra, Peter},
  journal={ACS Nano},
  volume={17},
  number={20},
  pages={20167--20178},
  year={2023},
  publisher={ACS Publications}
}

@article{phcSi,
  title={Single particle detection in CMOS compatible photonic crystal nanobeam cavities},
  author={Quan, Qimin and Floyd, Daniel L and Burgess, Ian B and Deotare, Parag B and Frank, Ian W and Tang, Sindy KY and Ilic, Rob and Loncar, Marko},
  journal={Optics Express},
  volume={21},
  number={26},
  pages={32225--32233},
  year={2013},
  publisher={Optical Society of America}
}

@article{slot1,
  title={Ultrasmall-V high-Q photonic crystal nanobeam microcavities based on slot and hollow-core waveguides},
  author={Yu, Ping and Qi, Biao and Jiang, Xiaoqing and Wang, Minghua and Yang, Jianyi},
  journal={Optics Letters},
  volume={36},
  number={8},
  pages={1314--1316},
  year={2011},
  publisher={Optical Society of America}
}

@article{slotBis,
  title={Slotted photonic crystal nanobeam cavity with an ultrahigh quality factor-to-mode volume ratio},
  author={Seidler, Paul and Lister, Kevin and Drechsler, Ute and Hofrichter, Jens and St{\"o}ferle, Thilo},
  journal={Optics Express},
  volume={21},
  number={26},
  pages={32468--32483},
  year={2013},
  publisher={Optical Society of America}
}

@article{TRN,
  title={Fundamental thermal noise limits for optical microcavities},
  author={Panuski, Christopher and Englund, Dirk and Hamerly, Ryan},
  journal={Physical Review X},
  volume={10},
  number={4},
  pages={041046},
  year={2020},
  publisher={APS}
}

@article{detectV,
  title={Detection of single nanoparticles and lentiviruses using microcavity resonance broadening.},
  author={Shao, Linbo and Jiang, Xue-Feng and Yu, Xiao-Chong and Li, Bei-Bei and Clements, William R and Vollmer, Frank and Wang, Wei and Xiao, Yun-Feng and Gong, Qihuang},
  journal={Advanced Materials (Deerfield Beach, Fla.)},
  volume={25},
  number={39},
  pages={5616--5620},
  year={2013}
}

@article{dark,
   author = {Mauranyapin, NP and Madsen, LS and Taylor, MA and Waleed, M and Bowen, WP},
   title = {Evanescent single-molecule biosensing with quantum-limited precision},
   journal = {Nature Photonics},
   volume = {11},
   number = {8},
   pages = {477-481},
   ISSN = {1749-4885},
   year = {2017},
   type = {Journal Article}
}

@article{surface2,
  title={Measuring nanoparticle polarizability using fluorescence microscopy},
  author={Cao, Wenhan and Chern, Margaret and Dennis, Allison M and Brown, Keith A},
  journal={Nano Letters},
  volume={19},
  number={8},
  pages={5762--5768},
  year={2019},
  publisher={ACS Publications}
}

@article{water,
  title={Linear refractive index and absorption measurements of nonlinear optical liquids in the visible and near-infrared spectral region},
  author={Kedenburg, Stefan and Vieweg, Marius and Gissibl, Timo and Giessen, Harald},
  journal={Optical Materials Express},
  volume={2},
  number={11},
  pages={1588--1611},
  year={2012},
  publisher={Optica Publishing Group}
}

@article{shift1,
  title={Shift of whispering-gallery modes in microspheres by protein adsorption},
  author={Arnold, S and Khoshsima, M and Teraoka, Iwao and Holler, S and Vollmer, F},
  journal={Optics Letters},
  volume={28},
  number={4},
  pages={272--274},
  year={2003},
  publisher={Optica Publishing Group}
}

@article{BSAinter,
  title={Thermodynamics of multilayer protein adsorption on a gold nanoparticle surface},
  author={Mishra, Akriti and Das, Puspendu Kumar},
  journal={Physical Chemistry Chemical Physics},
  volume={24},
  number={37},
  pages={22464--22476},
  year={2022},
  publisher={Royal Society of Chemistry}
}

@article{FerritinSalt,
  title={Electrostatic repulsion during ferritin assembly and its screening by ions},
  author={Sato, Daisuke and Takebe, Satsuki and Kurobe, Atsushi and Ohtomo, Hideaki and Fujiwara, Kazuo and Ikeguchi, Masamichi},
  journal={Biochemistry},
  volume={55},
  number={3},
  pages={482--488},
  year={2016},
  publisher={ACS Publications}
}

@article{ferritinPH1,
  title={Single-molecule level dynamic observation of disassembly of the apo-ferritin cage in solution},
  author={Maity, Basudev and Li, Zhipeng and Niwase, Kento and Ganser, Christian and Furuta, Tadaomi and Uchihashi, Takayuki and Lu, Diannan and Ueno, Takafumi},
  journal={Physical Chemistry Chemical Physics},
  volume={22},
  number={33},
  pages={18562--18572},
  year={2020},
  publisher={The Royal Society of Chemistry}
}

@article{inactive1,
  title={Adsorption and bioactivity of protein A on silicon surfaces studied by AFM and XPS},
  author={Coen, Martine Collaud and Lehmann, Roland and Gr{\"o}ning, Pierangelo and Bielmann, Michael and Galli, Carine and Schlapbach, Louis},
  journal={Journal of Colloid and Interface Science},
  volume={233},
  number={2},
  pages={180--189},
  year={2001},
  publisher={Elsevier}
}

@article{internal2,
  title={Single-molecule fluorescence probes dynamics of barrier crossing},
  author={Chung, Hoi Sung and Eaton, William A},
  journal={Nature},
  volume={502},
  number={7473},
  pages={685--688},
  year={2013},
  publisher={Nature Publishing Group UK London}
}

@article{memoryless,
  title={A fractional generalization of the Poisson processes},
  author={Mainardi, Francesco and Gorenflo, Rudolf and Scalas, Enrico},
  journal={arXiv preprint math/0701454},
  year={2007}
}

@article{roughnesss,
  title={Laser Photochemical Nanostructuring of Silicon for Surface Enhanced Raman Spectroscopy},
  author={Akb{\i}y{\i}k, Alp and Avishan, Nardin and Demirta{\c{s}}, {\"O}zge and Demir, Ahmet Kemal and Y{\"u}ce, Emre and Bek, Alpan},
  journal={Advanced Optical Materials},
  volume={10},
  number={14},
  pages={2200114},
  year={2022},
  publisher={Wiley Online Library}
}

@article{welleffect,
  title={Slow viscous motion of a sphere parallel to a plane wall—I Motion through a quiescent fluid},
  author={Goldman, Arthur Joseph and Cox, Raymond G and Brenner, Howard},
  journal={Chemical Engineering Science},
  volume={22},
  number={4},
  pages={637--651},
  year={1967},
  publisher={Elsevier}
}

@book{PBE,
  title={The Poisson-Boltzmann Equation: An Introduction},
  author={Blossey, Ralf},
  year={2023},
  publisher={Springer Nature}
}

@incollection{PBE2,
  title={Charged Membranes: Poisson--Boltzmann Theory, The DLVO Paradigm, and Beyond},
  author={Markovich, Tomer and Andelman, David and Podgornik, Rudolf},
  booktitle={Handbook of lipid membranes},
  pages={99--128},
  year={2021},
  publisher={CRC Press}
}

@article{Sicharge,
  title={The charge of glass and silica surfaces},
  author={Behrens, Sven H and Grier, David G},
  journal={arXiv preprint cond-mat/0105149},
  year={2001}
}

@article{d2oeta,
  title={Fuels, Propellants and Explosives},
  author={Cardarelli, Francois},
  journal={Materials Handbook: A Concise Desktop Reference},
  pages={999--1018},
  year={2008},
  publisher={Springer}
}

@article{sigma,
  title={Electrical double layer around a spherical colloid particle: The excluded volume effect},
  author={L{\'o}pez-Garc{\'\i}a, JJ and Aranda-Rasc{\'o}n, MJ and Horno, J},
  journal={Journal of Colloid and Interface Science},
  volume={316},
  number={1},
  pages={196--201},
  year={2007},
  publisher={Elsevier}
}

@book{jackson,
  title={Classical electrodynamics},
  author={Jackson, John David},
  year={2021},
  publisher={John Wiley \& Sons}
}

@article{protein,
  title={The determination of thickness and surface mass density of mesothick immunoprecipitate layers by null ellipsometry and protein 125iodine labeling},
  author={Benesch, Johan and Askendal, Agneta and Tengvall, Pentti},
  journal={Journal of Colloid and Interface Science},
  volume={249},
  number={1},
  pages={84--90},
  year={2002},
  publisher={Elsevier}
}

@article{density,
  title={Average protein density is a molecular-weight-dependent function},
  author={Fischer, Hannes and Polikarpov, Igor and Craievich, Aldo F},
  journal={Protein Science},
  volume={13},
  number={10},
  pages={2825--2828},
  year={2004},
  publisher={Wiley Online Library}
}

@article{Langevin,
  title={Physical processes as stochastic processes},
  author={Kubo, Ryogo and Toda, Morikazu and Hashitsume, Natsuki and Kubo, Ryogo and Toda, Morikazu and Hashitsume, Natsuki},
  journal={Statistical Physics II: Nonequilibrium Statistical Mechanics},
  pages={40--108},
  year={1985},
  publisher={Springer}
}

@article{PSD,
  title={Power spectrum analysis for optical tweezers},
  author={Berg-S{\o}rensen, Kirstine and Flyvbjerg, Henrik},
  journal={Review of Scientific Instruments},
  volume={75},
  number={3},
  pages={594--612},
  year={2004},
  publisher={American Institute of Physics}
}

@article{eqmotion,
  title={Input and output in damped quantum systems: Quantum stochastic differential equations and the master equation},
  author={Gardiner, Crispin W and Collett, Matthew J},
  journal={Physical Review A},
  volume={31},
  number={6},
  pages={3761},
  year={1985},
  publisher={APS}
}

@article{TPT,
  title={Transition path time distributions},
  author={Laleman, Michiel and Carlon, Enrico and Orland, Henri},
  journal={The Journal of Chemical Physics},
  volume={147},
  number={21},
  year={2017},
  publisher={AIP Publishing}
}

@article{Kramer,
  title={Direct observation of the entropic potential in a binary suspension},
  author={Kaplan, PD and Faucheux, Luc P and Libchaber, Albert J},
  journal={Physical Review Letters},
  volume={73},
  number={21},
  pages={2793},
  year={1994},
  publisher={APS}
}

@article{ferrR,
  title={Evidence for non-DLVO hydration interactions in solutions of the protein apoferritin},
  author={Petsev, Dimiter N and Vekilov, Peter G},
  journal={Physical Review Letters},
  volume={84},
  number={6},
  pages={1339},
  year={2000},
  publisher={APS}
}

@article{cataR,
   author = {Axelsson, Inge},
   title = {Characterization of proteins and other macromolecules by agarose gel chromatography},
   journal = {Journal of Chromatography A},
   volume = {152},
   number = {1},
   pages = {21-32},
   ISSN = {0021-9673},
   year = {1978},
   type = {Journal Article}
}

@inproceedings{monosalt,
  title={Driving forces for protein adsorption at solid surfaces},
  author={Norde, Willem},
  booktitle={Macromolecular Symposia},
  volume={103},
  number={1},
  pages={5--18},
  year={1996},
  organization={Wiley Online Library}
}

@article{surfacespe2,
  title={Mapping and identification of soft corona proteins at nanoparticles and their impact on cellular association},
  author={Mohammad-Beigi, Hossein and Hayashi, Yuya and Zeuthen, Christina Moeslund and Eskandari, Hoda and Scavenius, Carsten and Juul-Madsen, Kristian and Vorup-Jensen, Thomas and Enghild, Jan J and Sutherland, Duncan S},
  journal={Nature Communications},
  volume={11},
  number={1},
  pages={4535},
  year={2020},
  publisher={Nature Publishing Group UK London}
}

@book{thermalbook,
  title={Concepts in thermal physics},
  author={Blundell, Stephen J and Blundell, Katherine M},
  year={2010},
  publisher={Oup Oxford}
}

@article{enzymeengi,
  title={Optimization of conformational dynamics in an epistatic evolutionary trajectory},
  author={Gonz{\'a}lez, Mariano M and Abriata, Luciano A and Tomatis, Pablo E and Vila, Alejandro J},
  journal={Molecular Biology and Evolution},
  volume={33},
  number={7},
  pages={1768--1776},
  year={2016},
  publisher={Oxford University Press}
}

@article{proteindesign,
  title={Exploiting conformational dynamics to modulate the function of designed proteins},
  author={Rennella, Enrico and Sahtoe, Danny D and Baker, David and Kay, Lewis E},
  journal={Proceedings of the National Academy of Sciences},
  volume={120},
  number={18},
  pages={e2303149120},
  year={2023},
  publisher={National Academy of Sciences}
}

@article{fluoesem,
  title={Simple and Efficient Detection Scheme of Two-Color Fluorescence Correlation Spectroscopy for Protein Dynamics Investigation from Nanoseconds to Milliseconds},
  author={Sano, Yutaka and Itoh, Yuji and Kamonprasertsuk, Supawich and Suzuki, Leo and Fukasawa, Atsuhito and Oikawa, Hiroyuki and Takahashi, Satoshi},
  journal={ACS Physical Chemistry Au},
  volume={4},
  number={1},
  pages={85--93},
  year={2023},
  publisher={ACS Publications}
}

@book{wbook,
  title={Single molecule sensing beyond fluorescence},
  author={Bowen, Warwick and Vollmer, Frank and Gordon, Reuven},
  year={2022},
  publisher={Springer}
}

@article{hetero,
  title={The energy landscapes and motions of proteins},
  author={Frauenfelder, Hans and Sligar, Stephen G and Wolynes, Peter G},
  journal={Science},
  volume={254},
  number={5038},
  pages={1598--1603},
  year={1991},
  publisher={American Association for the Advancement of Science}
}

@article{onchip1,
  title={On-chip label-free biosensing based on active whispering gallery mode resonators pumped by a light-emitting diode},
  author={Kim, Yeseul and Lee, Hansuek},
  journal={Optics Express},
  volume={27},
  number={23},
  pages={34405--34415},
  year={2019},
  publisher={Optical Society of America}
}

@article{memory1,
  title={Generalized Langevin Equation with Fractional Gaussian Noise: Subdiffusion within a Single Protein Molecule},
  author={Kou, Samuel C and Xie, X Sunney},
  journal={Physical Review Letters},
  volume={93},
  number={18},
  pages={180603},
  year={2004},
  publisher={APS}
}

@article{ferritina1,
  title={A newly identified ferritin L-subunit variant results in increased proteasomal subunit degradation, impaired complex assembly, and severe hypoferritinemia},
  author={Shagidov, Dayana and Guttmann-Raviv, Noga and Cunat, S{\'e}verine and Frech, Liora and Giansily-Blaizot, Muriel and Ghatpande, Niraj and Abelya, Gili and Frank, Gabriel A and Aguilar Martinez, Patricia and Meyron-Holtz, Esther G},
  journal={American Journal of Hematology},
  volume={99},
  number={1},
  pages={12--20},
  year={2024},
  publisher={Wiley Online Library}
}

@article{ferritina2,
  title={Mutant L-chain ferritins that cause neuroferritinopathy alter ferritin functionality and iron permeability},
  author={McNally, Justin R and Mehlenbacher, Matthew R and Luscieti, Sara and Smith, Gideon L and Reutovich, Aliaksandra A and Maura, Poli and Arosio, Paolo and Bou-Abdallah, Fadi},
  journal={Metallomics},
  volume={11},
  number={10},
  pages={1635--1647},
  year={2019},
  publisher={Oxford University Press}
}

@article{chemicalb,
  title={Optical imaging of single-protein size, charge, mobility, and binding},
  author={Ma, Guangzhong and Wan, Zijian and Yang, Yunze and Zhang, Pengfei and Wang, Shaopeng and Tao, Nongjian},
  journal={Nature Communications},
  volume={11},
  number={1},
  pages={4768},
  year={2020},
  publisher={Nature Publishing Group UK London}
}

@article{atomicphc,
  title={Self-assembled photonic cavities with atomic-scale confinement},
  author={Babar, Ali Nawaz and Weis, Thor August Schimmell and Tsoukalas, Konstantinos and Kadkhodazadeh, Shima and Arregui, Guillermo and Vosoughi Lahijani, Babak and Stobbe, S{\o}ren},
  journal={Nature},
  volume={624},
  number={7990},
  pages={57--63},
  year={2023},
  publisher={Nature Publishing Group UK London}
}

@article{effTPT,
  title={Effect of memory and active forces on transition path time distributions},
  author={Carlon, Enrico and Orland, Hubert and Sakaue, Takahiro and Vanderzande, Carlo},
  journal={The Journal of Physical Chemistry B},
  volume={122},
  number={49},
  pages={11186--11194},
  year={2018},
  publisher={ACS Publications}
}

@article{markovianP,
  title={Automatic discovery of metastable states for the construction of Markov models of macromolecular conformational dynamics},
  author={Chodera, John D and Singhal, Nina and Pande, Vijay S and Dill, Ken A and Swope, William C},
  journal={The Journal of Chemical Physics},
  volume={126},
  number={15},
  pages={155101},
  year={2007},
  publisher={AIP Publishing}
}

@article{releakinetics,
  title={Engineering protein interfaces yields ferritin disassembly and reassembly under benign experimental conditions},
  author={Chen, Hai and Zhang, Shengli and Xu, C and Zhao, G},
  journal={Chemical Communications},
  volume={52},
  number={46},
  pages={7402--7405},
  year={2016},
  publisher={Royal Society of Chemistry}
}

@article{releakinetics2,
  title={Structure-guided rational design of ferritin nanocages unlocks thermoresponsive channels for accelerated drug encapsulation},
  author={Su, Hsiao-Ching and Huang, Chiun-Wei and Wang, Sheng-Hung and Chang, Chien-Yi and Lin, Jia-Yu and Tseng, Yi-Hsiang and Wu, Hung-Hsiang and Su, Nan-Wei and Huang, Feng-Ting},
  journal={International Journal of Biological Macromolecules},
  pages={150643},
  volume={347},
  year={2026},
  publisher={Elsevier}
}

@article{tptmeasure,
  title={Single-molecule enzymatic dynamics},
  author={Lu, H Peter and Xun, Luying and Xie, X Sunney},
  journal={Science},
  volume={282},
  number={5395},
  pages={1877--1882},
  year={1998},
  publisher={American Association for the Advancement of Science}
}

@article{phmemory,
  title={pH modulates friction memory effects in protein folding},
  author={Dalton, Benjamin A and Netz, Roland R},
  journal={Physical Review Letters},
  volume={133},
  number={18},
  pages={188401},
  year={2024},
  publisher={APS}
}

@article{amplify,
  title={Using receptor conformational change to detect low molecular weight analytes by surface plasmon resonance},
  author={Gestwicki, Jason E and Hsieh, Helen V and Pitner, J Bruce},
  journal={Analytical Chemistry},
  volume={73},
  number={23},
  pages={5732--5737},
  year={2001},
  publisher={ACS Publications}
}

@article{Tnoise,
  title={Fundamental thermal fluctuations in microspheres},
  author={Gorodetsky, Michael L and Grudinin, Ivan S},
  journal={Journal of the Optical Society of America B},
  volume={21},
  number={4},
  pages={697--705},
  year={2004},
  publisher={Optical Society of America}
}

@article{siliconN,
  title={Refractive index of silicon and germanium and its wavelength and temperature derivatives},
  author={Li, HH},
  journal={Journal of Physical and Chemical Reference Data},
  volume={9},
  number={3},
  pages={561--658},
  year={1980},
  publisher={American Institute of Physics for the National Institute of Standards and~…}
}

@article{mittag,
  title={Higher transcendental functions [volumes i-iii]},
  author={Bateman, Harry and others},
  year={1953}
}

@article{HMMmodel,
  title={Analysis of nanopore data using hidden Markov models},
  author={Schreiber, Jacob and Karplus, Kevin},
  journal={Bioinformatics},
  volume={31},
  number={12},
  pages={1897--1903},
  year={2015},
  publisher={Oxford University Press}
}

@article{tptmodel,
  title={Transition-event durations in one-dimensional activated processes},
  author={Zhang, Bin W and Jasnow, David and Zuckerman, Daniel M},
  journal={The Journal of Chemical Physics},
  volume={126},
  pages={074504},
  number={7},
  year={2007},
  publisher={AIP Publishing}
}

@article{tptmodel2,
  title={Transition path dynamics of non-Markovian systems across a rough potential barrier},
  author={Jangid, Pankaj and Chaudhury, Srabanti},
  journal={The Journal of Physical Chemistry A},
  volume={128},
  number={46},
  pages={10041--10052},
  year={2024},
  publisher={ACS Publications}
}

@article{QVequ,
  title={Detection of nanoparticles with a frequency locked whispering gallery mode microresonator},
  author={Swaim, Jon D and Knittel, Joachim and Bowen, Warwick P},
  journal={Applied Physics Letters},
  volume={102},
  number={18},
  pages = {183106},
  year={2013},
  publisher={AIP Publishing}
}

@article{powerlawmemory,
  title={Observation of a Power-Law Memory Kernel for Fluctuations within a Single Protein Molecule},
  author={Min, Wei and Luo, Guobin and Cherayil, Binny J and Kou, SC and Xie, X Sunney},
  journal={Physical Review Letters},
  volume={94},
  number={19},
  pages={198302},
  year={2005},
  publisher={APS}
}

@article{multiScience,
  title={Protein conformational dynamics probed by single-molecule electron transfer},
  author={Yang, Haw and Luo, Guobin and Karnchanaphanurach, Pallop and Louie, Tai-Man and Rech, Ivan and Cova, Sergio and Xun, Luying and Xie, X Sunney},
  journal={Science},
  volume={302},
  number={5643},
  pages={262--266},
  year={2003},
  publisher={American Association for the Advancement of Science}
}

@article{crossover1,
  title={Viscoelastic subdiffusion: From anomalous to normal},
  author={Goychuk, Igor},
  journal={Physical Review E—Statistical, Nonlinear, and Soft Matter Physics},
  volume={80},
  number={4},
  pages={046125},
  year={2009},
  publisher={APS}
}

@article{crossover2,
  title={Subdiffusion from competition between multi-exponential friction memory and energy barriers},
  author={Klimek, Anton and Dalton, Benjamin A and Netz, Roland R},
  journal={The European Physical Journal E},
  volume={48},
  number={8},
  pages={55},
  year={2025},
  publisher={Springer}
}

@article{li2022non,
  title={Non-ergodicity of a globular protein extending beyond its functional timescale},
  author={Li, Jun and Xie, JingFei and Godec, Alja{\v{z}} and Weninger, Keith R and Liu, Cong and Smith, Jeremy C and Hong, Liang},
  journal={Chemical Science},
  volume={13},
  number={33},
  pages={9668--9677},
  year={2022},
  publisher={Royal Society of Chemistry}
}

@article{grossman2015single,
  title={Single-molecule spectroscopy exposes hidden states in an enzymatic electron relay},
  author={Grossman, Iris and Yuval Aviram, Haim and Armony, Gad and Horovitz, Amnon and Hofmann, Hagen and Haran, Gilad and Fass, Deborah},
  journal={Nature Communications},
  volume={6},
  number={1},
  pages={8624},
  year={2015},
  publisher={Nature Publishing Group UK London}
}

@article{ye2018conformational,
  title={Conformational dynamics of a single protein monitored for 24 h at video rate},
  author={Ye, Weixiang and Götz, Markus and Celiksoy, Sirin and Tüting, Laura and Ratzke, Christoph and Prasad, Janak and Ricken, Julia and Wegner, Seraphine V and Ahijado-Guzmán, Rubén and Hugel, Thorsten and others},
  journal={Nano Letters},
  volume={18},
  number={10},
  pages={6633--6637},
  year={2018},
  publisher={ACS Publications}
}

@article{hungerNa,
  title={Tracking Brownian motion in three dimensions and characterization of individual nanoparticles using a fiber-based high-finesse microcavity},
  author={Kohler, Larissa and Mader, Matthias and Kern, Christian and Wegener, Martin and Hunger, David},
  journal={Nature Communications},
  volume={12},
  number={1},
  pages={6385},
  year={2021},
  publisher={Nature Publishing Group UK London}
}

@article{sch2021single,
  title={Single-molecule detection of ultrafast biomolecular dynamics with nanophotonics},
  author={Nüesch, Mark F and Ivanovic, Milos T and Claude, Jean-Beno{\^\i}t and Nettels, Daniel and Best, Robert B and Wenger, J{\'e}r{\^o}me and Schuler, Benjamin},
  journal={Journal of the American Chemical Society},
  volume={144},
  number={1},
  pages={52--56},
  year={2021},
  publisher={ACS Publications}
}

@article{TOC,
  title={Tantalum pentoxide integrated photonics: a promising platform for low-loss planar lightwave circuits with low thermo-optic coefficients},
  author={Liu, Zhenyu and Yao, Wenle and You, Mingjian and Yu, Xiaolun and Ding, Ning and Cheng, Weiren and Li, Zhengqi and Tang, Xingyu and Guo, Fei and Lu, Dan and others},
  journal={ACS Photonics},
  volume={12},
  number={2},
  pages={684--695},
  year={2025},
  publisher={ACS Publications}
}

@article{coronaReview,
  title={Nanoparticle protein corona: from structure and function to therapeutic targeting},
  author={Bashiri, Ghazal and Padilla, Marshall S and Swingle, Kelsey L and Shepherd, Sarah J and Mitchell, Michael J and Wang, Karin},
  journal={Lab on a Chip},
  volume={23},
  number={6},
  pages={1432--1466},
  year={2023},
  publisher={Royal Society of Chemistry}
}

@article{shift2,
  title={Theory of resonance shifts in TE and TM whispering gallery modes by nonradial perturbations for sensing applications},
  author={Teraoka, Iwao and Arnold, Stephen},
  journal={JOSA B},
  volume={23},
  number={7},
  pages={1381--1389},
  year={2006},
  publisher={Optica Publishing Group}
}

@article{ferrhistory,
  title={A Brief History of Ferritin, an Ancient and Versatile Protein},
  author={Arosio, Paolo and Cairo, Gaetano and Bou-Abdallah, Fadi},
  journal={International Journal of Molecular Sciences},
  volume={26},
  pages={206},
  year={2025},
  publisher={MDPI}
}

\section*{Acknowledgments} 

P.S. is the recipient of an Office of National Intelligence National Intelligence Postdoctoral Grant (project number 202310) funded by the Australian Government. The authors acknowledge the use of facilities at the Australian National Fabrication Facility (ANFF) and the Centre for Microscopy and Microanalysis (CMM), the University of Queensland.

\section*{Author contributions}

H.S. performed the experiments and data analysis. W.P.B. conceived and supervised the project. W.P.B., H.S., and S.C.S. wrote the manuscript.  P.S. performed mass photometry experiments. H.S. fabricated the devices with assistance from I.M. All authors discussed the results and provided feedback on the manuscript.



\section*{Funding}

This work was supported by the Australian Research Council Centre of Excellence in Quantum Biotechnology (QUBIC, CE230100021), and the Air Force Office of Scientific Research (AFOSR, FA9550-20-1-0391).

\section*{Competing interests}
The authors declare no competing interests.

\section*{Additional information}

Supplementary information

\end{document}


\section*{Supplementary information} 

\subsection*{Contents}
\noindent
\textbf{I. Slotted photonic crystal cavity for protein sensing} \\
\indent A. Protein signal and cavity noise \middotfill 3 \\
\indent \indent a. Protein signal \middotfill 3 \\
\indent \indent b. Noise spectral densities \middotfill 4 \\
\indent B. Slot design \middotfill 6 \\
\indent \indent a. Electrical double layer \middotfill 6 \\
\indent \indent b. Optical field uniformity \middotfill 8 \\
\indent C. Fabrication and characterization \middotfill 10 \\
\\
\textbf{II. Hard layer} \\
\indent A. Feasibility of hard layer formation \middotfill 11 \\
\indent B. Hard layer noise \middotfill 12 \\
\indent C. Effects of hard layer roughness \middotfill 12 \\
\indent D. Characteristic time estimation \middotfill 13 \\
\\
\textbf{III. Brownian motion fluctuations} \\
\indent A. Harmonic potential model  \middotfill 14 \\
\indent B. Loosely bound proteins  \middotfill 16 \\
\indent C. Optical trapping and loose binding   \middotfill 17 \\
\indent D. Catalase and BSA soft layers  \middotfill 19 \\
\indent E. Measurement bandwidth \middotfill 20 \\
\indent Appendix: Energy in the cavity \middotfill 21 \\
\\
\newpage
\noindent
\textbf{IV. Memroy dependent friction} \\
\indent A. Memory effects  \middotfill 22 \\
\indent B. Diffusion regimes  \middotfill 23 \\
\indent \indent a. Confined Brownian motion \middotfill 23 \\
\indent \indent b. Protein structural fluctuations \middotfill 24 \\
\indent C. Power spectral density for Markovian-like transtiion  \middotfill 26 \\
\\
\textbf{V. Stepping dynamics analysis} \\
\indent A. Hidden Markov model  \middotfill 27 \\
\indent B. Diffusion regimes in each state  \middotfill 27 \\
\indent C. Dwell time  \middotfill 28 \\
\indent D. Transition path time  \middotfill 29 \\
\indent F. Impact of fast motion bandwidth on resolving fast dynamics  \middotfill 31 \\
\\
\textbf{VI. Additional methods} \\
\indent A. Additional soft layer protein datasets  \middotfill 32 \\
\indent \indent a. Catalase and BSA \middotfill 32 \\
\indent \indent b. Apoferritin \middotfill 33 \\
\indent \indent a. Holoferritin \middotfill 34 \\
\indent B. Mass photometry  \middotfill 36 \\
\\
\textbf{Reference}  \middotfill 37 \\

\newpage
\section*{I. Slotted photonic crystal cavity for protein sensing}

\subsection*{A. Protein signal and cavity noise} 

We start by describing the dispersive sensing mechanism~\cite{shift1} for protein sensing, and then discuss thermal noise in the slot cavity.

\subsubsection*{a. Protein signal}

The entry of a single protein into the cavity modifies a local refractive index distribution, thereby perturbing the optical mode. This perturbation induces a resonance angular frequency shift $\delta \omega$, which can be estimated using first-order perturbation theory~\cite{shift1, shift2}
\begin{align}
{\delta \omega \over \omega} &= -{\alpha |\vec{E}^0(\vec{r}_p) |^2 \over 2 \int \epsilon(\vec{r}) |\vec{E}^0(\vec{r})|^2 dV}
\end{align}
where $\omega$ is the bare angular frequency, $\alpha$ is the excess polarizability of the protein, $\vec{E}^0$ is the electric field of the unperturbed mode, $\vec{r}_p$ is the position of the protein, and $\epsilon$ is the permittivity. The wavelength shift depends on the intensity of the electric field at the location of the protein. 
As the protein undergoes Brownian motion inside the slot, the wavelength shift therefore fluctuates according to the electric field distribution. These center of mass motion dependent fluctuations could potentially obscure fluctuations arising from changes in the protein excess polarizability. Thus, both protein entry into the high field region of the slot and a highly uniform field distribution within the slot are desirable for resolving polarizability changes. 

The mode volume $V_m$ is given by 
\begin{align}
V_m={\int \epsilon(\vec{r}) |\vec{E}^0(\vec{r})|^2 dV \over \mathrm{max}(\epsilon(\vec{r}) |\vec{E}^0(\vec{r})|^2)} \,.
\end{align}
At optical wavelengths, the excess polarizability of a spherical protein is given by~\cite{jackson}
\begin{align}
\alpha &= {3 m \epsilon_m \over \rho} {n_p^2-n_m^2 \over n_p^2+2n_m^2}
\end{align}
where $m$ is the mass of the protein, $\epsilon_m$ is the permittivity of the medium, $\rho$ is the density of the protein, $n_p$ is the refractive index of the protein, and $n_m$ is the refractive index of the medium ($n_{\mathrm{D_2O}} = 1.32$)~\cite{water}. We assume the refractive index of proteins~\cite{protein} is $n_p=$1.5, and the density of proteins~\cite{density} is $\rho=$1.35\,g/ml. Based on molecular weights (MWs), the expected excess polarizabilities are 21760, 11240, and 2990\,$\mathrm{\AA^3}$ for apoferritin (MW=480\,kDa), catalase (MW = 248\,kDa), and BSA (MW = 66\,kDa), respectively (see Section VI.B). For holoferritin, the excess polarizability is expected to range from 21760 to 45320\,$\mathrm{\AA^3}$, depending on the ferrihydrite content (see Section VI.B).

\subsubsection*{b. Noise spectral densities}

\begin{figure*}[ht!]
    \centering
    \includegraphics[width=0.5\textwidth]{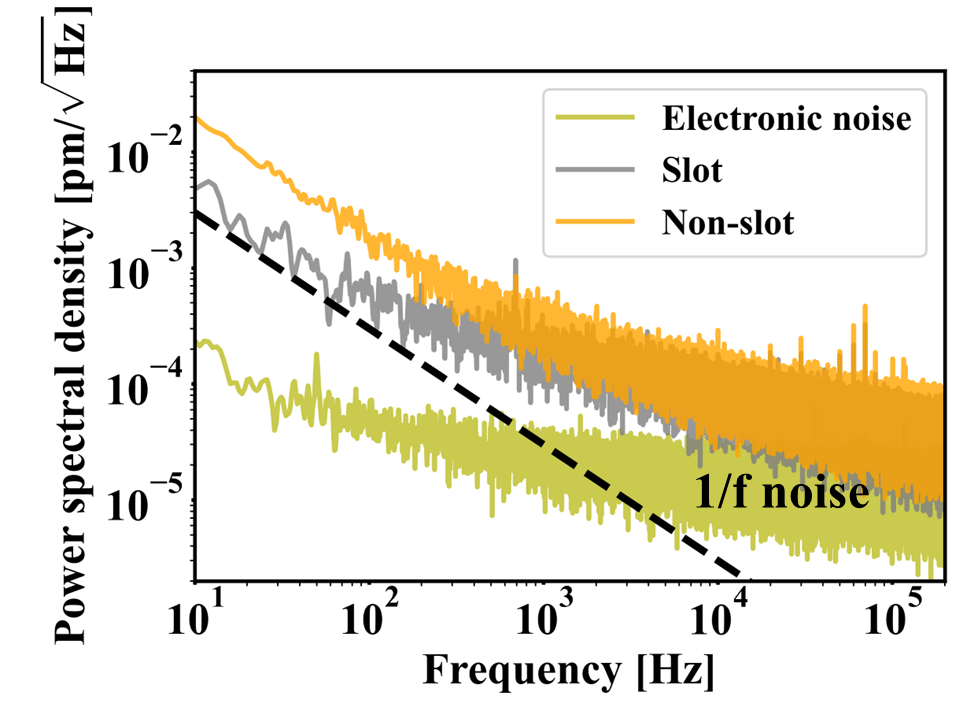}
    \caption{
    \textbf{Comparison of noise spectral densities.}
    The power spectral densities of non-slot and slot resonators in the absence of protein, and electronic noise. The black dashed line indicates the $1/f$ noise behavior.
    }
    \label{fig:noise}
\end{figure*}

We measure the noise on the red side of the resonance and obtain the noise spectral density (Fig.~\ref{fig:noise}), and then compare the signal-to-noise ratio (SNR) with and without a slot. The spectrum is dominated by thermal noise in the super-kilohertz range, whereas at low frequencies it is dominated by $1/f$ noise. A comparison between cavities with and without a slot reveals that slot photonic crystal cavities provide an improved signal to noise ratio (SNR), particularly in the sub-kilohertz regime. One possible explanation is that technical noise is due to thermal drifts, which scales with $(dn/dT)^2$~\cite{Tnoise}, where $n$ is the refractive index of the material sampled by the optical mode and $T$ is the temperature (here 293 K). The improvement arises because the slot design confines most of the optical field confine in heavy water rather than in silicon ($n_{\mathrm{silicon}}=3.47$)~\cite{siliconN}. Furthermore, once contributions from fiber and water motion are removed, measurements on a non-undercut device in air show no difference from those of the picked-up cavity. This indicates that the fiber-coupling technique does not introduce additional noise.

\newpage
\subsection*{B. Slot design}

In this subsection, we discuss the key considerations for the slot design in protein sensing, as introduced in Section I.A.a. We first examine the interaction between the silicon surface and proteins to facilitate protein entry into the slot, and then detail the requirement for a highly uniform optical field.  

\subsubsection*{a. Electrical double layer}

\begin{figure*}[ht!]
    \centering
    \includegraphics[width=0.9\textwidth]{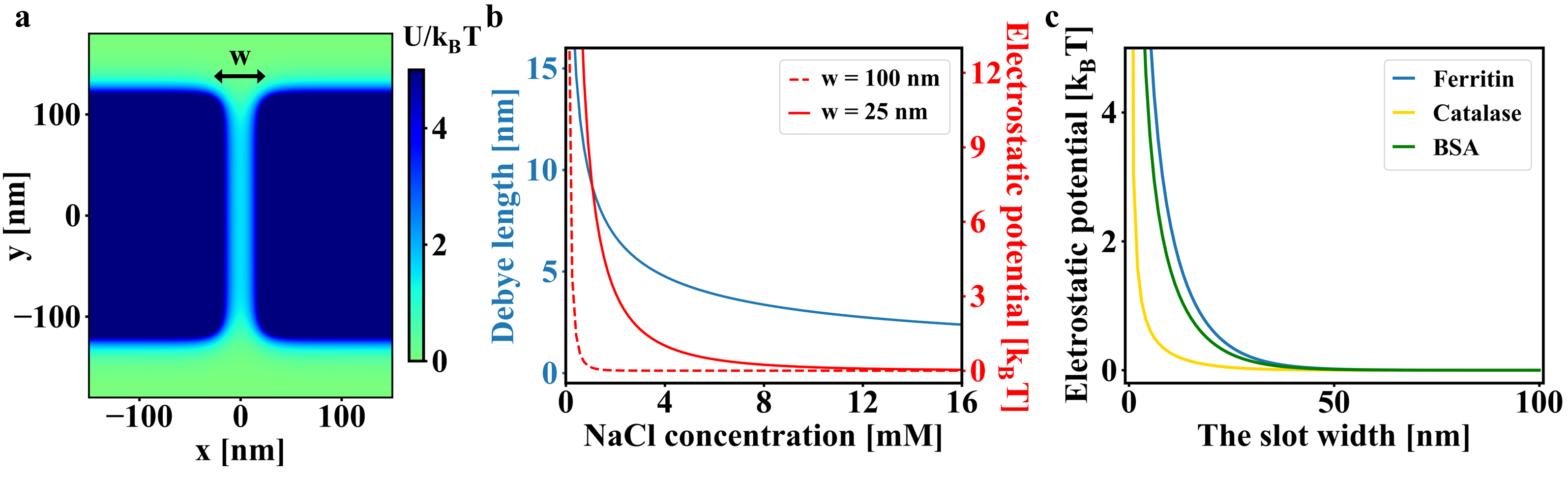}
    \caption{
    \textbf{Investigating salt and slot width effects.}
    \textbf{a},~The electrostatic potential distribution experienced by ferritin at 1\,mM NaCl across the slot cross-section.
    \textbf{b},~The Debye length and the electrostatic potential energy at the midplane ($x$=0) experienced by ferritin versus NaCl concentration.
    \textbf{c},~The electrostatic potential energy at the slot midplane versus slot width at 5\,mM NaCl for ferritin, catalase, and bovine serum albumin (BSA).
    }
    \label{fig:EDL}
\end{figure*}

Silicon surfaces become negatively charged in aqueous solutions and thereby interact with charged proteins~\cite{dark}, including ferritin, catalase, and bovine serum albumin (BSA). Salt plays an important role in electrolytic solutions by forming an electrical double layer on surface charges. At high salt concentrations, this layer facilitates the entry of proteins by screening the electrostatic interactions between charged proteins and surfaces, while also reducing the polarizability enhancement and thereby decreasing the protein signal~\cite{dark, surface2}. To identify the balance between these two effects, we evaluate the distributions of electrostatic potential energy $U_e=q \phi$, where $q$ is the net charge of the protein and $\phi$ is the electrostatic potential.

To calculate the electrostatic potential distributions, We perform finite element method (FEM) simulations based on the Poisson-Boltzmann equation, which considers the ion distributions in solution~\cite{PBE, PBE2}. We use a binary monovalent electrolyte (NaCl), so the Poisson-Boltzmann equation is given by
\begin{align}
\nabla \cdot (\epsilon \nabla \phi)  = 2 Iq_0 \sinh \Big({q_0 \phi \over k_{\mathrm{B}}T} \Big), \quad I=1000CN_{\mathrm{A}}
\end{align}
where $I$ is the ionic strength of the binary monovalent electrolyte, $q_0$ is the elementary charge, $k_{\mathrm{B}}$ is the Boltzmann constant, $C$ is the molar concentration of the electrolyte, and $N_{\mathrm{A}}$ is the Avogadro number. The surface charge density of silicon~\cite{Sicharge} is $\sigma_s = -10^{16}q_0\,\mathrm{m}^{-2}$, and the permittivity of heavy water~\cite{d2oeta} is $\epsilon_{\mathrm{D_2O}} = 78 \epsilon_0$, where $\epsilon_0$ is the vacuum permittivity. 
The surface potential $\phi_s$ is estimated by solving the Poisson-Boltzmann equation, $\phi_s = (2k_{\mathrm{B}} T/q_0) \sinh^{-1}(\sigma_s/\sqrt{8\epsilon k_{\mathrm{B}}T I})$. For $C\ge1$\,mM, the dimensionless potential remains relatively small ($|{q_0\phi/k_{\mathrm{B}}T}| < 0.8$), allowing the Poisson-Boltzmann equation to be approximately linearized. Based on the linearized Poisson-Boltzmann equation, the electrostatic potential at the midplane inside the slot is given by~\cite{PBE, PBE2}
\begin{align}
\phi_c &= {\lambda_d \sigma_s \over \epsilon \sinh({w \over 2 \lambda_d})}, \quad \lambda_d = \Big({\epsilon k_{\mathrm{B}} T \over 2Iq_0^2} \Big)^{1/2}
\end{align}
where $\lambda_d$ is the Debye length and $w$ is the slot width. The thickness of the electrical double layer is characterized by the Debye length, which decreases as the electrolyte concentration increases. 

To calculate the effective charge of the proteins, we measure their zeta potentials in solution (Zetasizer ZS90, Malvern Inc.). The measured zeta potentials $\psi$ in 5\,mM NaCl are -14.5, -2.3, and -22.5\,mV for ferritin, catalase, and BSA, respectively. From these zeta potentials, the diffuse layer surface charge density of the protein $\sigma_d$ is estimated using the semi-empirical equation~\cite{sigma}
\begin{align}
\sigma_d &= -{\epsilon k_{\mathrm{B}} T \over \lambda_d q_0} \Big( 2 \sinh({q_0 \psi \over 2 k_{\mathrm{B}} T}) +{4 \lambda_d \over r_0}\tanh({q_0 \psi \over 4 k_{\mathrm{B}} T})  \Big)
\end{align}
where $r_0$ is the physical radius of the protein. The effective charge is approximately calculated as $q_{\mathrm{eff}} \approx 4\pi r_h^2 \sigma_d$, where $r_h$ is the hydrodynamic radius. The hydrodynamic radii of ferritin~\cite{ferrR}, catalase~\cite{cataR}, and BSA~\cite{cataR} are 6.3, 5.12, and 3.48\,nm, respectively. Using these values, we calculate the electrostatic potential energy distributions as $U_e \approx q_{\mathrm{eff}} \phi$.

The charged silicon forms a repulsive potential distribution within the slot (Fig.~\ref{fig:EDL}a), whose magnitude decreases as the NaCl concentration increases (Fig.~\ref{fig:EDL}b). A previous study reported protein detection at up to 5\,mM NaCl~\cite{dark}, where the Debye length is approximately 4\,nm. We similarly utilize a Debye length of roughly 3-5\,nm, corresponding to a NaCl concentration range of 4-12\,mM (Fig.~\ref{fig:EDL}b). 
Because the electrostatic potential energy of the three proteins at the slot midplane ($x$=0) becomes nearly zero at 5\,mM NaCl for slot widths above 50\,nm (Fig.~\ref{fig:EDL}c), we select a slot width of 55\,nm. 

\subsubsection*{b. Optical field uniformity}

\begin{figure*}[ht!]
    \centering
    \includegraphics[width=0.8\textwidth]{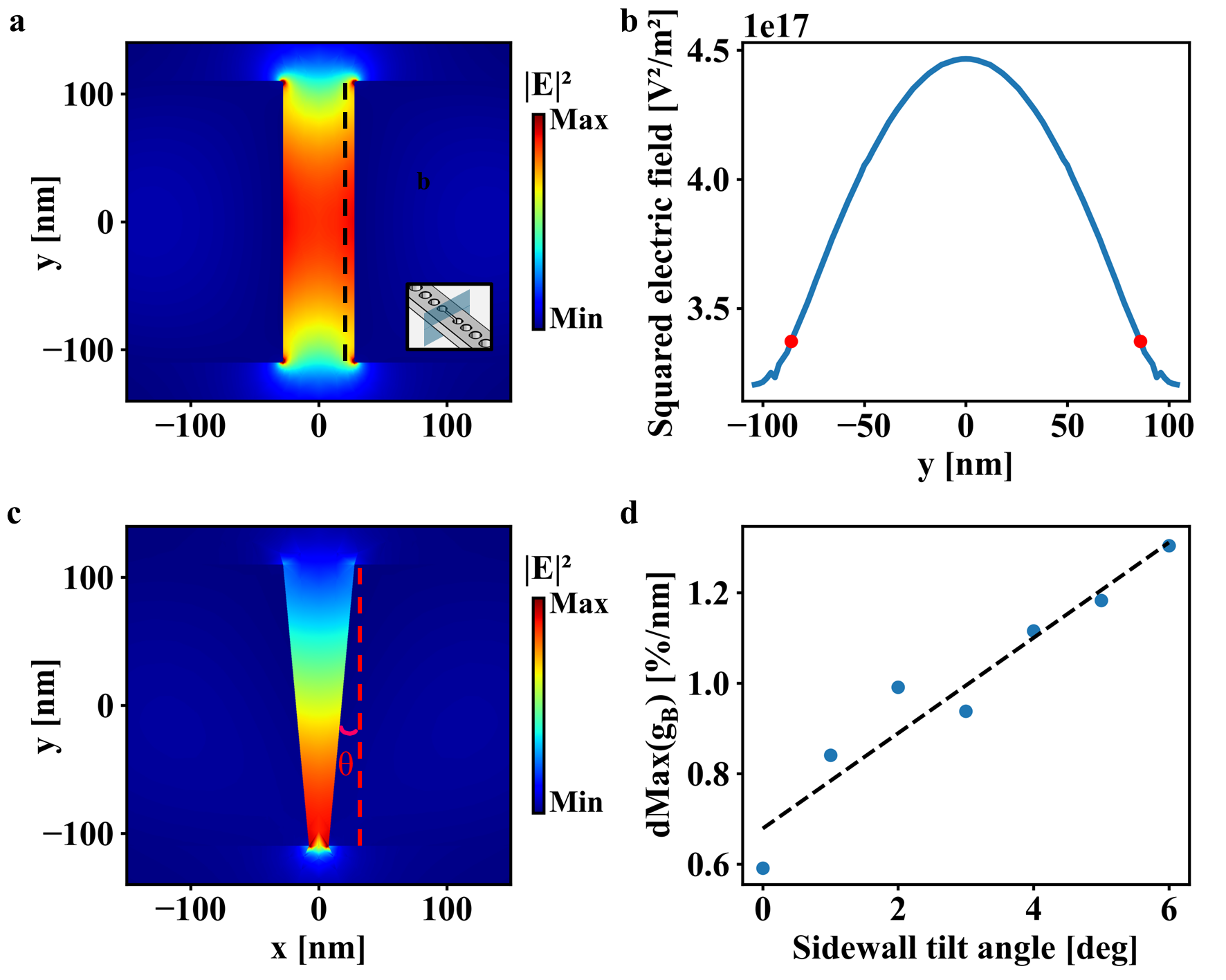}
    \caption{
    \textbf{Optical field uniformity on the cavity surface.}
    \textbf{a},~The optical field distribution for a vertical sidewall slot. \textbf{b},~The corresponding optical field profile along the black dashed lines in \textbf{a}.
    \textbf{c},~The optical field distribution for a slot tilted by an angle $\theta$.
    \textbf{d}, The maximum $g_{\mathrm{B}}$ versus sidewall tilt angle. 
    } 
    \label{fig:verti}
\end{figure*}

We next investigate how to achieve high optical field uniformity in slot cavities to suppress the translational Brownian noise. To examine the field uniformity, we simulate the optical field distributions and define the Brownian coupling strength $g_{\mathrm{B}}=|d(\ln |E|^2) /dx|$, which represents the local fractional signal variation per unit displacement of the protein. To avoid meshing artifacts at the slot edges, we evaluate $g_{\mathrm{B}}$ over the range from -95 to 95\,nm along the direction parallel to the surface. For a slot with perfectly smooth and vertical sidewalls (Fig.~\ref{fig:verti}a), the mean $g_{\mathrm{B}}$ over the slot surface is approximately 0.3$\%$/nm, while the maximum value reaches nearly 0.6$\%$/nm (red points in Fig.~\ref{fig:verti}d). The field uniformity is degraded by tilted sidewalls (Fig.~\ref{fig:verti}b). The maximum $g_{\mathrm{B}}$ increases approximately linearly with the sidewall tilt angle $\theta$ at a rate of about $0.1\%/\mathrm{deg}$ (Fig.~\ref{fig:verti}e). This tolerance remains permissive because, even near closure of the opposite slot width ($\theta \approx 6^{\circ}$), the maximum $g_{\mathrm{B}} \approx 1.3\%/\mathrm{nm}$ remains more than one order of magnitude lower than plasmonic approaches ($\approx$ 20$\%$ per nanometre)~\cite{nano}. 

\newpage
\subsection*{C. Fabrication and characterization}

\begin{figure*}[ht!]
    \centering
    \includegraphics[width=0.95\textwidth]{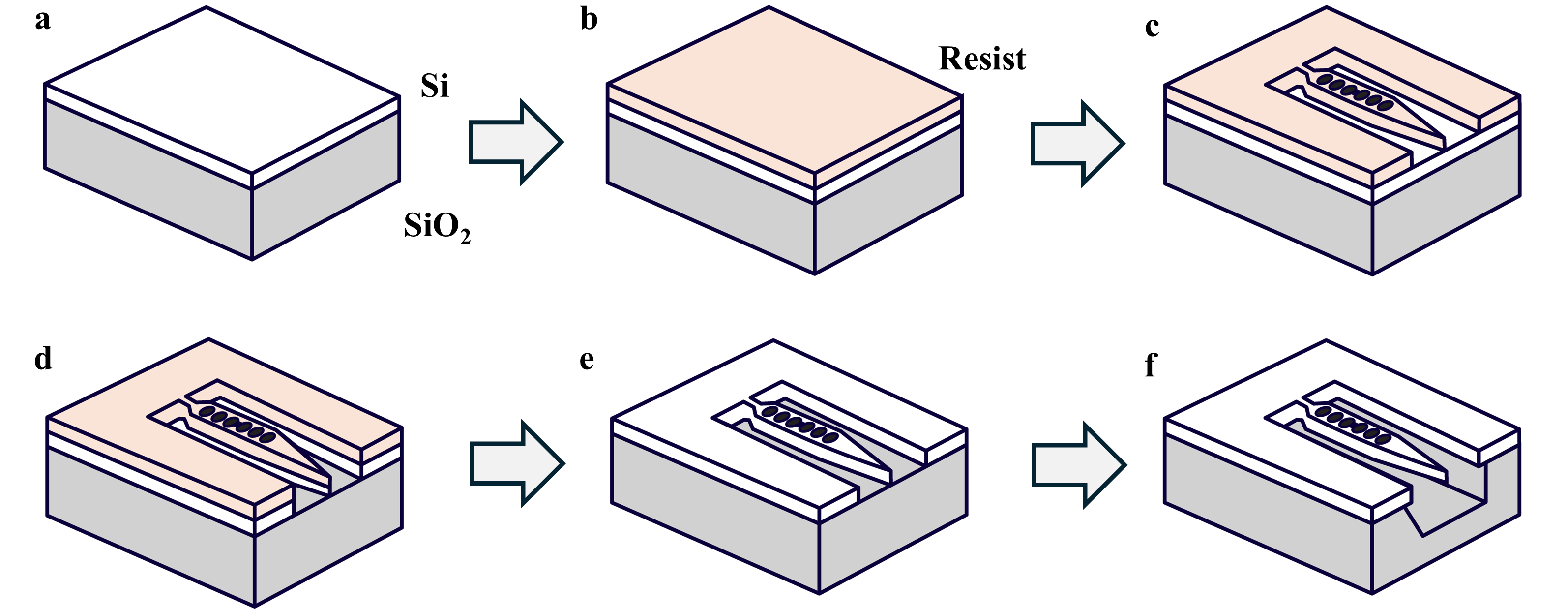}
    \caption{
    \textbf{Representation of the fabrication process.}
    \textbf{a},~Silicon on insulator (SOI) chip. 
    \textbf{b},~Spin coating and baking of ARP e-beam resist.
    \textbf{c},~Electron-beam exposure and development.
    \textbf{d},~Inductively coupled plasma reactive-ion etching of silicon.
    \textbf{e},~Resist removal.
    \textbf{f},~Undercutting of the SiO${}_2$ layer by vapor HF etching.
    }
    \label{fig:fab}
\end{figure*}

\begin{figure*}[ht!]
    \centering
    \includegraphics[width=0.95\textwidth]{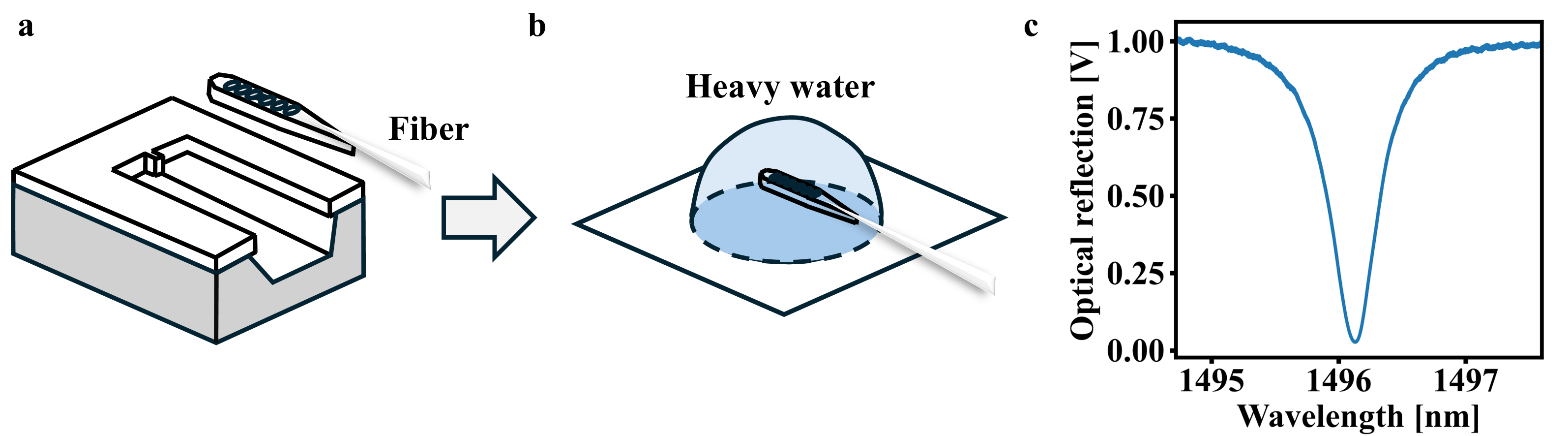}
    \caption{
    \textbf{Schematic of the pickup process.}
    \textbf{a},~A tapered optical fiber sticks to the end of the optical cavity device. By swinging the fiber, the tether is broken and the device is lifted.
    \textbf{b},~The optical cavity device is immersed in a droplet of heavy water.
    \textbf{c},~The optical reflection spectrum of the fabricated optical cavity device in heavy water.
    }
    \label{fig:pick}
\end{figure*}

This subsection presents simplified schematics of the fabrication procedure (Fig~\ref{fig:fab}) and the picking up process (Fig.~\ref{fig:pick}c) for the slot device, as described in the Methods section of the main text. The detached slot device is characterized in a droplet of heavy water (Fig.~\ref{fig:pick}b). We find that the quality factor $Q$ is approximately 3000 (Fig.~\ref{fig:pick}c).

\clearpage
\newpage
\section*{II. Hard layer}

Beyond facilitating protein entry into the slot, salt plays an important role in protein-surface interactions~\cite{monosalt}. By screening the surface potential, salt reduces electrostatic repulsion and promotes stable adsorption of negatively charged proteins onto the silicon surface. We term this protein layer the `hard layer' in analogy to the `hard protein corona'~\cite{hardcorona}. In this section, we first estimate the protein concentration required for hard layer formation. We then analyze the noise arising from the formed hard layer, the optical field nonuniformity induced by hard layer roughness, and the characteristic diffusion time on the hard layer surface.

\subsection*{A. Feasibility of hard layer formation}

We evaluate the feasibility of hard layer formation by estimating the number of proteins required to fully cover the outer surface of the slot device. The total surface area of the device, with dimensions of $220\times 600 \times 7400\,\mathrm{nm}^3$, is approximately $A_{\mathrm{slot}} \approx 12.4\,\mu \mathrm{m}^2$. Assuming a simple cubic packing of protein on the surface, the effective footprint of the single protein is defined as $A_{\mathrm{protein}} = d_h^2$, where $d_h$ is the hydrodynamic diameter. For ferritin ($d_h \approx 12.6$\,nm), the number of proteins required for full coverage is estimated to be $N_{\mathrm{cover}} = A/A_{\mathrm{protein}} \approx 8 \times10^4$. In a solution volume of $V_s \approx 500\,\mu$L with a protein concentration of $C_{\mathrm{protein}}=$1\,nM, the total number of proteins is estimated as $N_{\mathrm{protein}} \approx N_A  V_s C_{\mathrm{protein}} \approx 3\times 10^{11}$, which exceeds $N_{\mathrm{cover}}$ by more than six orders of magnitude. Protein adsorption may also occur on the coverslip surface in contact with the solution. The wetted coverslip area is approximated as a circular footprint with a droplet diameter of 12\,mm ($A_{\mathrm{coverslip}} \approx 120\,\mathrm{mm}^2$). Even if adsorption occurs across the entire wetted coverslip area, the number of proteins in the solution is sufficient to achieve a surface coverage on the order of 40$\%$.

\newpage
\subsection*{B. Hard layer noise} 

\begin{figure*}[ht!]
    \centering
    \includegraphics[width=0.5\textwidth]{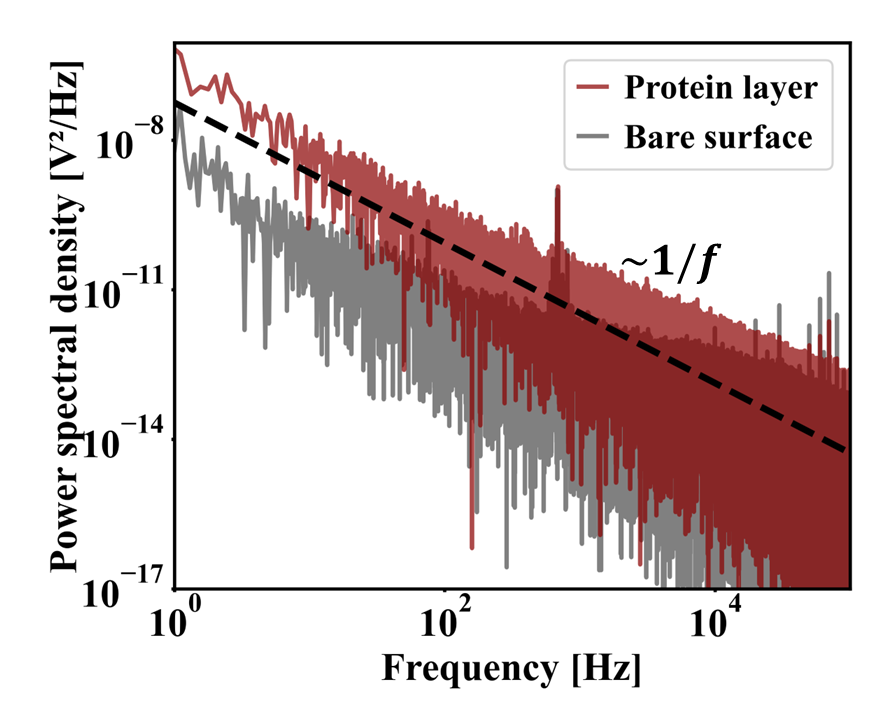}
    \caption{
    \textbf{Comparison of noise levels.} 
    The power spectral density measured before and after hard ferritin layer formation. The black dashed line indicates the $1/f$ fit.
    }
    \label{fig:mononoise}
\end{figure*}

We observe that the noise level increases after hard layer formation. A typical comparison of the power spectral densities (PSDs) before and after ferritin layer formation is shown in Fig.~\ref{fig:mononoise}. At low frequency, the noise is dominated by hard layer noise characterized by $1/f$ behavior, but above approximately 100\,kHz, it becomes comparable to thermal noise.

\subsection*{C. Effects of hard layer roughness} 

\begin{figure*}[ht!]
    \centering
    \includegraphics[width=0.5\textwidth]{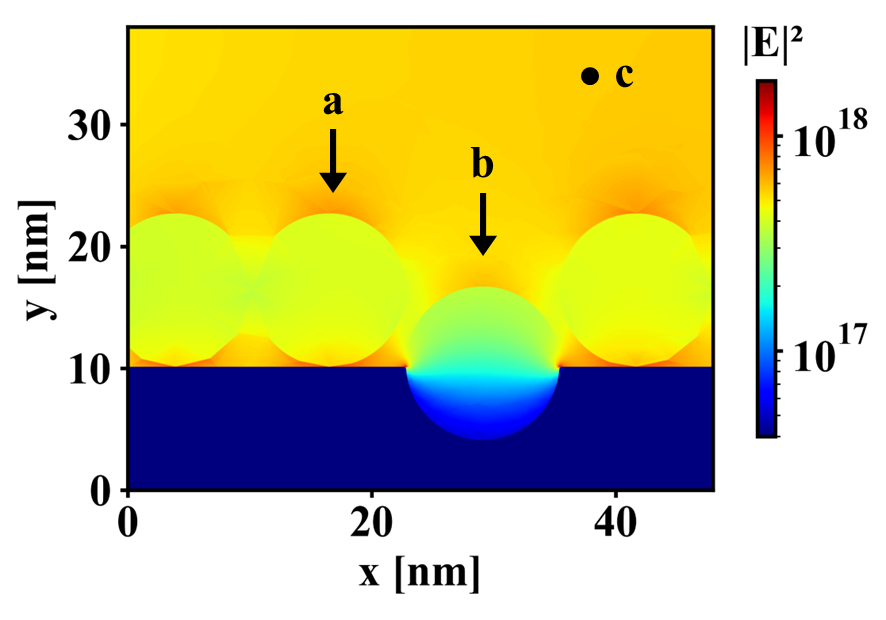}
    \caption{
    \textbf{Effect of surface roughness on the optical field distribution.} The squared electric field distribution showing local field enhancement induced by surface roughness. \textbf{a} indicates a point above ferritin adsorbed on the flat surface. \textbf{b} indicates a point above ferritin located inside the nanoscale pit. \textbf{c} indicates a point in the surrounding medium far from the hard layer.
    }
    \label{fig:hotspot}
\end{figure*}

To investigate the effect of surface roughness arising from the hard layer and the silicon surface, we perform FEM simulations. Nanoscale pits with sizes comparable to that of ferritin ($r_h$ = 6.3\,nm) are introduced on the silicon surface and assumed to be filled with ferritin~\cite{corona1}. The squared electric field intensity above ferritin located inside the pit (Fig.~\ref{fig:hotspot}b) becomes similar to that in the buffer region far from the hard layer (Fig.~\ref{fig:hotspot}c). The field intensity in these two cases is approximately 25$\%$ lower than that above ferritin adsorbed on the flat surface (Fig.~\ref{fig:hotspot}a). As a result, this surface heterogeneity accounts for only a 1.25-fold difference in the signals between the two states.

\subsection*{D. Characteristic time estimation} 

Diffusion across these nanoscale pits, or the rotational motion of a tethered protein around a pit, may generate two state signals. We estimate the characteristic timescales for these two cases using the relations $\tau_D \sim l^2/D$ and $\tau_{D_r} \sim (\Delta \theta)^2/D_r$, respectively, where $l$ is the distance to a pit, $\Delta \theta$ is the rocking angle, $D$ is the diffusion coefficient, and $D_r$ is the rotational diffusion coefficient. For a spherical protein, the drag coefficient is given by Stokes' law $\gamma = 6\pi \eta r_h$, where $\eta$ is the viscosity of the medium ($\eta_{\mathrm{D_2O}} = 1.25$\,mPa$\cdot$s)~\cite{d2oeta}. Based on the Stokes--Einstein and Stokes--Einstein--Debye relations, diffusion and rotational diffusion coefficients are given by
\begin{align}
D &= \frac{k_B T}{6 \pi \eta r_h}
\end{align}
and
\begin{align}
D_r &= \frac{k_B T}{8 \pi \eta r_h^3} \, .
\end{align}
For ferritin, $D$ and $D_r$ are approximately 30\,$\mu$m$^2$/s and 5\,$\mu$s$^{-1}$, respectively. Assuming $l \approx$ 6-18\,nm and $\Delta \theta \approx$ 1-3\,rad, the characteristic time scales for both processes are estimated to be on the order of 1-10\,$\mu$s. These time scales may become longer due to wall effects~\cite{welleffect} or protein-protein interactions.

\clearpage
\newpage
\section*{III. Brownian motion fluctuations}

In this section, we compare the fluctuations arising from the translational Brownian motion under optical trapping and loosely bound conditions.

\subsection*{A. Harmonic potential model}

The squared electric field inside the slot follows an $x^2$ dependence, as shown in Fig~\ref{fig:verti}d. The Brownian motion of a protein in a harmonic potential is described by the Langevin equation~\cite{Langevin}, given by
\begin{align}
m \ddot{x} (t)  &= - \gamma \dot{x} (t) - k x(t) + \sqrt{2 k_{\mathrm{B}} T \gamma} \xi(t)
\end{align}
where $k$ is the spring constant and $\xi$ is the Gaussian white noise term with zero mean. Since the characteristic time for velocity relaxation due to friction is much shorter than the sampling resolution,  we assume that the system is overdamped ($m \approx 0$). Because the optical field distribution is approximately radially symmetric, the wavelength shift in the optical potential is expressed as $\delta \lambda (t) = - \beta a_e r^2(t) + \delta \lambda_{\mathrm{max}}$ where $\beta = \pi c \alpha / (\omega \int_V \epsilon(\vec{r}) |\vec{E}^0(\vec{r})|^2 dV)$ and $\delta \lambda_{\mathrm{max}} = \beta b_e$. To convert the simulation results to physical values, the squared electric field is scaled using a field normalization constant $c_e$, such that $E_{\mathrm{phys}}^2 = c_e E^2$, where $E^2(x) = -a_ex^2 + b_e$, and $a_e$ and $b_e$ are fitting parameters (see Section III.appendix).

Given the quadratic dependence of the wavelength shift, the autocorrelation function is given by 
\begin{align}
C_s(\tau)&= \langle (\delta \lambda (t) - \delta \lambda_{\mathrm{mean}})(\delta \lambda (t+\tau) - \delta \lambda_{\mathrm{mean}}) \rangle \nonumber \\
&=4\beta^2 a_e^2 {k_{\mathrm{B}}^2 T^2 \over k^2} \mathrm{e}^{-{ 2k \over \gamma} \tau}
\end{align}
where $\tau$ is the lag time. The corresponding PSD is obtained by Fourier transforming the autocorrelation function, giving
\begin{align}
S_s(\omega) &= \int^{\infty}_{-\infty} C_s(\tau)\mathrm{e}^{-i\omega \tau} d\tau \nonumber \\
&=  {4 \beta^2 a_e^2 D^2 \over \omega_c^3 (1+ {\omega^2 / 4 \omega_c^2})} 
\end{align}
where $\omega_c = k/\gamma$ is the corner frequency. The PSD of confined Brownian motion is expected to exhibit a characteristic Lorentzian shape~\cite{PSD}. Based on Boltzmann statistics $p = C\mathrm{e}^{-U/k_{\mathrm{B}} T}$, where $C$ is the normalization constant and $U$ is the potential, the radial probability density distribution in the optical potential $U=-\alpha E^2/4$ is given by 
\begin{align}
p_s(r) = {\alpha a_e r \over 2 k_{\mathrm{B}} T} \exp\left(-{\alpha a_e \over 4k_{\mathrm{B}} T}r^2\right)
\end{align}
where $p_s(r)$ is defined according to $\int p_s(r) dr = 1$. The wavelength shift is bounded as $\delta \lambda \leq \delta \lambda_{\mathrm{max}}$, since $r=\sqrt{( \delta \lambda_{\mathrm{max}} - \delta \lambda) / \beta a_e }$. The corresponding distribution for the wavelength shift is derived as
\begin{align}
p_s(\delta \lambda) &= p_s(r) \Big |{dr \over d\delta\lambda} \Big| \nonumber\\
&= {\alpha c_e\over 4 k_{\mathrm{B}} T \beta} \exp\left( {\alpha c_e\over  4 k_{\mathrm{B}} T \beta} (\delta \lambda-\delta \lambda_{\mathrm{max}}) \right ) \, .
\end{align}
The wavelength shift histogram in optical trapping shows an asymmetric distribution, increasing exponentially and sharply dropping to zero at the maximum wavelength shift. 

\newpage
\subsection*{B. Loosely bound proteins}

Proteins interacting with the hard layer are expected to be loosely bound within a nanometer scale region~\cite{inactive1, BSAinter}. We call this layer formed by loosely bound proteins the `soft layer'~\cite{hardcorona}. When these short range interactions occur within the optical potential, the wavelength shift can be approximated by a linear relation, $\delta \lambda (t)\approx \beta a_l(x(t)-x_0) + \delta \lambda_{\mathrm{mean}}$. Given the linear dependence of the wavelength shift on the position, the autocorrelation function is given by 
\begin{align}
C_l(\tau)=\beta^2 a_l^2 {k_{\mathrm{B}} T \over k_{\mathrm{eff}}} \exp\left(-{ k_{\mathrm{eff}} \over \gamma} \tau\right)
\end{align}
where $k_{\mathrm{eff}}$ is the effective spring constant. The PSD for loosely bound proteins is therefore given by
\begin{align}
S_l(\omega) 
&= \int^{\infty}_{-\infty} C_l(\tau)\mathrm{e}^{-i\omega \tau} d\tau \nonumber \\
&= \beta^2 a_l^2{2D  \over \omega^2 + \omega_c^2} \, .
\end{align}
The PSD exhibits a Lorentzian shape, and the variance of the signal is $\sigma_l^2 = C_l(0) = (\beta a_l)^2{k_{\mathrm{B}} T / k_{\mathrm{eff}}}$. Under the local harmonic approximation, $U_l(x) \approx U_l(x_0) + k_{\mathrm{eff}}(x-x_0)^2/2$, the position probability distribution is given by
\begin{align}
p_l(x) \approx \sqrt{k_{\mathrm{eff}} \over 2 \pi k_{\mathrm{B}} T} \exp\left({-{k_{\mathrm{eff}} \over 2k_{\mathrm{B}} T}(x-x_0)^2}\right) \, .
\end{align}
By applying the variable transformation, the corresponding distribution for wavelength shift is derived as
\begin{align}
p_l(\delta \lambda) &= p_l(x) \Big |{dx \over d\delta\lambda} \Big| \nonumber\\
&= {1 \over \sqrt{2 \pi \sigma_l^2}} \exp\left(-{(\delta \lambda-\delta \lambda_{\mathrm{mean}})^2 \over 2 \sigma_l^2}\right) \, .
\end{align}
The wavelength shift follows a Gaussian distribution. The distribution is slightly skewed when the protein is bound near the center of the optical potential or when it moves over a broad region within the optical potential well. By applying Boltzmann statistics to the wavelength shift distribution, the corresponding effective potential is given by
\begin{align}
U_l(\delta \lambda) &={k_B T\over 2\sigma_l^2}(\delta \lambda-\delta \lambda_{\mathrm{ref}})^2 \,.
\end{align}

\newpage
\subsection*{C. Optical trapping and loose binding}

\begin{figure*}[ht!]
    \centering
    \includegraphics[width=\textwidth]{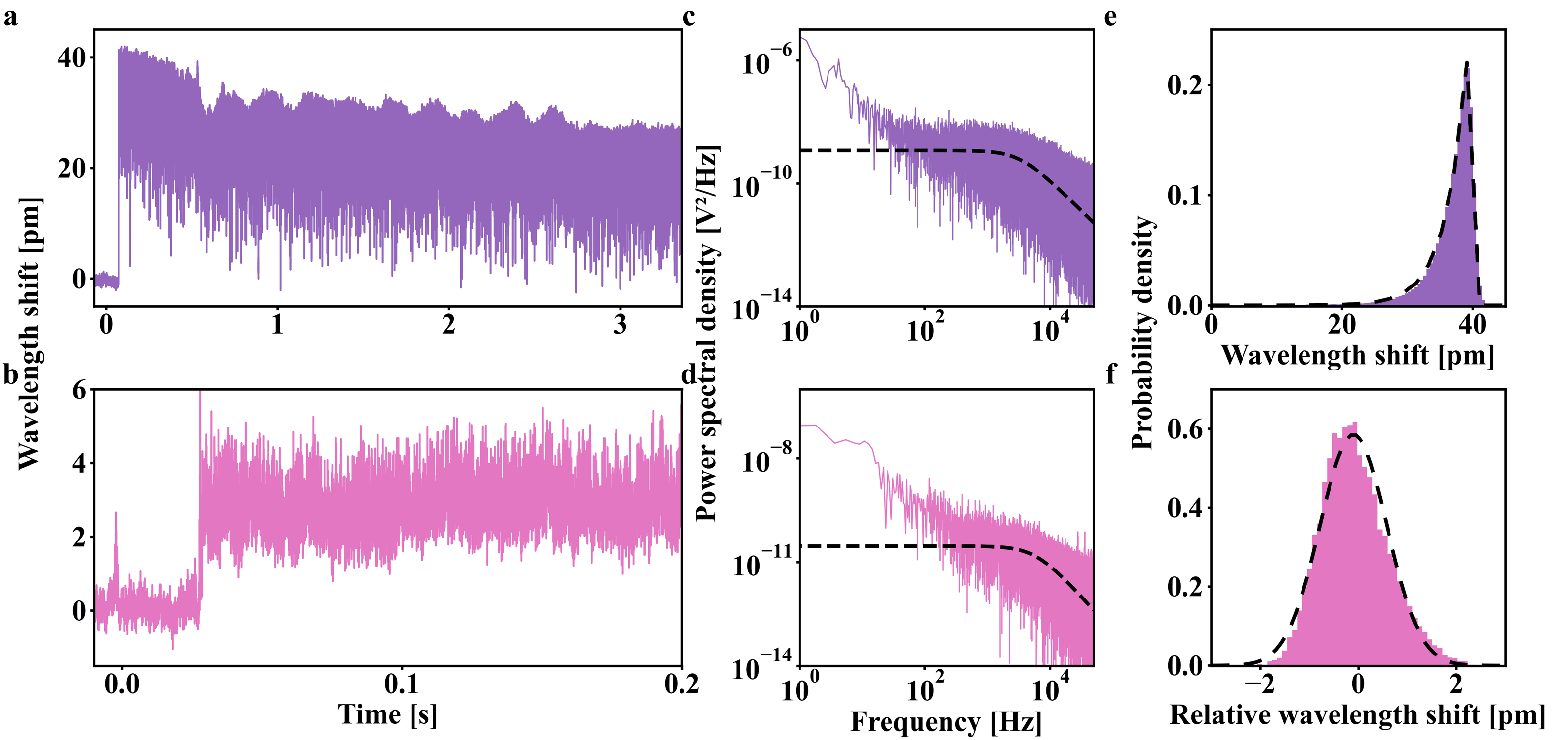}
    \caption{
    \textbf{Two types of signals from a 15\,nm radius polystyrene particle.} 
    Time traces showing (\textbf{a})~optical trapping and (\textbf{b})~interacting to the hard layer. We use a laser power of 10\,mW in (\textbf{a}) and 5\,mW in (\textbf{b}).
    \textbf{c-d},~The corresponding power spectral densities, with the black dashed lines representing Lorentzian fits.
    \textbf{e-f},~The corresponding histograms of the wavelength shifts, with the black dashed lines representing theoretical fits.
    A 10\,Hz high-pass filter is applied before generating histograms. 
    }
    \label{fig:optipoly}
\end{figure*} 

Protein signals depend on both the centre of mass motion of the protein and fluctuations in its polarizability induced by structural changes. To evaluate the suppression of Brownian motion noise within the soft layer, we use protein-scale polystrene particles. We observe two distinct types of signals from a 15\,nm radius polystyrene particle, as shown in Fig.~\ref{fig:optipoly}. The experiments are conducted in a solution comprising $20\%$ Dulbecco’s phosphate buffered saline (DPBS) and $80\%$ heavy water. The refractive index of polystyrene is assumed to be $n_{\mathrm{poly}}=1.59$.

\begin{figure*}[ht!]
    \centering
    \includegraphics[width=\textwidth]{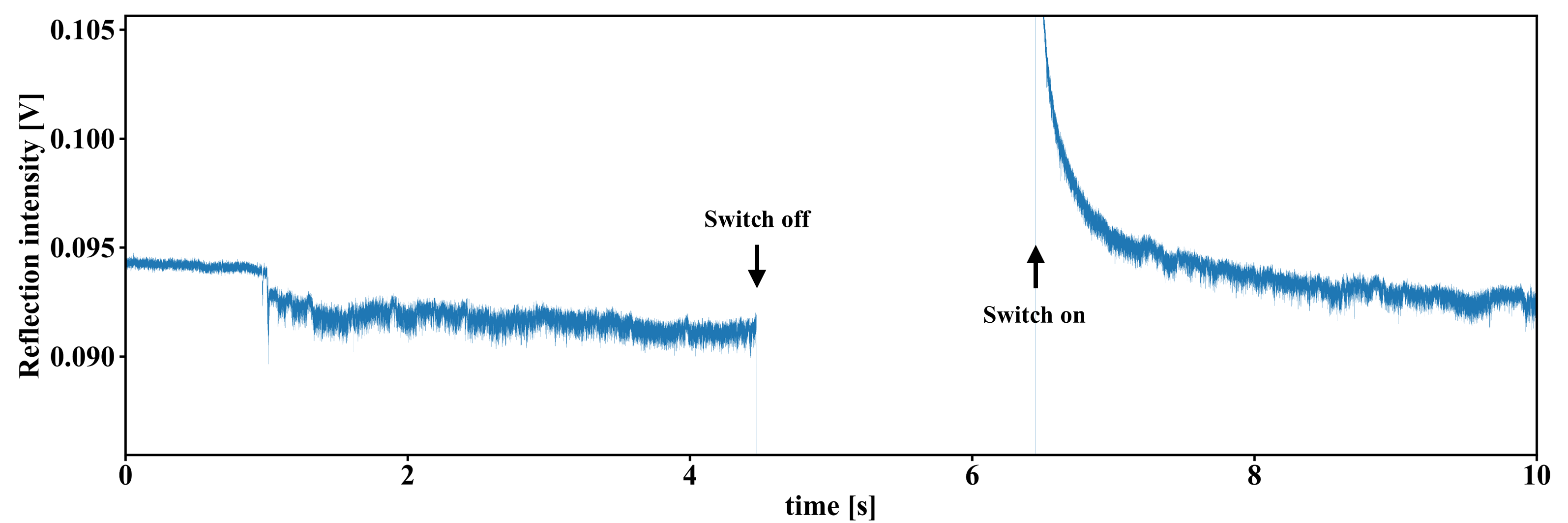}
    \caption{
    \textbf{Signal remaining after switching the optical power off and on.} 
    Time trace of ferritin-ferritin interactions before and after optical power switching.
    }
    \label{fig:switch}
\end{figure*} 

As shown in Fig.~\ref{fig:optipoly}a-d, these signals differ from the adsorption events observed during the hard layer formation, characterized by enhanced fluctuations. The first type of trace (Fig.~\ref{fig:optipoly}a) exhibits a clear Lorentzian shape in the corresponding PSD (Fig.~\ref{fig:optipoly}c), while its wavelength shift distribution exhibits an exponential increase up to its maximum shift value (Fig.~\ref{fig:optipoly}e). These features are consistent with optical trapping. From fitting the distribution, we estimate the particle radius to be 13.8\,nm. By contrast, the other type of signal (Fig.~\ref{fig:optipoly}b) does not follow this trend. Although its PSD also exhibits a Lorentzian shape (Fig.~\ref{fig:optipoly}d), the signal distribution is a nearly Gaussian distribution (Fig.~\ref{fig:optipoly}f). In addition, this type of signal remains even after the laser is turned off (Fig.~\ref{fig:switch}). These observations suggest that the second type of signal arises from a loosely bound protein. We therefore attribute this signal to proteins loosely bound to the hard layer. Compared with the signal distribution produced by optical trapping, this signal exhibits an approximately fivefold suppression of Brownian motion noise.

\newpage
\subsection*{D. Catalase and BSA soft layers}

\begin{figure*}[ht!]
    \centering
    \includegraphics[width=\textwidth]{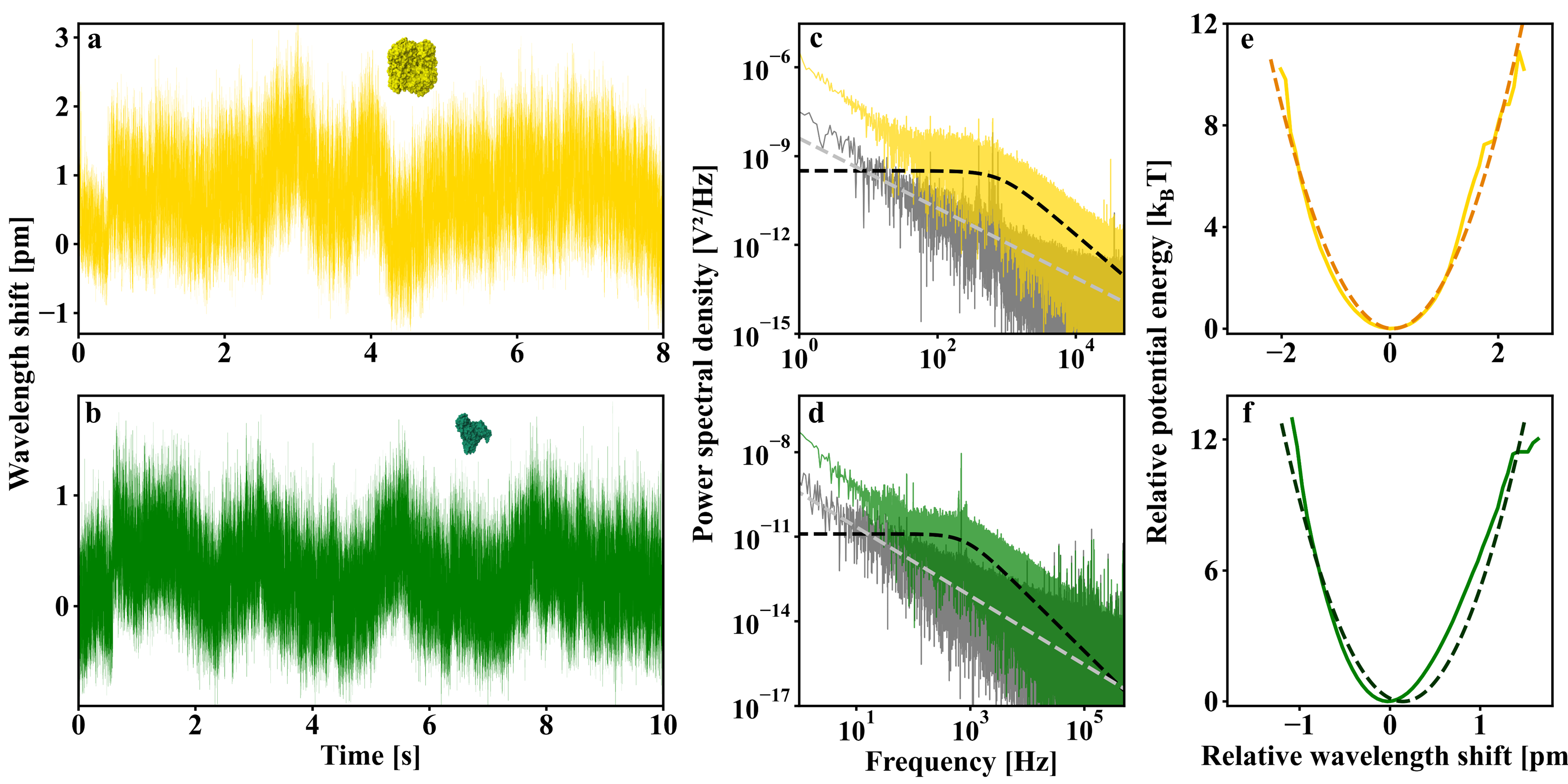} \\
    \caption{
        \textbf{Interprotein interactions in catalase and BSA.}
         Time traces of protein-protein interactions for (\textbf{a})~catalase (yellow) and (\textbf{b})~BSA (green). To ensure single protein detection, the protein concentration is maintained at low level ($\le$1\,nM).
         In \textbf{a}~and \textbf{b}, a 0.4\,Hz high-pass filter is applied for clarity. 
         \textbf{c-d},~The corresponding power spectral density. The grey traces represent recorded noise under the same settings but without proteins. The black dashed lines show Lorentzian fits, and grey dashed lines indicate linear fits to the background noises.
         \textbf{e-f},~The corresponding relative potential energy landscape. The dashed lines show the theoretical potential fits. A 10\,Hz high-pass filter is applied before constructing the energy landscape.
        }
    \label{fig:bindings2}
\end{figure*}

We observe loose binding of both catalase and BSA, each exhibiting a single potential well (Fig.~\ref{fig:bindings2}).

\newpage
\subsection*{E. Measurement bandwidth}

\begin{figure*}[ht!]
    \centering
    \includegraphics[width=\textwidth]{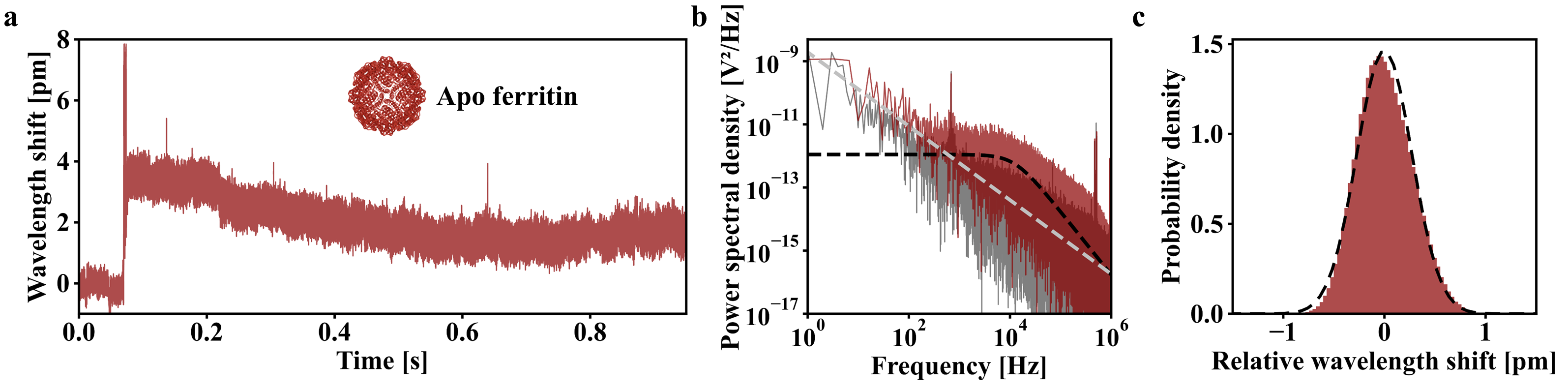}
    \caption{
    \textbf{Interprotein interactions of apo ferritin.} 
    \textbf{a}, Time trace of protein-protein interactions for apo-ferritin at 150\,mM NaCl.
    \textbf{b},~The corresponding power spectral densities, with the black dashed line representing Lorentzian fits. The gray traces indicate background noise, and the gray dashed line represents a linear fit to the noise. 
    \textbf{c},~The corresponding histograms of the wavelength shifts, with the black dashed lines representing theoretical fits.
    A 10\,Hz high-pass filter is applied before generating histograms. 
    }
    \label{fig:bandwidth}
\end{figure*}

This subsection demonstrates that the measurement bandwidth of our sensor reaches 1\,MHz (\ref{fig:bandwidth}b).

\newpage
\subsection*{Appendix: Energy in the cavity}

Unlike loosely bound proteins, optically trapped proteins exhibit wavelength shift distributions that strongly depend on the energy in the cavity. In this appendix, we describe how the energy within the slot is estimated to convert the simulated optical field distribution into the corresponding physical field distribution.

The Heisenberg equation of motion for the optical mode is  \cite{eqmotion}
\begin{align}
\dot{\hat{a}} &= -i \Delta \hat{a} - {\kappa \over 2} \hat{a} + \sqrt{\kappa_{\mathrm{in}}} \hat{a}_{\mathrm{in}} 
\end{align}
where $\hat{a}$ is the intracavity field, $\Delta = \omega_l - \omega_0$ is the detuning between the laser angular frequency $\omega_l$ and the cavity resonance angular frequency $\omega_0$, $\kappa = \omega_0 / Q$ is the total cavity decay rate, $\kappa_{\mathrm{in}} = \kappa \eta_c$ is the input coupling rate, with the coupling efficiency $\eta_c$, and $\hat{a}_{\mathrm{in}}$ is the input field. Since a protein moves slower than the dynamics of the cavity, we assume the steady state of the optical field, $\dot{\hat{a}} = 0 $. The number of photons in the cavity is then given by
\begin{align}
\hat{n} &= |\hat{a}|^2 \nonumber \\
&= {\kappa_{\mathrm{in}} \hat{a}_{\mathrm{in}}^2 \over ({\kappa \over 2})^2 + \Delta^2} \, .
\end{align}
With the laser power $P_{\mathrm{in}} = \hbar \omega_l a_{\mathrm{in}}^2 $, the energy in the cavity is given by
\begin{align}
\hat{H} &= \hbar \omega_l \hat{n} \nonumber \\
&= {\kappa_{\mathrm{in}} \over ({\kappa \over 2})^2 + \Delta^2} P_{\mathrm{in}} \, .
\end{align}

\indent
When the coupling efficiency is $\eta_c = 0.10$, the quality factor is $Q \approx 3000$, the resonance frequency is $f_0 \approx 200$\,THz (1500\,nm), the laser power is $P_{\mathrm{in}}$ = 5-10\,mW, and the detuning of the laser $\Delta$ is about -46.6\,GHz (0.35\,nm), the energy stored in the cavity is estimated to be $H=2.42$\,fJ. The electric field intensity is normalized using a normalization constant $c_e=H/H_s$, where $H_s$ =231\,fJ is the simulated energy in the cavity. In our experiment setup, the field normalization constant is $c_e \approx$ 0.007-0.014. 

\newpage
\section*{IV. Memory dependent friction}
Protein structural dynamics can induce fluctuations in the wavelength shift through changes in the protein polarizability. If the observed signal is dominated by the structural fluctuations, it is expected to exhibit memory effects~\cite{multiScience, powerlawmemory, memory1}, a characteristic feature of protein internal dynamics but not of Brownian center of mass motion. In this section, we detail various diffusion regimes in signals dominated by confined Brownian motion and in signals arising from protein structural fluctuations. 

\subsection*{A. Memory effects}

When protein excess polarizability fluctuates around equilibrium under restoring stiffness $k_s$, the dynamics can then be described by the generalized Langevin equation with a power law memory kernel $K(t-t')$~\cite{multiScience, powerlawmemory, memory1}
\begin{align*}
m_{\alpha}\ddot{\alpha}(t) &= -\gamma \int^t_0 K(t-t') \dot{\alpha}(t')dt' - k_s \alpha(t) + \xi(t)
\end{align*}
where $m_{\alpha}$ is the effective mass. Similarly to the memoryless case, because the sampling resolution is much longer than the inertial relaxation time, the system becomes overdamped
\begin{align*}
\gamma \int^t_0 K(t-t') \dot{\alpha}(t')dt'  &= - k_s \alpha(t) + \xi(t)
\end{align*}
The autocorrelation function is then given by~\cite{memory1}
\begin{align*}
C_{a}(\tau) &={k_B T \over k_s} E_a (- (\omega_m \tau)^a), \quad \omega_{m} = \Big({k_s \over \gamma \Gamma(3-a)} \Big)^{1/a}
\end{align*}
where $E_{a}(z) = \sum^{\infty}_{i=0}(z^i / \Gamma(a i+1))$ is the Mittag-Leffler function~\cite{mittag} and $\omega_m$ is the crossover frequency. 
To investigate memory dependent behavior, we analyze the time-averaged mean squared displacement, tMSD($\tau$) $= \langle (\alpha(t+\tau) - \alpha(t))^2 \rangle_t$, as it is less sensitive to noise due to temporal averaging. The corresponding tMSD is given by
\begin{align*}
\mathrm{tMSD}(\tau) &= {2 k_B T \over k_s} (1 - E_{a} (- (\omega_m \tau)^a) )
\end{align*}
For protein structural dynamics ($0<a<1$), the tMSD is expected to initially exhibit $\tau^{a}$ growth, followed by a power law approach to a plateau for $\tau \gg \tau_m$, where $\tau_m=1/\omega_m$ is the crossover time. In contrast, for confined Brownian motion ($a=1$), the tMSD is given by
\begin{align*}
\mathrm{tMSD}(\tau) &= {2 k_B T \over k_s} (1 - \mathrm{e}^{- \omega_c \tau} ) \, .
\end{align*}
In this case, the tMSD grows linearly with $\tau$ and then approaches a plateau exponentially.

\newpage
\subsection*{B. Diffusion regimes}

\subsubsection*{a. Confined Brownian motion}

\begin{figure*}[ht!]
    \centering

    \begin{minipage}[c]{0.42\textwidth}
        \centering
        \includegraphics[width=\linewidth]{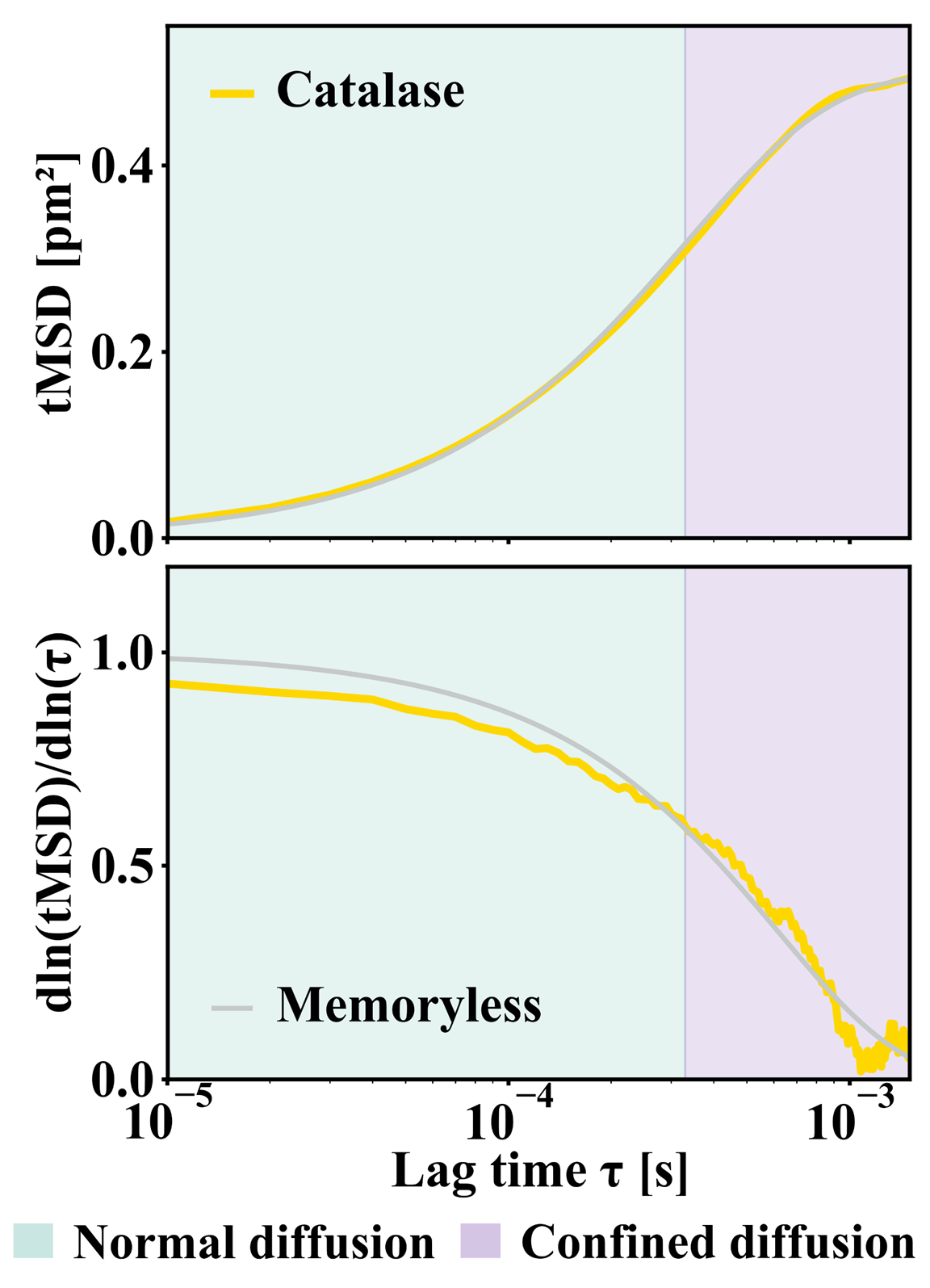}
    \end{minipage}
    \hfill
    \begin{minipage}[c]{0.52\textwidth}
        \caption{
        \textbf{Diffusion regimes of confined Brownian motion dynamics.}
        \textbf{a}, Time-averaged mean squared displacement of catalase. 
        \textbf{a}, The corresponding time dependent exponent $a_{\tau} (\tau)$. The silver curves represent fits to memoryless dynamics. The pastel mint and purple regions represent regimes dominated by normal diffusion and confined diffusion, respectively.
        }
        \label{fig:bMSD}
    \end{minipage}

\end{figure*}

To identify the diffusion regime at different time scales, we examine the time dependent exponent~\cite{crossover2}
\begin{align*}
a_{\tau}(\tau) = {d \ln \mathrm{tMSD} \over d \ln \tau} \, .
\end{align*}
For signals dominated by the center of mass motion of a loosely bound protein, two principal diffusion regimes are expected: normal diffusion ($a_{\tau} = 1$) at short timescales, followed by confined diffusion ($a_{\tau}$ approaches zero exponentially) once the protein experiences confinement due to interprotein interactions. Catalase exhibits a tMSD dominated by confined Brownian motion, as $a_{\tau}(\tau)$ is initially close to 1 ($a_{\tau} \approx 0.93$) and then decays exponentially toward zero (Fig.~\ref{fig:bMSD}) beyond the corner time $\tau_c = \ 1/\omega_c$. This behavior suggests that the signal fluctuations are dominated by the Brownian motion of catalase within the soft layer.

\clearpage
\subsubsection*{b. Protein structural fluctuations}

\begin{figure*}[ht!]
    \centering
    \includegraphics[width=\textwidth]{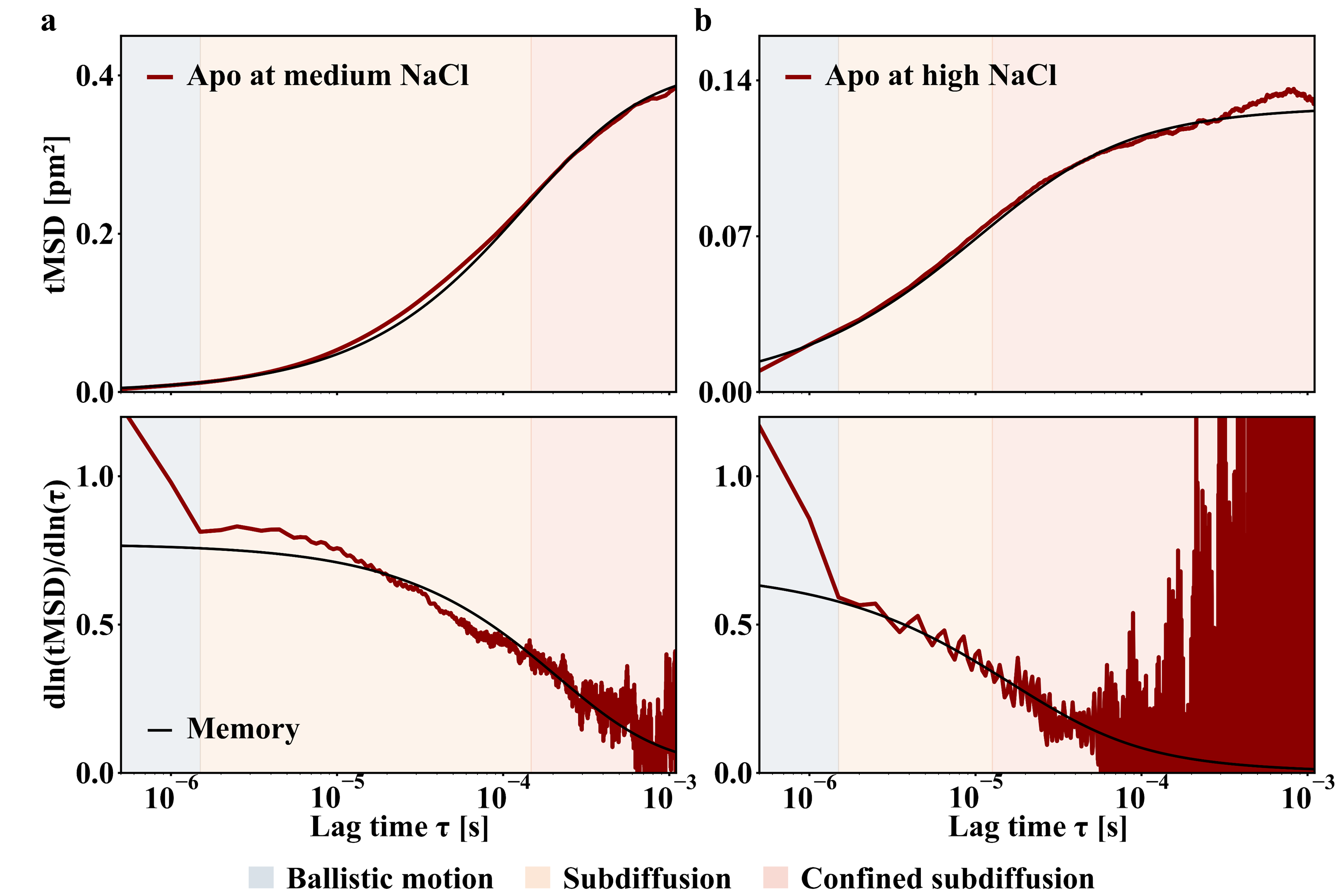}
    \caption{
    \textbf{Diffusion regimes of protein structural fluctuations.}
    Time-averaged mean squared displacements of apoferritin at (\textbf{a}) medium (10\,mM) and (\textbf{a}) high (150\,mM) NaCl concentrations, with the corresponding time dependent exponents. The black curves represent fits to memory dominated dynamics. The pastel gray, orange, and pink regions represent regimes dominated by ballistic motion, subdiffusion, and confined subdiffusion, respectively.
    }
    \label{fig:aMSD}
\end{figure*}

When signals are dominated by protein structural fluctuations, memory dependent friction leads to subdiffusion ($0<a_{\tau}<1$), followed by confined subdiffusion ($a_{\tau}$ approaches zero as a power law). Apoferritin measured at both medium (10\,mM) and high (150\,mM) NaCl exhibits subdiffusion, forming temporal plateaus below zero ($a_{\tau} = 0.77$ and $a_{\tau} = 0.68$, respectively), followed by a power law decay toward zero (Fig.~\ref{fig:aMSD}). This memory dependent behavior becomes dominant after the short time ballistic regime, during which inertia dominates over memory friction~\cite{crossover2}. Power law memory in protein structural fluctuations is an effective approximation to the multiexponential decay arising from multiple conformational states~\cite{multiScience}, with each exponential component characterized by a distinct memory time. Beyond the longest memory time, a crossover from a memory dependent to a memoryless regime is expected~\cite{crossover1, crossover2}. For apoferritin, however, this crossover is not observed within 1\,ms under either the medium or high NaCl concentrations.

\begin{figure*}[ht!]
    \centering
    \includegraphics[width=\textwidth]{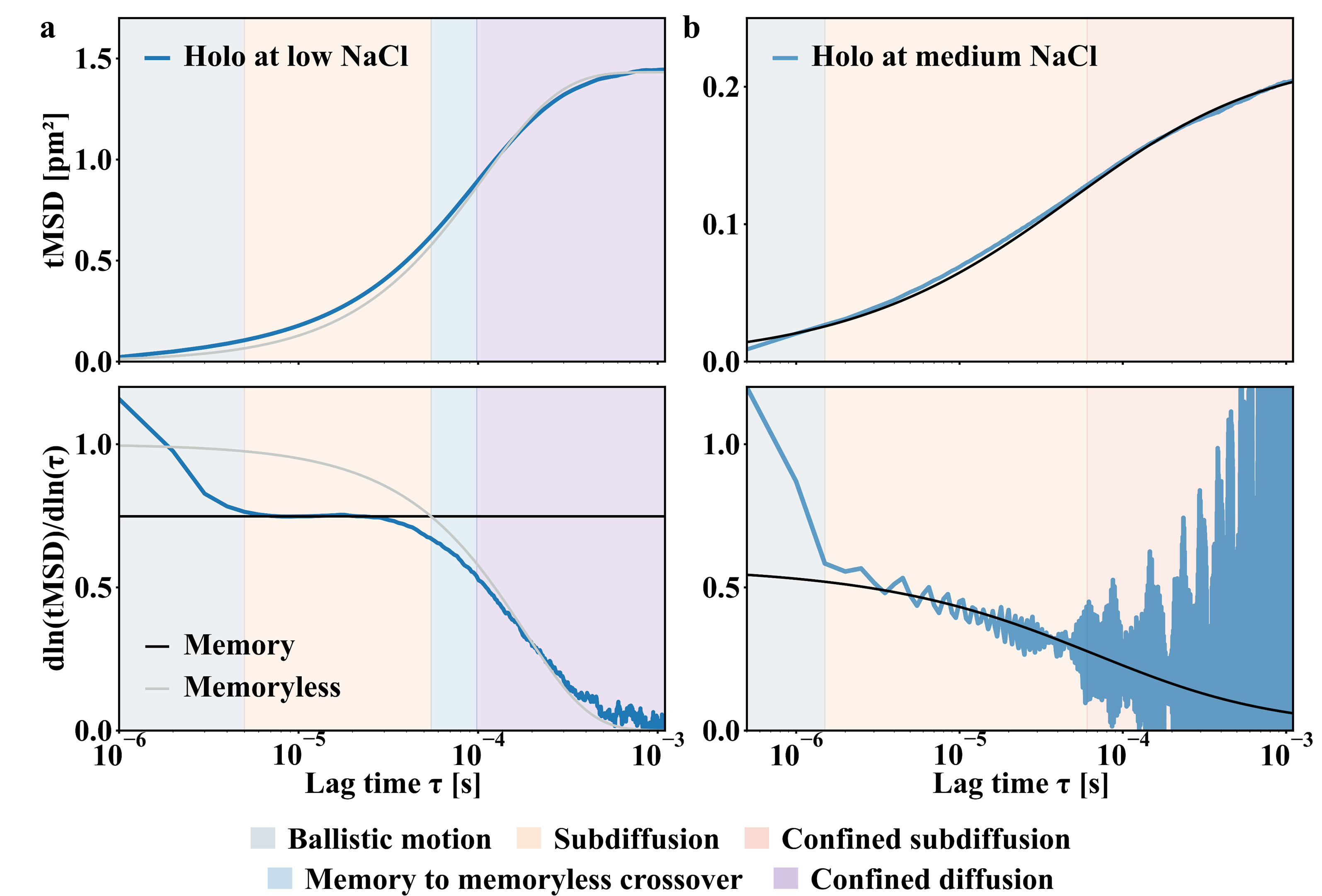}
    \caption{
    \textbf{Effects of weakened internal interactions on the memory dominated regime.}
    Time-averaged mean squared displacements of holoferritin at (\textbf{a}) low (4\,mM) and (\textbf{a}) medium (10\,mM) NaCl concentrations, with the corresponding time dependent exponents. The silver curves represent fits to memoryless dynamics. 
    The pastel gray, orange, pink, blue, and purple regions represent regimes dominated by ballistic motion, subdiffusion, confined subdiffusion, memory to memoryless crossover, and confined diffusion, respectively.
    }
    \label{fig:wMSD}
\end{figure*}

Similar behavior is observed with holoferritin at medium NaCl, subdiffusive motion ($a_{\tau} = 0.57$) followed by a power law decay up to 1\,ms (Fig.~\ref{fig:wMSD}b). By contrast, holoferritin at low (4\,mM) NaCl exhibits distinct behavior (Fig.~\ref{fig:wMSD}a). After the ballistic regime, it likewise exhibits subdiffusion ($a_{\tau} = 0.75$), but in the long time regime, it undergoes an exponential decay towards zero. This behavior suggests a crossover from a memory dependent to a memoryless regime. Because we observe a shorter memory time at lower NaCl concentrations, we attribute this reduction in memory time to weakened internal interactions caused by enhanced electrostatic repulsion. This is consistent with the reduced memory time resulting from weakened salt-bridge interactions at lower pH~\cite{phmemory}. We estimate the memory to memoryless crossover time to be $54.7 \pm 0.3$\,$\mu$s, obtained from the intersection of extended temporal plateau with the memoryless model.

\subsection*{C. Power spectral density for Markovian-like transition}

When the memory dominant region is shortened by weakened internal interactions, the transitions can be treated as approximately Markovian, provided that the dwell time is sufficiently long for the system to lose memory of its previous state~\cite{markovianP}. When protein structural fluctuations dominate the protein signal, the wavelength shift can be approximated as $\delta \lambda (t) \approx \beta' \alpha (t)$ where $\beta' = \pi c E_{\mathrm{phys}}^2 / (\omega \int_V \epsilon(\vec{r}) |\vec{E}^0(\vec{r})|^2 dV)$. For a two polarizability states system,  the autocorrelation function is given by 
\begin{align}
C_{\alpha}(\tau)= \beta'^2\sigma_{\alpha}^2 \mathrm{e}^{- \omega_{\alpha} |\tau|}
\end{align}
where $\sigma_{\alpha}^2$ is the variance of the polarizability and $\omega_{\alpha}$ is the relaxation rate of two state polarizability.
The corresponding PSD is given by 
\begin{align}
S_{\alpha}(\omega) 
&= \int^{\infty}_{-\infty} C_{\alpha}(\tau)\mathrm{e}^{-i\omega \tau} d\tau \nonumber \\
&= \beta'^2 \sigma^2_{\alpha}{2\omega_{\alpha}  \over \omega^2 + \omega_{\alpha}^2} \, .
\end{align}
The PSD of internal protein dynamics also exhibit a Lorentzian shape.

\clearpage
\newpage
\section*{V. Within and between state dynamics}

\subsection*{A. Hidden Markov model}

Unlike catalase and BSA, ferritin exhibits a double well wavelength shift distribution at low salt concentration. To separate the two states, we employ a Gaussian hidden Markov model (HMM). It is effective at distinguishing two states because it uses the full sequence of signal transitions, rather than a single value~\cite{HMMmodel}. The HMM identifies the parameters that maximize the likelihood function P$(O\,|\,\mathbf{T}, \boldsymbol{\pi}, \mathbf{e})$, where $O$ is the observed wavelength shift time trace, $\mathbf{T}$ is the transition probability matrix between the two hidden states, $\boldsymbol{\pi}$ is the vector of initial state probabilities, and $\mathbf{e}=\{(\mu_i, \sigma^2_i) \}^2_{i=1}$ are the emission parameters, where $\mu_i$ and $\sigma^2_i$ are the mean and the variance of the Gaussian emission distribution for each state $i$.

\subsection*{B. Diffusion regimes in each state}

\begin{figure*}[ht!]
    \centering
    \includegraphics[width=\textwidth]{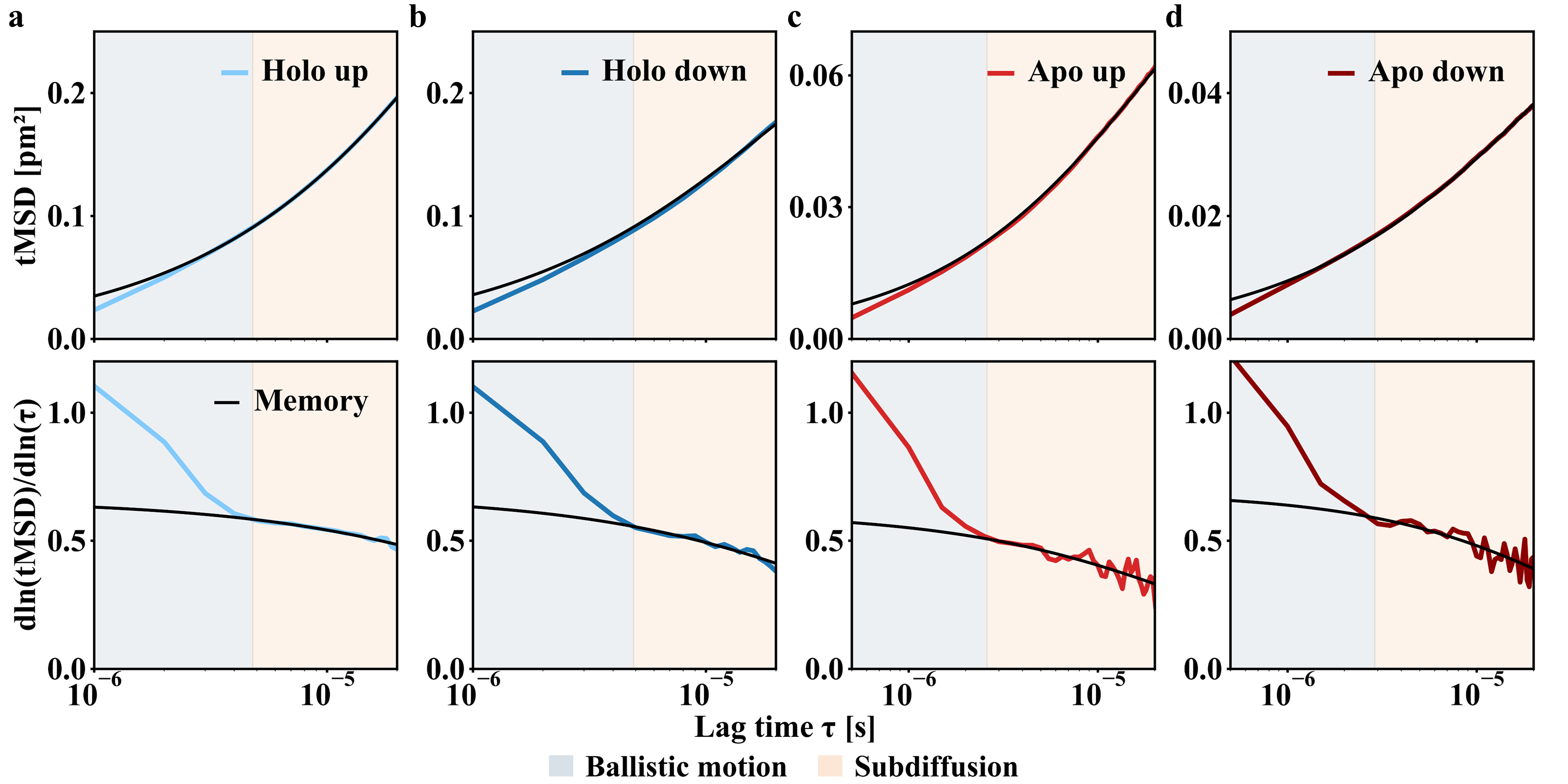}
    \caption{
    \textbf{Evaluating diffusion regimes in each state.}
    Time-averaged mean squared displacements of apoferritin at (\textbf{a}) medium (10\,mM) and (\textbf{a}) high (150\,mM) NaCl concentrations, with the corresponding time dependent exponents. The black curves represent fits to memory dominated dynamics. The pastel gray and orange regions represent regimes dominated by ballistic motion and subdiffusion, respectively.
    }
    \label{fig:sMSD}
\end{figure*} 

We first characterize the diffusion regimes using the tMSDs for each state. Similar to the results from the analysis of the full measurement traces, we observe a short time ballistic regimes followed by subdiffusion (Fig.~\ref{fig:sMSD}). For the up and down states of holoferritin, $a_{\tau} = 0.66$ and $a_{\tau} = 0.68$, respectively. For the up and down states of apoferritin, $a_{\tau} = 0.69$ and $a_{\tau} = 0.61$, respectively. Due to the relatively short trajectories within each state, only the ballistic and subdiffusion regimes are observed. We estimate the timescales of baliistic regimes from the intersection between the memory dominant model and the data. The ballistic regimes of the up and down states remain up to $4.8\pm 1.7$\,$\mu$s and $4.9\pm 0.2$\,$\mu$s, for holoferritin, respectively, and $2.6\pm 0.1$\,$\mu$s and $2.88\pm 0.04$\,$\mu$s for apoferritin, respectively.

\newpage
\subsection*{C. Dwell time}

We next analyze the dwell times in each state, which reflect the kinetic stability of the corresponding free-energy wells~\cite{Kramer}. Moreover, the dwell time distributions exhibit differ depending on whether the dynamics are dominated by memory effects~\cite{memoryless}. For a memory-dependent non-Markovian process, such as subdiffusion, the dwell time $t_d$ distributions can be described by a Mittag-Leffler distribution~\cite{memoryless} 
\begin{align}
p_d(t_d) = {1 \over \tau_m^a}t_d^{a-1} E_{a, a} (-(t_d/\tau_m)^a) , \quad E_{a, a}(z) = \sum^{\infty}_{i=0} {z^i \over \Gamma(a i + a)}
\end{align}
where $\tau_m$ is the characteristic dwell time. In long time limit, $p_d(t_d) \propto t_d^{-(1+a)}$, resulting in a power law tail. By contrast, for a memory-less Markovian process, the dwell time $t_d$ follows an exponential distribution~\cite{memoryless}
\begin{align}
p_d(t_d) &= {1 \over \tau_d}\mathrm{e}^{-t_d/\tau_d}
\end{align}
where $\tau_d$ is the mean dwell time.

\subsection*{C. Transition path time}

To investigate conformational transition dynamics, we examine transition path times, defined as the time spent along a trajectories spent crossing from one state to the next state. To calculate the transition path time, we evaluate the instantaneous slope at each state transition point. Using the average wavelength shift on both sides of the transition, we estimate each transition path time for each event.

In a double well potential with the high energy barrier ($\Delta U \gg k_{\mathrm{B}}T$), transition path time distributions are typically derived under the assumption that recrossing the initial position are highly unlikely~\cite{tptmodel}. However, the energy barriers of holo- and apoferritin are low ($\approx k_{\mathrm{B}} T$), so we instead estimate the transition path time distribution using an eigenfunction expansion~\cite{tptmodel2}, with the eigenvalue problem solved numerically by a finite difference method. 
\begin{align}
p_{\mathrm{tpt}}(t) = {k \over \gamma_a}t^{a-1}E_{a,a}({k \over \gamma_a}t^a){\sum^N_{n=1} Y'_n(-x_0)Y'_n(x_0) E_{a, 1}({k \over \gamma_a}t^a)^{-r_n-1} \over \sum^N_{n=1} (Y'_n(-x_0)Y'_n(x_0))/r_n}
\end{align}
where $\gamma_a = \gamma \Gamma(3-a)$, $r_n = -\lambda_n k_{\mathrm{B}}T/k$, $Y_n(x)$ is the $n$-th eigenfunction, and $\lambda_n$ is the $n$-th eigenvalue. The eigenfunctions and eigenvalues are obtained by solving the position dependent differential equation
\begin{align}
{d^2 Y_n (x) \over dx^2} - \Big({k \over 2 k_{\mathrm{B}} T} + \Big({k x \over 2 k_{\mathrm{B}} T}\Big)^2 \Big) Y_n(x) = \lambda_nY_n(x)
\end{align}
with the absorbing boundary conditions $Y_n(-x_0)=Y_n(x_0) = 0$. We estimate $k$ and $x_0$ from the measured double well potentials.

\begin{figure*}[ht!]
    \centering
    \includegraphics[width=0.5\textwidth]{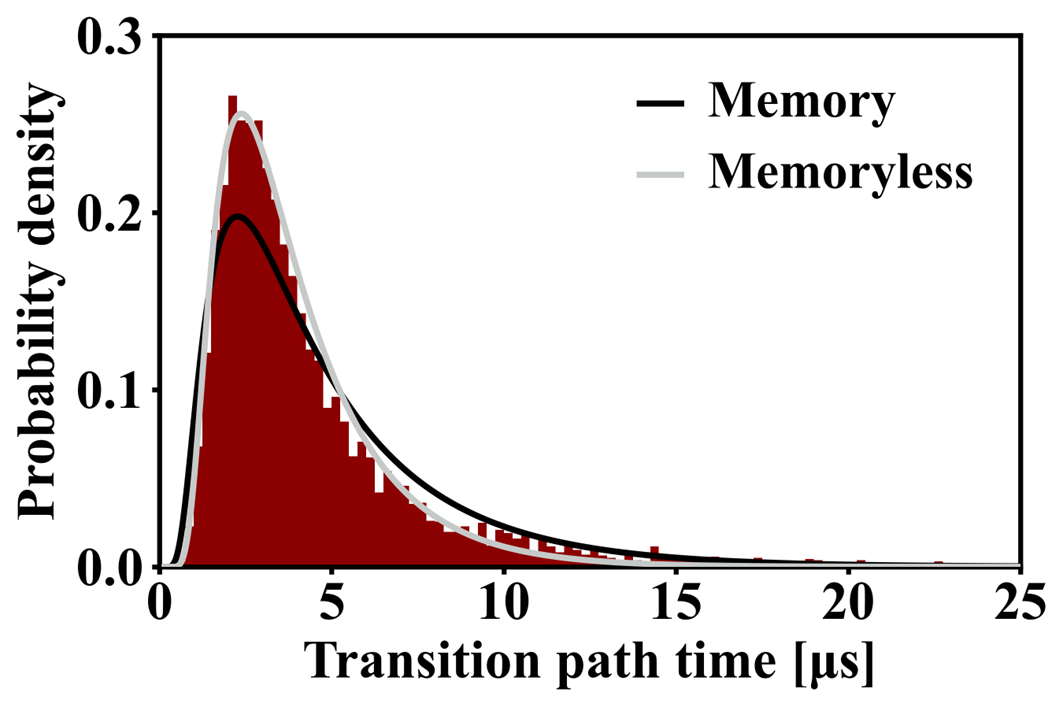}
    \caption{
    \textbf{Memory effects in transition path time distributions.} 
    Transition path time distribution for downward transitions of apo-ferritin at 10\,mM.
    The black and silver curves represent fits to memory dependent ($a=0.7$) and memoryless models, respectively.
    }
    \label{fig:tpt}
\end{figure*}

We find that the transition path time distributions exhibit memoryless behavior (Fig.~\ref{fig:tpt}), consistent with previous observations of protein dynamics~\cite{tptmeasure}. One possible explanation is that inertia effects dominate over memory friction at short timescales (Section IV.B.b), whereas the rapid increase in the barrier crossing velocity suppresses memory effects in long timescales~\cite{effTPT}.

\clearpage
\subsection*{F. Impact of fast motion bandwidth on resolving fast dynamics}

\begin{figure*}[ht!]
    \centering
    \includegraphics[width=0.5\textwidth]{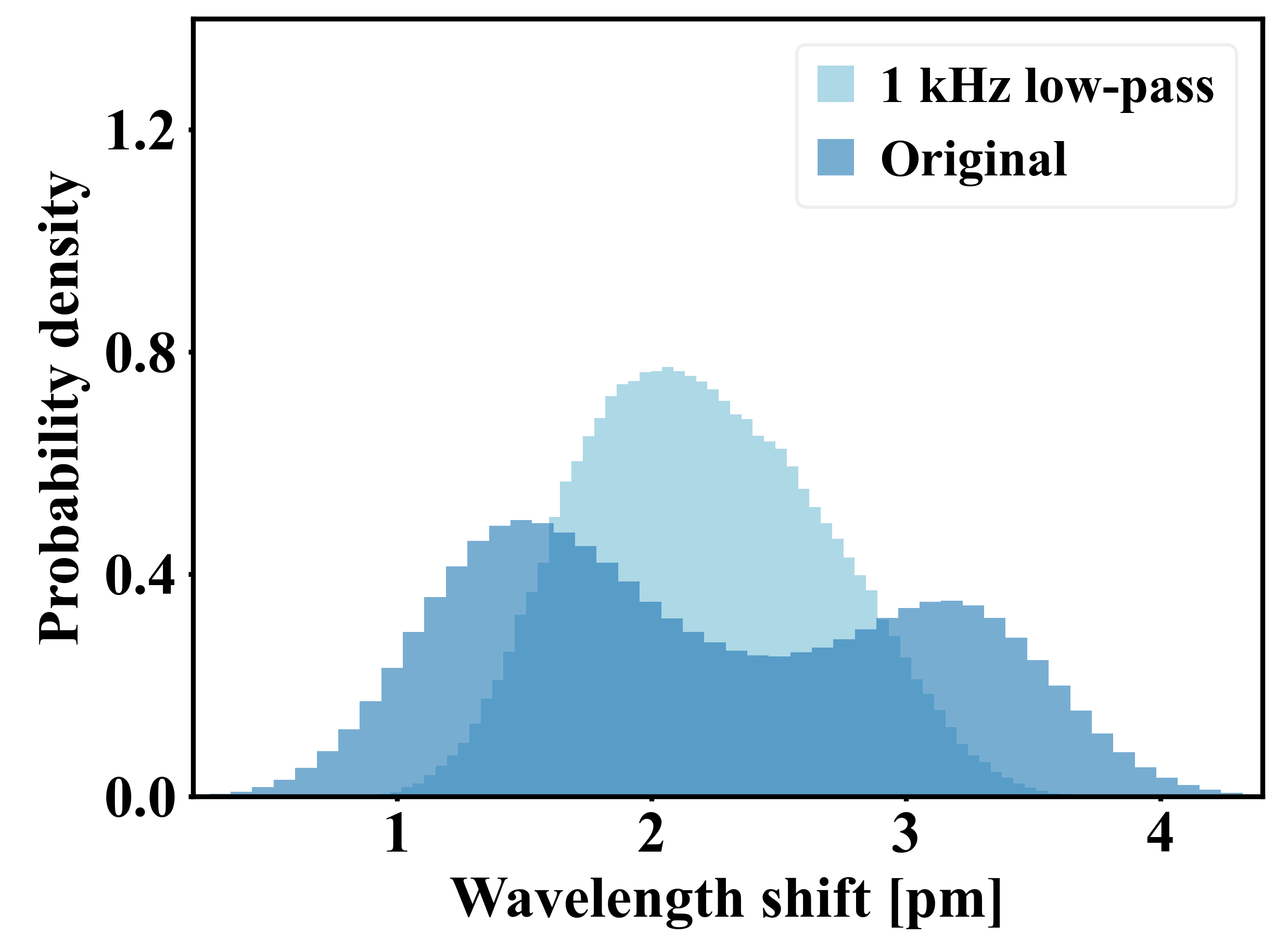}
    \caption{
    \textbf{Effect of fast motion bandwidth on resolving fast dynamics.} 
    Comparison of histograms before and after 1\,kHz low-pass filtering. A 10\,Hz high-pass filter is applied in both cases.
    }
    \label{fig:1kHzdis}
\end{figure*} 

Applying a 1\,kHz low-pass filter, corresponding to the effective bandwidth of the double nanohole measurements of conformational fluctuations~\cite{ferritinPH2}, suppresses fast motion and prevents the two states from being resolved (Fig.~\ref{fig:1kHzdis}).

\clearpage
\newpage
\section*{VI. Additional data}

\subsection*{A. Additional soft layer protein datasets}

This subsection presents the measured data for the soft layer. 
We observe stepping dynamics in holoferritin at 4\,mM NaCl and apoferritin at 10\,mM NaCl.

\subsubsection*{a. Catalase and BSA}
\begin{figure*}[ht!]
    \centering
    \includegraphics[width=\textwidth]{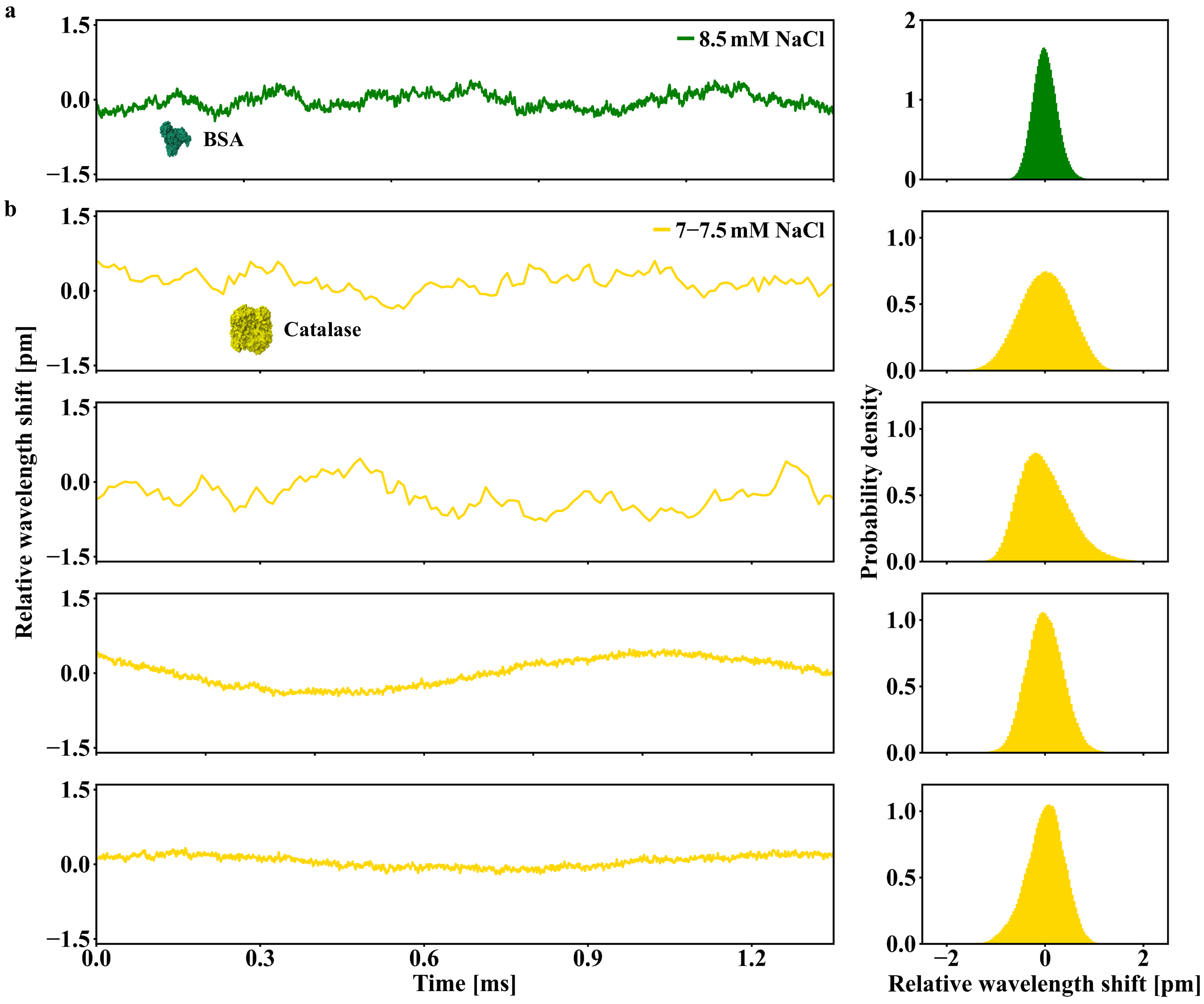}
    \caption{
    \textbf{Time-domain traces of catalase-catalase and BSA-BSA interactions.}  
    Short timescale views of a bound (\textbf{a})~catalase and (\textbf{b})~BSA, with the corresponding the probability densities.
    A 10\,Hz high-pass filter is applied for clarity.
    }
    \label{fig:catalaseBSA}
\end{figure*} 

\subsubsection*{b. Apoferritin}

\begin{figure*}[ht!]
    \centering
    \includegraphics[width=\textwidth]{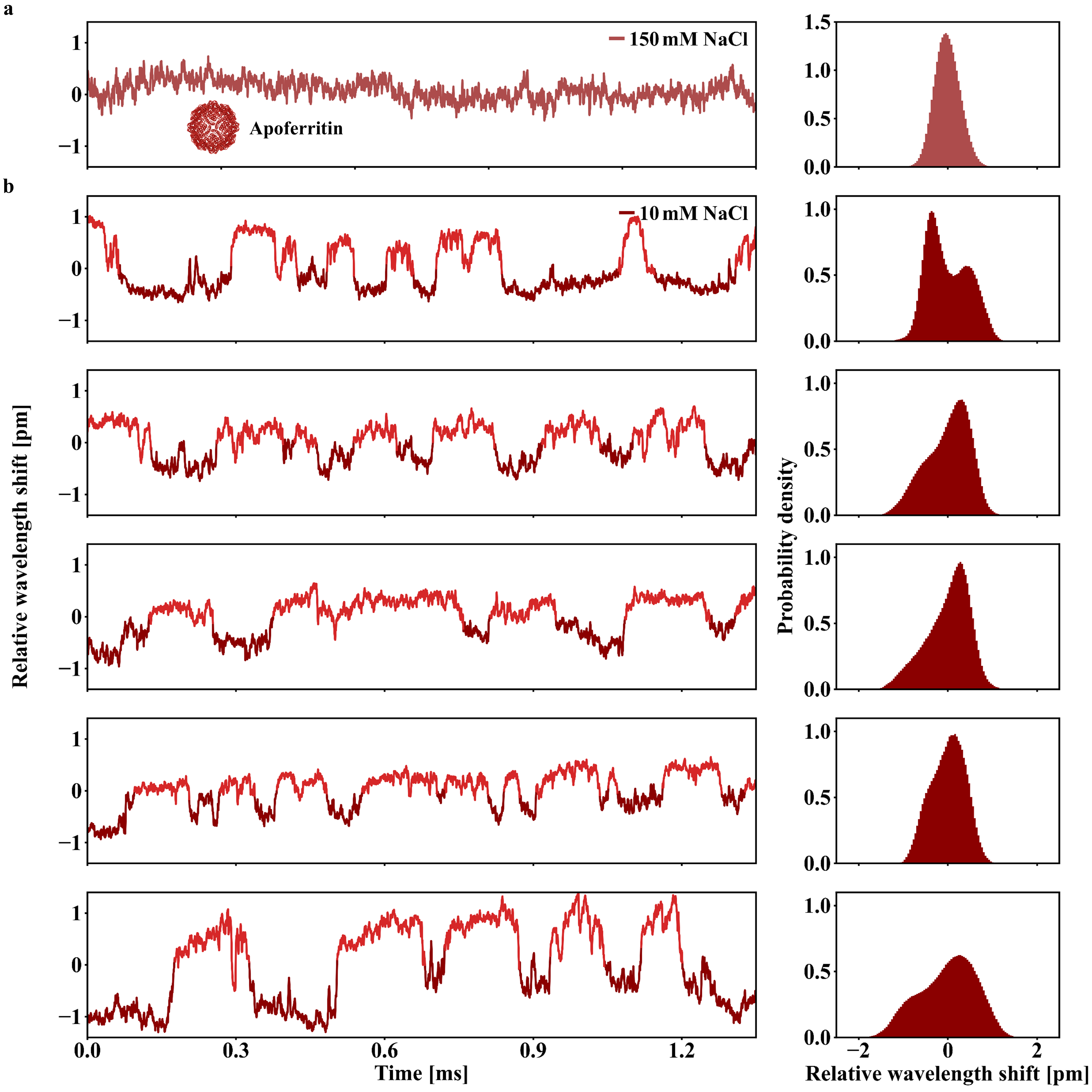}
    \caption{
    \textbf{Time traces of apo-ferritin-apo-ferritin interactions.}  
    Short timescale views of a bound apo-ferritin with the corresponding the probability densities at (\textbf{a})~150\,mM NaCl and (\textbf{b})~10\,mM NaCl.
    A 10\,Hz high-pass filter is applied for clarity.
    }
    \label{fig:apo}
\end{figure*} 

\newpage
\subsubsection*{c. Holoferritin}

\begin{figure*}[ht!]
    \centering
    \includegraphics[width=\textwidth]{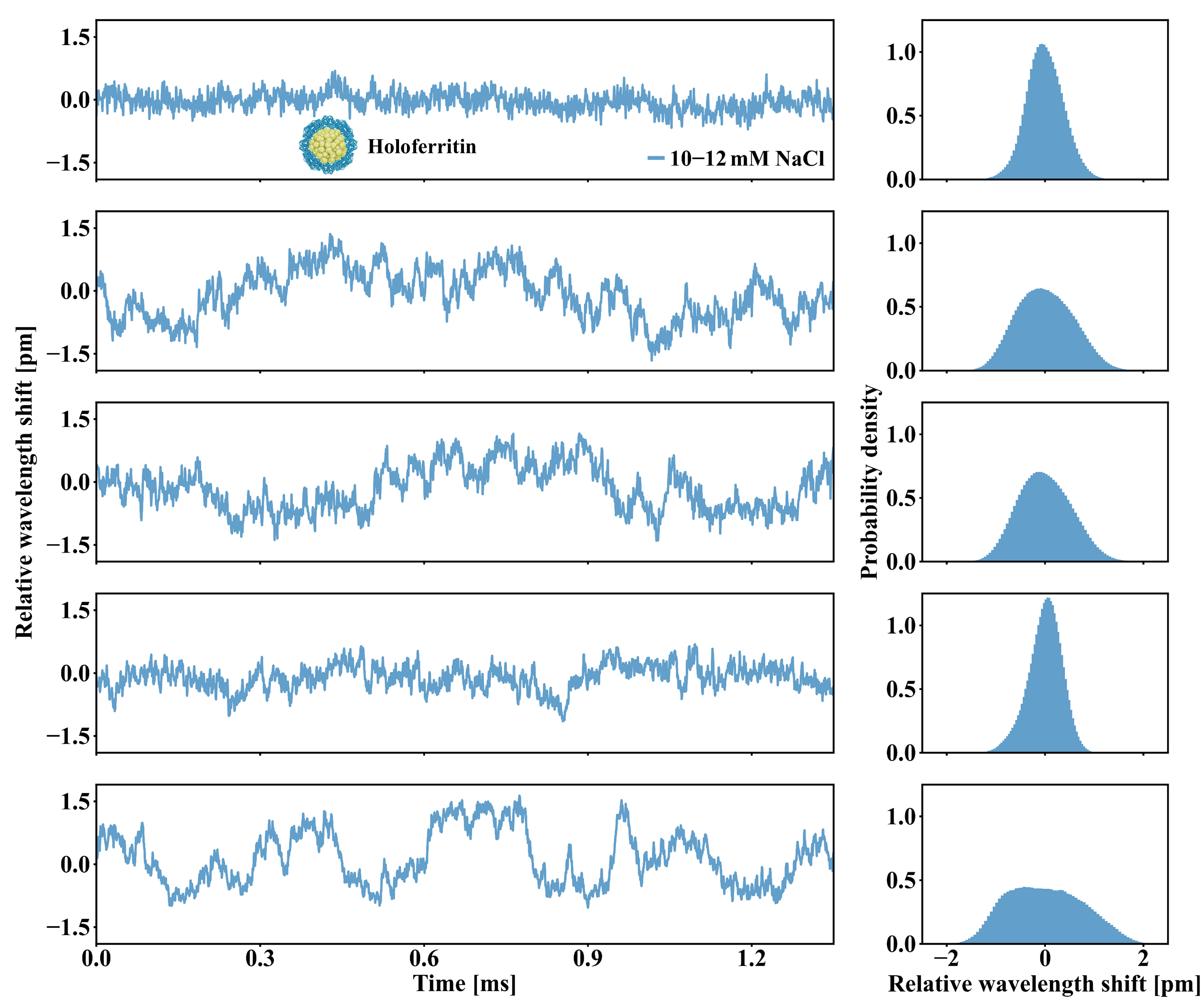}
    \caption{
    \textbf{Time traces of holo-ferritin-holo-ferritin interaction.}  
    Short timescale views of a bound holo-ferritin with the corresponding the probability densities.
    A 10\,Hz high-pass filter is applied for clarity.
    }
    \label{fig:holo}
\end{figure*} 

We observe holoferritins that may undergo stepping in 10\,mM NaCl (Fig.~\ref{fig:holo}). Based on mass photometry measurements, holoferritin contains heterogeneous amounts of iron (see Section VI.A). Some holoferritins exhibit molecular weights similar to those of apoferritin, suggesting that they contain very little iron. These nearly empty holoferritin may be relatively unstable compared to holoferritin containing sufficient iron.

\begin{figure*}[ht!]
    \centering
    \includegraphics[width=\textwidth]{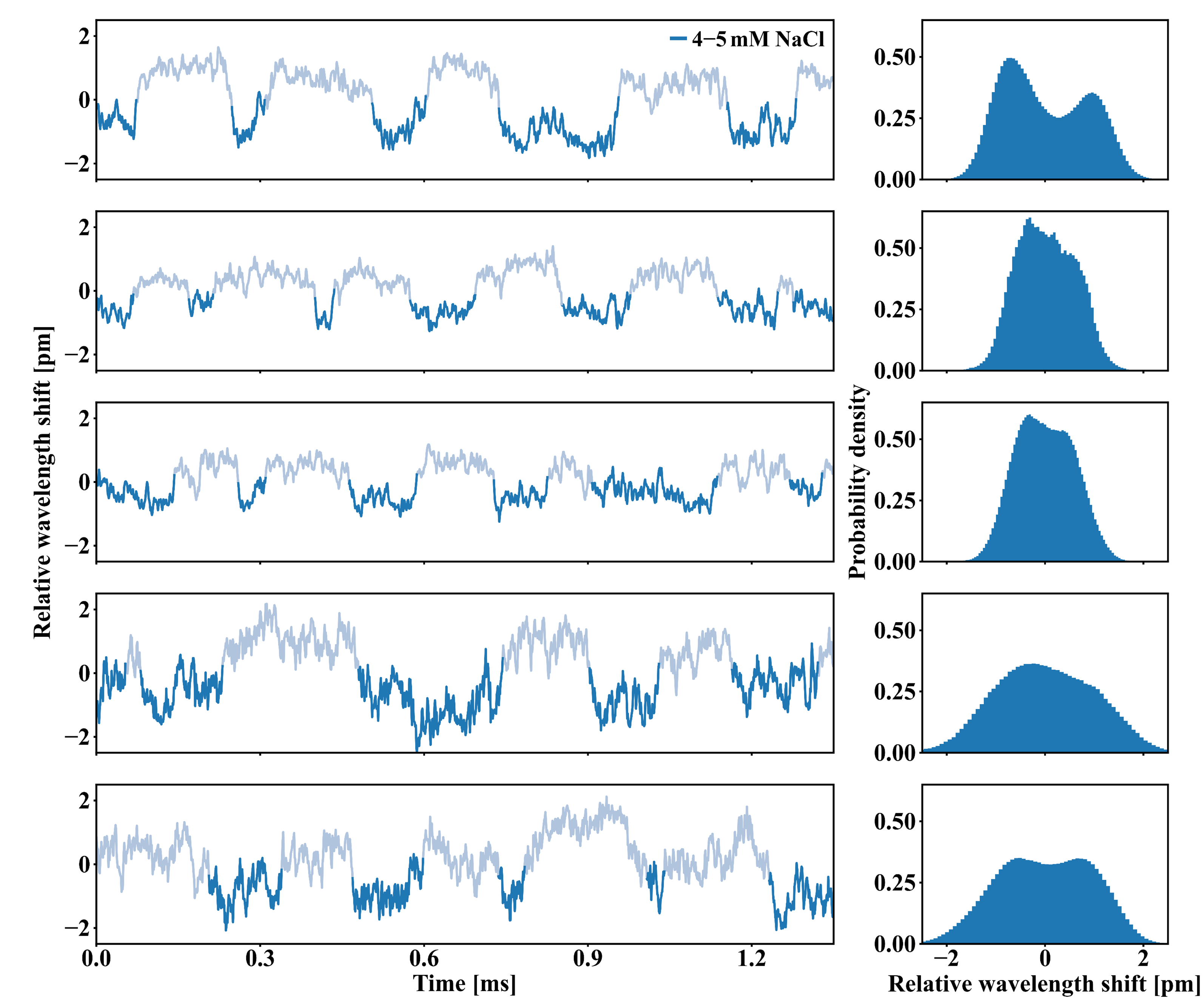}
    \caption{
    \textbf{Time traces of holoferritin-holoferritin interaction at lower salt concentration.}  
    Short timescale views of a bound holo ferritin with the corresponding the probability densities at lower salt concentration.
    A 10\,Hz high-pass filter is applied for clarity.
    }
    \label{fig:holoL}
\end{figure*} 

\clearpage
\newpage
\subsection*{B. Mass photometry}
\begin{figure*}[ht!]
    \centering
    \includegraphics[width=\textwidth]{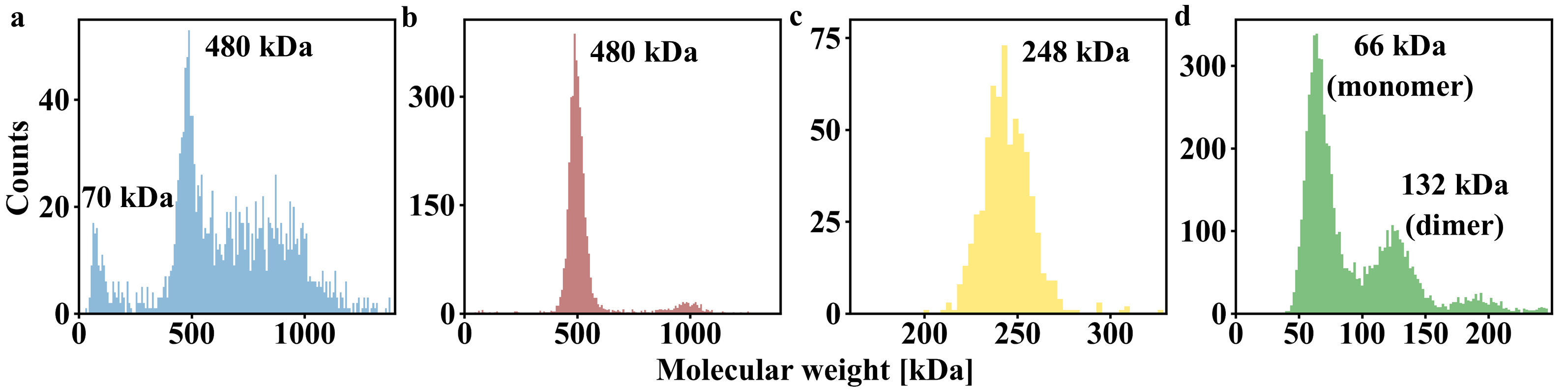}
    \caption{
    \textbf{Verifying the species present in the buffer using mass photometry.}
    Mass photometry results for (\textbf{a})~holoferritin, (\textbf{b})~apoferritin, 
    (\textbf{c})~catalase, and (\textbf{d})~BSA.
    }
    \label{fig:mass}
\end{figure*}

To evaluate the composition of the buffer solution, including the presence of proteins and potential impurities, mass photometry is employed. Mass photometry data for holo- and apo-ferritin, catalase, and BSA are acquired in Refeyn OneMP (Refeyn Ltd). Briefly, 15\,$\mu$L of 7-10\,mM NaCl in D${}_2$O is loaded on silicon gasket wells to acquire the background ratiometric contrast. The protein sample, 5\,$\mu$L of a 20\,nM protein in 7-10\,mM NaCl in D${}_2$O is added and mixed well. The data are analysed using Refeyn DiscoverMP software. A 3-point calibration is performed using protein standards with known masses of 66, 132 and 198\,kDa. 

As shown in Fig.~\ref{fig:mass}a-d, we confirm that the molecular weights of holo- and apoferritin, catalase, and BSA are 480, 248, and 66\,kDa, respectively, from the peak in the mass photometry results. As shown in Fig.~\ref{fig:mass}a, holoferritin shows a distinct fragment around 70\,kDa and a broad peak that varies with iron (ferrihydrite) content. In contrast, apoferritin shows a single peak around 480\,kDa, with a minor population corresponding to the dimerization of two apoferritins near 1000\,kDa. In our BSA solution, approximately 20$\%$ of the BSA is present as a dimer, as shown in Fig~\ref{fig:mass}c.

Catalase (C3556) and bovine serum albumin (BSA, A7030) were purchased from Sigma-Aldrich and stored according to the manufacturer's instructions. Catalase from human erythrocytes was supplied in a Tris-buffered solution and used without further purification. BSA was supplied as a lyophilized powder. Lyophilized BSA was reconstituted in heavy water containing 10\,mM NaCl prior to each experiment.

\newpage
\section*{References}
\bibliography{reference}